\documentclass[fleqn,usenatbib]{mnras}

\usepackage{mathptmx}

\usepackage[caption = false]{subfig}
\usepackage{pdflscape}	
\usepackage{longtable}	

\usepackage[T1]{fontenc}

\usepackage{comment}

\DeclareRobustCommand{\VAN}[3]{#2}
\let\VANthebibliography\thebibliography
\def\thebibliography{\DeclareRobustCommand{\VAN}[3]{##3}\VANthebibliography}

\usepackage{graphicx}	
\usepackage{amsmath}	
\usepackage{amssymb}	
\usepackage{xcolor}

\newcommand{\MK}{MeerKAT }
\renewcommand{\deg}{^{\circ}} 

\defcitealias{Johnston2019}{JK19}

\title[The TPA pulsar population census]{The Thousand-Pulsar-Array program on MeerKAT -- IX. The time-averaged properties of the observed pulsar population}

\author[B. Posselt et al.]{B. Posselt$^{1,2}$\thanks{E-mail: bettina.posselt@physics.ox.ac.uk},
A. Karastergiou\,$^{1,3}$,
S. Johnston\,$^{4}$, 
A.~Parthasarathy\,$^{5}$,
L. S. Oswald\,$^{1,6}$,\newauthor
R.~A.~Main\,$^{5}$
A.~Basu\,$^{7}$
M.~J.~Keith\,$^{7}$,
X. Song\,$^{7}$, 
P. Weltevrede\,$^{7}$,
C. Tiburzi\,$^{8}$,
M.~Bailes\,$^{9,10}$,\newauthor
S.~Buchner\,$^{11}$,
M.~Geyer\,$^{12,13}$,
M.~Kramer\,$^{5,7}$,
R.~Spiewak\,$^{7,9,10}$,
V.~Venkatraman~Krishnan\,$^{5}$
\\
{$^{1}$ Department of Astrophysics, University of Oxford, Denys Wilkinson Building, Keble Road, Oxford OX1 3RH, UK}\\
{$^{2}$ Department of Astronomy \& Astrophysics, Pennsylvania State University, 525 Davey Lab, 16802 University Park, PA, USA}\\
{$^{3}$ Department of Physics and Electronics, Rhodes University, PO Box 94, Grahamstown 6140, South Africa}\\
{$^{4}$ Australia Telescope National Facility, CSIRO Space and Astronomy, PO~Box~76, Epping NSW~1710, Australia}\\
{$^{5}$ Max-Planck-Institut f\"{u}r Radioastronomie, Auf dem H\"{u}gel 69, D-53121 Bonn, Germany}\\
{$^{6}$ Magdalen College, University of Oxford, Oxford OX1 4AU, UK}\\
{$^{7}$ Jodrell Bank Centre for Astrophysics, Department of Physics and Astronomy, University of Manchester, Manchester M13 9PL, UK}\\
{$^{8}$ INAF, Osservatorio Astronomico di Cagliari, Via della Scienza 5, 09047 Selargius, Italy}\\
{$^{9}$ Centre for Astrophysics and Supercomputing, Swinburne
University of Technology,
PO Box 218, Hawthorn, 
 Vic, Australia, 3122}\\
{$^{10}$ ARC Centre
of Excellence for Gravitational Wave Discovery (OzGrav)}\\
{$^{11}$ South African Radio Astronomy Observatory, Cape Town, South Africa}\\
{$^{12}$ South African Radio Astronomy Observatory, 2 Fir Street, Black River Park, Observatory 7925, South Africa}\\
{$^{13}$ Department of Astronomy, University of Cape Town, Rondebosch, Cape Town, 7700, South Africa}
\\
}

\date{Accepted XXX. Received YYY; in original form ZZZ}

\pubyear{2022}

\begin{document}
\label{firstpage}
\pagerange{\pageref{firstpage}--\pageref{lastpage}}
\maketitle

\begin{abstract}
We present the largest single survey to date of average profiles of radio pulsars, observed and processed using the same telescope and data reduction software. Specifically, we present measurements for 1170 pulsars, observed by the Thousand Pulsar Array (TPA) programme at the 64-dish SARAO MeerKAT radio telescope, in a frequency band from 856 to 1712 MHz. We provide rotation measures (RM), dispersion measures, flux densities and polarization properties. The catalogue includes 254 new RMs that substantially increase the total number of known pulsar RMs. Our integration times typically span over 1000 individual rotations per source. We show that the radio (pseudo)luminosity has a strong, shallow dependence on the spin-down energy, proportional to $\dot{E}^{0.15\pm0.04}$, that contradicts some previous proposals of population synthesis studies. In addition, we find a significant correlation between the steepness of the observed flux density spectra and $\dot{E}$, and correlations of the fractional linear polarization with $\dot{E}$, the spectral index, and the pulse width, which we discuss in the context of what is known about pulsar radio emission and how pulsars evolve with time. 
On the whole, we do not see significant correlations with the estimated surface magnetic field strength, and the correlations with $\dot{E}$ are much stronger than those with the characteristic age.
This finding lends support to the suggestion that magnetic dipole braking may not be the dominant factor for the evolution of pulsar rotation over the lifetimes of pulsars. A public data release of the high-fidelity time-averaged pulse profiles in full polarization accompanies our catalogue.\end{abstract}



\section{Introduction}
\label{intro}
The observed neutron star population shows astonishing diversity. On the one hand, 
neutron star properties are commonly assumed to be governed by a limited number of physical parameters such as their rotational properties, their magnetic field configuration, their internal structure, and perhaps their environment. 
On the other hand, even if one restricts to the population of non-recycled radio pulsars, there is a large variation in observed quantities such as the range of the spin period $P$ and period derivative $\dot{P}$, the characteristics of flux density spectra, temporal features such as the pulse profile, its components and polarization, and phenomena on single-pulse timescales and shorter.
The large parameter space of measured observables is a challenge and at the same time an opportunity to constrain the physics of neutron stars.\\

There are, it seems, three ways in which the field has made progress. 
First, through looking at individual sources that exhibit some phenomenon in an extreme way in order to try and reduce the dimensionality of the problem. Examples are PSR~J0738$-$4042 where the polarization changes with time in a way that proves we are observing the incoherent sum of orthogonally polarized modes \citep{0738kar}, and J1906+0746 where general relativistic precession allows mapping of the radio beam \citep{1906des}. 
Secondly, through monitoring specific populations of neutron stars with common features (such as $\gamma$-ray emission, or frequent glitching) and attempting overarching explanations (e.g., \citealt{Lower2021,radiogamma}).
Finally, by measuring the observables of a large population in order to identify unifying and distinguishing properties with the goal of understanding how sub-classes of neutron stars relate to each other or evolve from one another.
Examples of interesting connections revealed in that way are the observation that the angle between rotation and magnetic axes decreases on long time scales, i.e. with age \citep{Tauris1998}, or the realization that the radio luminosity must decrease with age \citep{Arzoumanian2002}.\\

We emphasize that the true age is actually a very difficult quantity to determine for the vast majority of pulsars, and typical age estimates are usually based on specific model assumptions.
Associating particular population phenomena to true ages would allow one, for example, to constrain various mechanisms for pulsar spin-down (magnetic braking, winds, glitches, fallback disks, alignment of spin and magnetic axes, etc.) and their relative strengths at different times. In contrast to the true age, the (current) loss rate of kinetic energy, $\dot{E}$ (or spin-down power) is well constrained because the instantaneous spin-down of pulsars with time can be measured with high precision.
It is determined as $\dot{E} = 4\pi^2 I \dot{P} P^{-3}$ where $I$ is the moment of inertia of the pulsar (generally taken to be $10^{45}$\,g\,cm$^2$), $P$ its spin period and $\dot{P}$ its spin-down rate. In population studies, $\dot{E}$ is thus not only an important parameter of the available kinetic energy, but it is also used as proxy for the age.
The amplitude of and time between glitches are related to $\dot{E}$ (e.g., \citealt{Lower2021}), and single pulse phenomena such as nulling and drifting also have an $\dot{E}$ dependence \citep{Song2022} ({\emph{submitted}}). A relationship between the degree of linear polarization and $\dot{E}$ has also been established (e.g., \citealt{Weltevrede2008b}).\\ 

Pulsar radio emission is influenced by many other physical properties. 
The observed integrated pulse profiles are highly stable, indicating that they can be used to probe the global properties of the pulsar magnetosphere.
It is widely accepted that the radio emitting regions are confined to the open polar cap region inside the
light-cylinder radius \citep{Ruderman1975}, and at a height of a few hundred kilometers. 
The canonical pulsar is assumed to be a centered, rotating, inclined magnetic dipole. Some recent models also consider multipoles or offset dipoles (e.g., \citealt{Petri2019,Petri2016}), leading to potentially very different interpretations of the observables. 
Based on the pulse profiles, different topologies of the emission beams have been introduced for quantification and modeling.  
Whilst \citet{Lyne1988} prefer a patchy, random emission beam, \citet{Rankin1990,Rankin1993} proposed a structured beam with inner ``core'' and outer ``cone'' emission components.  
The beam maps from \citet{1906des} for pulsar J1906+0746 showed a patchy or fan-like beam structure that is inconsistent with a cone beam though.
Variations in the combination of radio beam topologies and sightline effects complicate the seemingly simple picture of the canonical pulsar.\\ 

The polarization information in pulse profiles can be used to understand the geometry of the pulsar, typically with estimates based on the ``rotating vector model'' (RVM) by \citet{Radhakrishnan1969}. The RVM can be used to fit polarization data in up to 50\% of pulsars where this is attempted (e.g., Johnston et al. (\emph{in prep.}), \citealt{1906des,Rookyard2015, Johnston2006,Everett2001}).
In order to explain typical polarization behaviour in the pulse profiles, two co-existent orthogonally polarized modes (OPMs; \citealt{Backer1976}) were proposed. The radio emission is then described by a superposition of the two OPMs, with the overall degree of linear polarization depending on the relative contribution of each OPM at a specific pulse phase (e.g., \citealt{Stinebring1984}).
Considering relativistic effects, \citet{Blaskiewicz1991} illustrated that the height
of the radio emission can be constrained as well. In general, the so-called radius-to-frequency mapping assumes that radio emission from lower frequencies arises from higher up in the magnetosphere than the emission at higher frequencies (e.g., \citealt{Komesaroff1970,Cordes1978}). 
Other models, however, suggest emission at the same frequency from different heights due to different beam structures (e.g., in the form of a fan beam, \citealt{Wang2014,Gupta2003}).
Despite decades of study, there remain many open questions regarding the radio emission process, the pulsar magnetosphere and the neutron star properties in general. 
A robust assessment of the validity of the proposed models requires a large, homogeneous pulsar survey that is sensitive enough to enable the necessary polarization and flux measurements even for the fainter members of the rotation-powered pulsar population.\\

Here, we report on the MeerTime Thousand Pulsar Array (TPA) programme \citep{Johnston2020} targeting nearly half of the non-recyled, ``slow'' (rotation periods above 30\,ms) pulsar population with the MeerKAT telescope. The data originate from a single telescope with very similar 
observing parameters, and the measurements presented here result from a single data reduction pipeline, resulting in a uniform data set.
This removes some of the difficulties in using data from disparate surveys, as is often the case with measurements in the ATNF pulsar catalogue \citep{Manchester2005}.
The ultimate goal of the TPA is to accelerate the progress in the field on all three lines of investigation outlined above.  
This work focuses on the fundamental observables obtained from time-averaged pulse profiles in full polarization. We present a catalogue of measurements accompanying the public data release of the underlying processed TPA data for over 1200 pulsars.
We note that we do not differentiate between individual profile components in this work.
The TPA data is the largest homogeneous data set of the non-recyled pulsar population to date, enabling a large variety of statistially meaningful constraints on the radio emission properties of pulsars.
Here, we restrict to a few statistical example studies to demonstrate the capability of this new resource.\\

The paper is organized as follows. 
We give an overview of the observations and methods in Section~\ref{obssec}. In $\S$~\ref{sec:cor}, we outline how the example correlations are identified and investigated.
Section~\ref{results} presents the catalogue of the TPA census.
$\S$~\ref{sample} shows the representativeness of the TPA pulsar sample, 
$\S$~\ref{sec:dmrm} covers parameters related to the Interstellar Medium (ISM) such as the Faraday Rotation Measures (RMs) and Dispersion Measures (DMs), whilst 
$\S$~\ref{sec:fluxsi} and $\S$~\ref{sec:polOV} present parameters related to flux and polarization measurements.
Section~\ref{sec:discussion} discusses a few example correlations between pulsar parameters of the TPA census.
Correlations with the spectral index, luminosity and polarization fractions are discussed in  $\S$~\ref{sec:SI} to $\S$~\ref{sec:pol}. 
Section~\ref{sec:sum} is a brief summary, whilst Section~\ref{sec:thedata} describes where the data and the catalogue are available and how they can be accessed.

\section{Observations \& Methods}
\label{obssec}

\subsection{Observing programme}
The TPA pulsar sample encompasses the non-recycled pulsar population with a rotation period range from 30\,ms to 6\,s.
The survey sources  were chosen in 2019 to include as many pulsars as possible with declination lower than $25\deg$. Pulsars with
positional uncertainties greater than 2 arcsec were excluded.\\ 

\noindent The 64-dish SARAO MeerKAT (MK) radio telescope has carried out over 10,000 TPA pulsar observations since commissioning in February 2019. 
Here, we present pulsar observations with the L-band receiver that was centred at a frequency of $\nu=1284$\,MHz, for more details of pulsar observations with \MK  in general, see \citet{Bailes2020}.
The channelized time series were processed by the four Pulsar Timing User Supplied Equipment (PTUSE) machines. 
We use a total bandwidth of 775\,MHz.
Observing times for the pulsars were chosen based on pulse profile fidelity as discussed by \citet{Song2021}.
They inferred that the optimal observation lengths for the TPA pulsars should cover at least 200\,rotation periods for a $\sigma_{\rm shape}=0.1$ uncertainty of the shape parameter which measures the flux differences of two pulse profile bins of interest.
Since some of our targets did not have recorded 1.4\,GHz fluxes (e.g., because they were discovered at lower frequencies), the first TPA observations aimed for verification of the pulsar detection and establishing the total flux level at that time. Subsequent observations were obtained to (i) increase the overall signal-to-noise ratio, (ii) account for variability due to intrinsic pulsar variability or scintillation, and (iii) obtain the aimed-for $\sim 1000$ pulses for each target. The multiple-epoch observations for about 500 pulsars are also used to investigate time-dependent behaviours in single-pulse trains. 
Overall 1279 unique pulsars were observed between 2019 March 8 and 2021 November 3 (10,248 observations), partly during the commissioning phase of the MeerKAT telescope.
Here, we only consider fold-mode observations obtained with the full array ($>58$ out of the 64 antennas).\\ 

\noindent  
Pulsars can exhibit emission changes and the variability analysis will be the topic of a separate paper. Here, we restrict to one observation epoch for a pulsar, in the following the  TPA-census data set. The chosen epoch is typically the longest TPA observation covering most pulsar rotations ($\sim 1000$ rotations) except  for the rare cases where technical issues rendered the longest observation not useful. 

\subsection{Processing and calibration}
The pipeline to process the TPA data is described by \citet{Parthasarathy2021}, and it is available at \href{https://github.com/aparthas3112/meerpipe}{https://github.com/aparthas3112/meerpipe} together with the instructions on how to use it.
The effects of radio frequency interference (RFI) are removed within the pipeline following a modified procedure decribed by \citet{Lazarus2016}. \\
The resulting data have 1024 frequency channels, sub-integration times of 8\,s, and all 4 Stokes parameters.
We consider in the following  frequency-averaged (FA) profiles, eight frequency channels (8ch) for our analyses. 
The polarization calibration uses the procedures described in detail in \citet{Serylak2021}.

\subsubsection{Flux calibration}
Flux calibration was obtained via a bootstrap method. The radiometer equation can be used to compute the expected root mean square (rms) in a single channel via
\begin{equation}
\sigma_{c}= \frac{G \, (T_{\rm sys} + T_{\rm sky})}{N_{\rm A} \, \sqrt{2 \times  {\triangle\nu_{\rm bw}} \, N^{-1}_{\rm ch} \, t_{\rm obs} \,  N^{-1}_{\rm bin}}}
\label{equ:expsig}
\end{equation}
Here, $N_{\rm A}$ is the number of antennas, ${\triangle\nu_{\rm bw}}$ is the total bandwidth in MHz,  $N_{\rm ch}$ is the number of frequency channels,  $t_{\rm obs}$ is the observing time (corrected for RFI removal),  $N_{\rm bin}$ is the number of phase bins. We take $T_{\rm sys}$ to be $17.1$\,K (which includes small contributions from spillover and the atmosphere) and the gain of a single antenna is  $G = 19.0$\,Jy\,K$^{-1}$ at 1390\,MHz \citep{MKspeci}.
This leaves $T_{\rm sky}$ as an unknown but critical parameter as $T_{\rm sky}$ on the Galactic plane can often exceed the $T_{\rm sys}$ contribution. We initially used the sky map of \citet{Calabretta2014} to compute $T_{\rm sky}$ but the large disparity in the resolution of the \citet{Calabretta2014} data (14 arcmin) compared to the MeerKAT beam (6 arcsec) meant that the $T_{\rm sky}$ was systematically overestimated in most cases. We were therefore also forced to bootstrap $T_{\rm sky}$ directly from the data. We did this by observing a high latitude pulsar and assuming that $T_{\rm sky} = 3.37$~K at that location. By comparing the rms (in digitiser counts) for the high latitude pulsar with all other pulsars observed in that session we could obtain $T_{\rm sky}$ for all the pulsars. Armed with this knowledge we then again used the radiometer equation to convert digitiser counts into mJy.
The actual rms, $\sigma_{\rm obs}$, is obtained from the median value of the rms in 20 (RFI-free) channels centered at 1390\,MHz.
The data are then flux-calibrated by scaling them with the factor $ C_{F}=\sigma_{c}/\sigma_{\rm obs}$.\\

To assess the agreement with existent flux density measurements, we considered 261 TPA-pulsars that are covered by frequent Parkes observations.
Measuring fluxes in several frequency bands (8 for TPA) with the same method (see Section~\ref{profilemeas}) for a similar total frequency bandwidth, we determined fluxes at 1264\,MHz, $F_{1264}$, and  power law (PL) slopes by (unweighted) regression.
Without accounting for any poor measurements (e.g., due to scintillation or large flux uncertainties), the distributions of the PL slopes for the TPA and Parkes data are very similar, whilst the flux ratios, $F^{\rm TPA}_{1264}/F^{\rm Parkes}_{1264}$, showed a slight offset from 1 with a mean and standard deviation of $0.79 \pm 0.31$.
Restricting our comparison to those pulsars in both data sets that have spectral indices, i.e., PL indices with uncertainties less than 0.2 (i.e. are well measured) and agreement of their PL slopes within their $3\sigma$ uncertainties (i.e., no obvious scintillation effects or effects due to pointing offsets), the number of comparison pulsars is reduced to 48. Their flux ratio distribution has a mean and standard deviation of $0.90 \pm 0.14$. 
We checked for a dependency of the flux ratio on the flux at 1264\,MHz (or the logarithm of that flux), but found it to be statistiscally insignificant for the 48 comparison objects.
Overall, we regard the agreement of the pulsar fluxes from the TPA and Parkes data as good, especially since one needs to account for  systematic uncertainties, too. One such systematic, for example, may result from the position uncertainty. For the narrow MeerKAT beam ($\sim 6\arcsec$) a slight pointing offset from the actual position can reduce the measured flux and impact the derived PL index.

\subsection{Template generation, ephemerides, DMs and RMs}
\label{sec:eph}
The iterative process of updating ephemerides and DMs uses profile templates. 
Noise-free standard templates were generated automatically from the Gaussian-Process (GP) models described in Section~\ref{profilemeas}. The smoothed profiles were cut to the GP-defined on-pulse region of the profile   
or 2\% of the peak height (whichever is smaller). To avoid a sharp cut-off, the tails of the profile were then modelled by fitting a von-Mises function to the profile edges, resulting in a smooth, periodic, noise-free template. These templates were all individually inspected, and in a small fraction (13\%) of cases replaced with a manually generated template where the automatic template was judged to be of insufficient quality. This typically occurred in pulsars with long scattering tails or where interference signals contaminated the profile. Manually generated templates were created using \texttt{PSRCHIVE}, or in the case of heavily scattered profiles, using custom python code to fit the scattering tail.\\

The pulsar signal is maximized by refining the pulsar ephemeris from the ATNF catalog. 
During the work on the single-pulse observations of the TPA pulsars (\citealt{Song2022} {\emph{submitted}}), we noticed a few pulsars for which the listed ATNF rotation frequency was actually a harmonic with the real period being a factor two longer (PSRs J0711$+$0931, J0836$-$4233, J1427$-$4158, J1618$-$4723, J1639$-$4604, J1739$-$3951, J1802$+$0128, J1848$-$1150, J1946$-$1312), or a factor three (PSRs J0211$-$8159, J1714$-$1054) longer.\\ 

The determination of pulsars' dispersion measures (DMs) depends to some extent on the chosen templates, the frequency range considered, frequency-dependent profile evolution and scattering.
For the TPA data set, we use our profile templates and the respective frequency--time folded pulse profiles.
We derived new DM values and their respective uncertainties using {\sc tempo2} \citep{Hobbs2006tempo2,Edwards2006tempo2} and manual outlier rejection. 
Using the times of arrival (TOAs) from {\sc tempo2} for 32 frequencies, we also explored the systematic error range employing an independent fit algorithm with automatic outlier rejection. 
Using about 200 test pulsars, we obtained a median DM offset of 0.1 between the results from the two methods. We add the 0.1 as a systematic error estimate to our statistical error estimates for the DM. Some of our fits were formally not good, but the averaged frequency aligned pulse profiles look reasonable. 
We emphasize that DM values determined in this way (fitting TOAs for a $\nu^{-2}$ dependency) can be affected by frequency-dependent profile evolution and scattering, thus may be different if these effects are modeled at the same time.\\

We determine the rotation measures (RMs) for all TPA pulsars.
Three main methods are regularly used to determine RMs, typically based on polarization angles that are averaged over phase space. The RM synthesis, developed by \citet{Brentjens2005}, 
uses the Fourier-like relationship between the complex polarization intensity vector as a function of wavelength squared. The RM is obtained from    
maximizing the Fourier Transform on the polarization angle across the whole available frequency-squared space. 
The method by \citet{Noutsos2008} directly fits for the quadratic relationship between polarization angle and frequency using frequency bins. 
Another method is to maximize the S/N of the linear polarization of the average polarization angle across frequency. This is implemented in {\sc psrchive} - \texttt{rmfit}  \citep{vanStraten2012}.
Since the TPA pulsars cover a large range in S/N of the total intensity as well as of the linear polarization, we employ the {\sc psrchive} - \texttt{rmfit} method, maximizing the linear polarization, refined by a crude fit to the polarization angle. The searched RM ranges could encompass values from $-1000$ to 1000. Depending on the pulsar, searches were done with different sampling of the RM range.
We visually checked the significance and goodness of the RM fits and the resulting polarization profiles.
Since the application of the correct RM maximizes the linear polarization, the S/N of the linear polarization, $S/N_{LP}$\footnote{For the RM evaluation, only the statistical noise of the linear polarization is considered}, is directly related with the achievable accuracy of the RM estimate. Thus, we determine our RM uncertainties by the empirical formula $\sigma_{\rm RM} = 2 + 30 /  (S/N_{LP})$ after evaluation of our RM-fit results for pulsars with $S/N_{LP}>10$. 
We note that the choice of method can affect the RM measured, and so can the used DM value. \citet{Oswald2020} showed that DM and RM corrections are covariant when they measured both DM and RM in a consistent way. 

\subsection{Measurements on pulsar profiles}
\label{profilemeas}
We use the method presented by \citet{Johnston2019} and \citet{Brook2019} to derive a Gaussian Process (GP) for pulse profiles, using the Python GP-package \texttt{George}\footnote{\href{https://george.readthedocs.io}{https://george.readthedocs.io}} by \citet{georgepython}.
The GP allows us to determine a smooth noiseless profile, as well as the noise variances, $\sigma_{\rm GP}$, of the input (observed) pulse profiles. The general features and applicability of the GP are described by \citet{Roberts2012} and \citet{Rasmussen2006}.
The method was already described and used by \citet{Posselt2021} to measure pulse widths of the TPA pulsar sample.
For our flux measurements, we use $\sigma_{\rm GP}$ to define a signal to noise ratio (S/N) for each point in the observed data profile. Aiming to collect also the flux of faint wide pulse wings, we define the on-pulse as the region around the maximum of the profile with S/N$>1$ with $n_{\rm onP}$ bins. Outside of the on-pulse region, we double check whether the already performed baseline-subtraction needs corrections. We measure the (mean) flux density\footnote{Note that we abbreviate the flux density with "flux" for easier reading throughout the text.} by adding the flux bins in the on-pulse region of the observed data and dividing by the total number of bins in the profile, $n_{\rm allP}$. The flux uncertainty is calculated by scaling $\sigma_{\rm GP}$ with the on-pulse profile region (scale factor: $(n_{\rm onP})^{0.5} \sigma_{\rm GP} / n_{\rm allP}$). The fluxes measured in this way do not depend on a pulse-profile template, and thus, they are more reliable, in particular for strongly scattered pulsars or those with profile evolution. Since we chose to maximize the on-pulse region, our statistical flux uncertainties are very conservative estimates. For very few faint pulsars (e.g., for the frequency-averaged data two out of 1172 pulsars) this results in the significance of the flux measurement to be below $3\sigma$. We remove these pulsars from further analysis.\\  

Following \citet{Lorimer2012}, we estimate (pseudo) luminosities in the L-band as:
\begin{equation}  
\label{eq:lum} 
L_{\rm L} \approx 7.4 \times 10^{27} {\rm erg \, s}^{-1} \times \left(\frac{d}{{\rm kpc}}\right)^{2} \times \frac{F_t}{{\rm mJy}}, 
\end{equation} 
where $d$ is the distance and $F_t$ is the continuum-equivalent flux measured in the frequency-averaged total power profile. The factor $7.4 \times 10^{27}$\,erg\,s$^{-1}$ is an approximation for some average assumptions regarding the pulse duty cycle, beaming fraction, and a PL spectral index over a large range of radio frequencies. Note that we use mean fluxes for our luminosity estimates, i.e., the duty cycle is already taken into account.

Utilizing the flux measurements and their uncertainties, we further carry out weighted PL-fits and determine spectral indices, SI. Note that only $\sim 80$\% of the pulsar radio spectra can be described with PLs with the rest showing curved spectra, spectral breaks, or low-frequency turnovers (e.g., \citealt{Jankowski2018}). A detailed characterisation of all the TPA pulsar spectra is, however, beyond the scope of this paper. In order to gauge the goodness of the PL-fit, we define a pseudo modulation index, $m_{\rm SI}$, as:
\begin{equation}  
\label{eq:modSI1} 
m_{\rm SI} = \large( \sigma^2_{F_{\rm{SIres}}} - \overline{\sigma_{F}}^2 \large)^{0.5} \large( \overline{F_{\rm{SIres}}} \large)^{-1}, 
\end{equation} 
with
\begin{equation}  
\label{eq:modSI2} 
F_{\rm{SIres},i} = F_i-F^{\rm SI}_i + \overline{F^{\rm SI}},
\end{equation} 
where $F_{\rm{SIres},i}$ describes the residual flux in frequency channel $i$ with respect to the mean flux, $\overline{F^{\rm SI}}$, of all PL-fit flux values, and its standard deviation, $\sigma_{F_{\rm{SIres}}}$, considering all frequency channels. 
Here, $F_i$ and  $F^{\rm SI}_i$ are the measured and the fitted flux value in the respective frequency channel $i$.
$\overline{\sigma_{F}}$ is the average flux uncertainty of $F_i$ considering all frequency channels. It is used as a bias correction. If it is larger than $\sigma_{F_{\rm{SIres}}}$, only the latter is used in equation~\ref{eq:modSI1}. 
Large $m_{\rm SI}$ can either indicate insufficent PL-fits to the spectral shape, or the presence of relevant scintillation that affects the flux measurements.
Based on visual inspection of the PL-fits and visual screening of the data with the full frequency resolution (928 channels) for scintillation patterns, we decided to use $m_{\rm SI} <0.13$ as the criterion to select pulsars that have spectra well described by a PL.\\ 

We use the GP to derive noise variances and smooth noiseless pulse profiles for all Stokes parameters ($I, Q, U, V$), the linear polarization, $L$, and circular and (absolute) circular polarization, $V$ ($|V|$) for the frequency-averaged as well as the eight frequency channels where possible.  
Following \citet{Everett2001}, we remove the positive bias of the measured linear polarization, $L_{m,i}=\sqrt{Q_i^2+U_i^2}$ in pulse phase bin $i$ of the data by calculating the true value of the linear polarization $L_i$ as 
\begin{equation}
L_i = \left\{ 
\begin{array}{ll}
\sigma_{I} \sqrt{  (\frac{L_{m,i}}{\sigma_{I}})^2 - 1}  & \mbox{if $\frac{L_{m,i}}{\sigma_{I}} \geq 1.57$} \\
0 & \mbox{otherwise.}\end{array} 
\right.
\end{equation}
Here, $\sigma_{I}$ is the GP-based noise variance of the Stokes I profile.
Similarly, the measured absolute value of the circular polarization $V_m$, a half-normal distribution, is corrected for the bias of the folded normal distribution by its expectation value as
\begin{equation}
|V_i|=\left\{ 
\begin{array}{ll}
|V_{m,i}| - \sigma_{I} \sqrt{\frac{2}{\pi}} & \mbox{if $|V_{m,i}| > \sigma_{I} \sqrt{\frac{2}{\pi}}$} \\ 
0 & \mbox{otherwise.}\end{array} \right.
\end{equation}
For this, the noise distribution of $V$ is assumed to be Gaussian and its mean is assumed to be 0 since the data are baseline-subtracted.
The total polarizations ($L=\sum{L_i}$, $V=\sum{V_i}$, $|V|=\sum{|V_i|}$) are derived from the sum over pulse phases where the GP-I profile of the frequency-averaged data fulfills S/N$>3$. Polarization fractions (LPF$=L/I$, CPF$=V/I$, aCPF$=|V|/I$) are calculated with respect to sum in $I_i$ in the same (on-pulse) phase region of the data.
For the linear polarization, we list in addition to the de-biased continuum-added $L$ also the vector-added $L^*=\sqrt{(\sum{Q_i})^2+(\sum{U_i})^2}$. 
Listed uncertainties consider the (propagated) statistical errors to which we add a conservative estimate of 3 per cent for systematic effects, as done similarly by \citet{Serylak2021}.
While the above polarization measures employ directly the data, our GP-noiseless pulse profiles enable additional estimates that are not prone to large statistical fluctuations in individual phase bins. Using the available GP-noiseless pulse profiles of the linear and absolute circular polarization, the fraction of the pulse profile that is above a certain threshold can be easily obtained.   
We estimate the fractions of the pulse that have linear (circular) polarization values above 30\%, 60\%, or 90\% (10\%, 25\%, or 40\%). The pulse profile in $I$ with S/N$_{\rm bin} > 10$ defines the 100\% region for this purpose. The $1\sigma$ statistical unertainties of the polarization fractions are then considered in each bin, and new measurements are done of the fractions of the pulse that have linear (circular) polarization values above the mentioned percentages. These measurements are used to determine the uncertainty ranges for the threshold-based values.\\ 

Due to the updated DMs, some pulse widths changed with respect to the TPA sample in \citet{Posselt2021}. We follow the same pulse width measurement procedure for the TPA-census FA data set. The obtained $W_{50}$ and $W_{10}$ describe where the noiseless pulse profile is 50\% and 10\% of its peak value respectively.
Since \citet{Posselt2021} used combined data from several observations per pulsar, their measurements have usually a higher significance than those based on the TPA-census data. Thus, we only update the pulse width values if their difference with respect to the old values is more than $1\sigma$ \emph{and} the difference in DM is either larger than $5$\,cm$^{-3}$\,pc or larger than 1\% of the old DM value. This results in the (slight) revision of 15 $W_{50}$ and 10 $W_{10}$ values.
\citet{Song2021} derived a metric to judge the fidelity of the pulse profiles. Equipped with many new TPA measurements for $W_{50}$, mean flux densities, modulation indices from single pulse studies \citep{Song2022} ({\emph{submitted}}), and a few updated periods, we use formulas (7 and 8) by \citet{Song2021} to update the nominal optimal observing length with MK, $t_{\rm nom}$  to achieve $\sigma_{\rm shape} =0.1$ for each pulsar.  

\subsection{Correlation analysis}
\label{sec:cor}
We use the Spearman rank correlation coefficient and its $p$-value to non-parametically check for possible relations between measured TPA quantities (e.g., spectral indices, $W_{10}$, luminositites, polarization fractions) and other pulsar observables (e.g., $P$, $\dot{P}$, $\dot{E}$, $\tau$, $B$, distance $d$, Galactic longitude $Gl$ and latitude $Gb$, transverse spatial velocity $v_{\rm trans}$). 
Since $\dot{E}$, $\tau$, and $B$ can be expressed as functions of $P^a$, $\dot{P}^b$, we also calculated the 2D surface of the Spearman rank and $p$-values in the $a-b$ plane for the respective sample. The ranks or $p$-values are symmetric and vary very little along wedge-like sections of the $a-b$ plane. In addition, although high ranks and low $p$ values indicate correlations, they can depend on the sample selection and are difficult to assess in terms of uncertainties. For this reason, we use the Spearman rank analysis just as guidance for a correlation which we follow up with distribution plots and simple fits. For this paper, we use only linear ordinary least-square (OLS) fits. They are partly carried out in common logarithmic space (base 10). 
We use Bootstrap to estimate uncertainties. Fit results (and their uncertainties) for $\log {P}$, $\log \dot{P}$, $\log \dot{E}$, $\log \tau$, $\log {B}$ can be expressed via the $a,b$ parameters, allowing us to plot these results together in a single plot.

\section{Results}
\label{results}
\begin{figure}
\includegraphics[width=8.5cm]{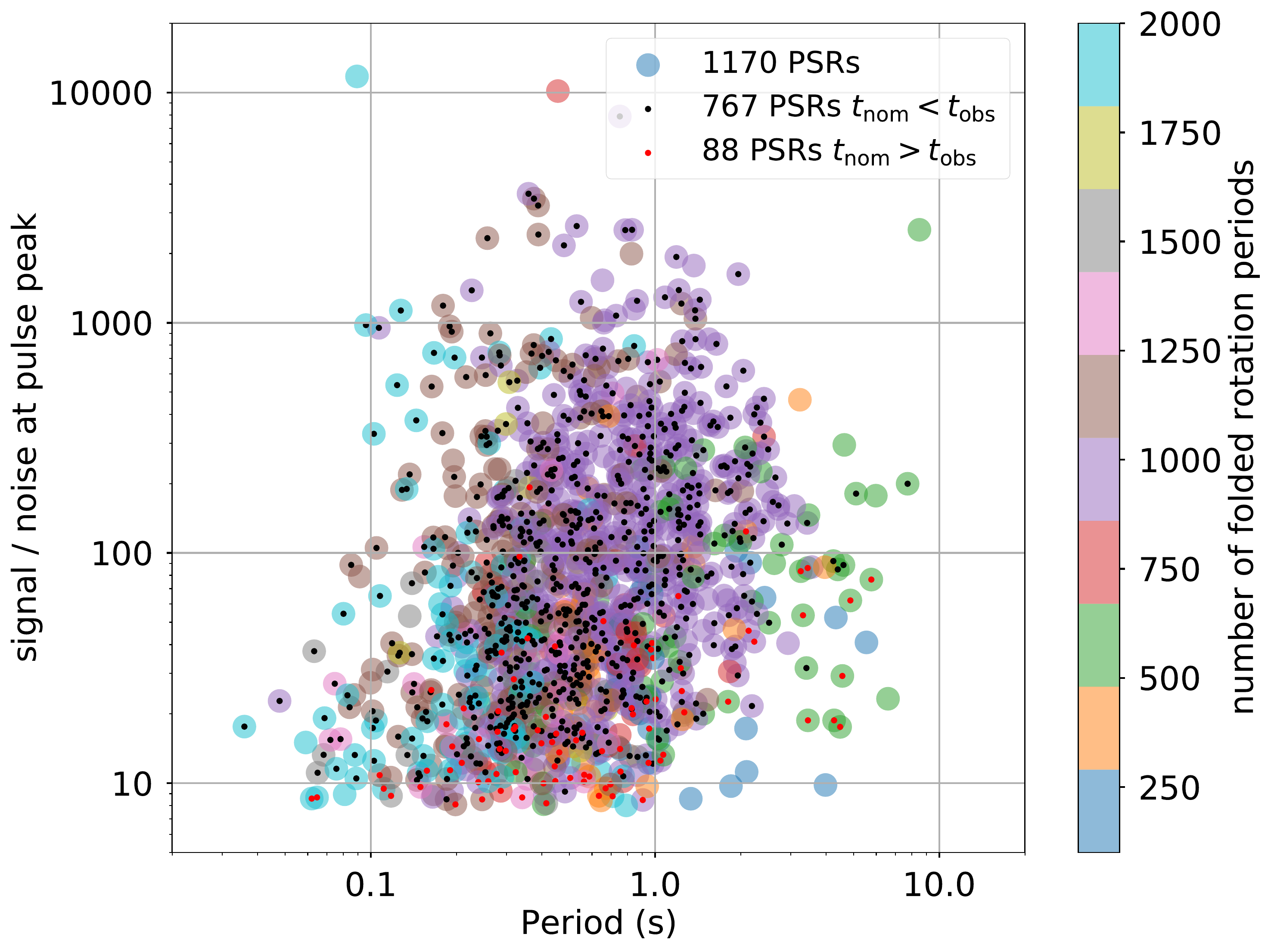}
\caption{The reached signal-to-noise ratio at the maximum of the pulse profile (Stokes I, frequency-averaged) over the rotation period of the TPA pulsars. The colour indicates the number of rotation periods covered in the reference epoch for the TPA census.
Black and red small inner dots indicate $t_{\rm nom} < t_{obs}$ and $t_{\rm nom} > t_{obs}$, respectively. The 100 pulsars that lack such inner points did not have all the parameters needed to estimate $t_{\rm nom}$ according to \citet{Song2021} (e.g., did not have a $W_{50}$). }
\label{fig:SNR_P}
\end{figure}

The general properties of the TPA pulsars and the census observations are listed in Table~\ref{tab:PSRs}. It includes the nominal optimal observing lengths, $t_{\rm nom}$,  according to \citet{Song2021} together with the actual TPA census observation duration. 
Table~\ref{tab:PSRs} also includes results from previous TPA works, in particular, scattering time-scales $\tau_{\rm scat}$ and the scattering spectral index $\alpha_{\rm scat}$ for 84 single-component TPA pulsars from \citet{Oswald2021}, and pulse widths $W_{10}$ (measured at 10\% of the pulse peak) from \citet{Posselt2021}. 
If not otherwise noted we do not select for specific pulsars (for instance, we do not explicitly remove scattered pulsars).
There are 1275 observed census pulsars with a signal-to-noise ratio at the peak position of the total power of the frequency-averaged data $S/N>3$.
From these, 1271 have available $P$ and $\dot{P}$ parameters (PSR\,J0837$-$2454, PSR\,J1402$-$5124, PSR\,J1653$-$4518 are not in the ATNF catalog (version 1.67), PSR\,J1357$-$62 does not have a listed $\dot{P}$).
We concentrate in this paper on the 1170 pulsars with $S/N>8$ (peak position) and $S/N>3$ (flux). Their $S/N$ distribution is shown in Figure~\ref{fig:SNR_P} together with the number of covered rotation periods in the folded profile. For the large majority of the census pulsars (83\%) the number of covered rotation periods in the reference observation is larger than 1000, 99\% have at least 200 roation periods covered. Figure~\ref{fig:SNR_P} also indicates which pulsars fulfil $t_{\rm nom} < t_{\rm obs}$ (90\% of those where $t_{\rm nom}$ could be estimated). 

Depending on the analysis, we have different numbers of pulsars. For example, out of the 1170 pulsars with the significance cut in the frequency-averaged data, we only determine a spectral index for the 1143 pulsars that have at least four frequency channels with significant ($>3\sigma$) flux measurements. 
For the polarization analysis, we analyse 1057 (1054 with listed $\dot{P}$) sources after removal of any pulsars for which our rotation measures are badly determined. We note that among these removed pulsars (i) there are some non-zero polarization measurements, (ii) some pulsars may have intrinsically a very low linear polarization hampering the determination of the rotation measures.

\begin{table*}
  \caption{The properties of the TPA census pulsars. For each pulsar, we list the period, $P$, the logarithm of the spin-down energy, $\log{\dot{E}}$, the modified Julian date, MJD, of the census pulsar observation, its duration $t_{obs}$, the nominal optimal observing length with MK, $t_{\rm nom}$ according to \citet{Song2021}, the reached rms in folded pulse profile in Stokes I, $\sigma_{I}$, the signal-to-noise ratio, S/N, of the pulse peak position in Stokes I, the  dispersion measure, DM, and rotation measure, RM, the time scale of the temporal broadening of pulses $\tau$ due to scattering and the scattering spectral index $\alpha_{\rm scat}$ from \citet{Oswald2021}, the  $W_{10}$ from \citet{Posselt2021} and updates as described in Section~\ref{profilemeas}. For some DM values, our error estimates failed, and we left $\sigma_{\rm DM}$ blank as an indication. The full table is online.}
  \label{tab:PSRs}
  \begin{tabular}{cccccccc|cc|cc|cc|cc|cc}
    \hline
    PSRJ & $P$ & $\log{\dot{E}}$ & MJD & $t_{obs}$ & $t_{\rm nom}$ &  $\sigma_{I}$ & S/N & DM &$\sigma_{\rm DM}$ & RM &$\sigma_{\rm RM}$ & $\tau_{\rm scat}$ & $\sigma_{\tau}$& $\alpha_{\rm scat}$ & $\sigma_{\alpha}$ & $W_{10}$ & $\sigma_{\rm W_{10}}$\\
         & s & erg\,s$^{-1}$    & days&  s & s & mJy & & cm$^{-3}$\,pc & & rad\,m$^{-2}$ &  & (ms) &  & & & $\deg$  \\
    \hline
J0034-0721 & 0.9430 &    31.3 & 58774.9 &  981 &     401 &  0.23 &   356 &   14.2 &   0.2 &     9.9 &   2.9 &        &        &        &          &  44.65 &   0.70 \\
 J0038-2501 & 0.2569 &    30.3 & 58823.8 &  602 &         &  0.26 &    18 &    6.1 &   0.1 &    14.0 &   3.9 &        &        &        &          &        &        \\
 J0045-7042 & 0.6323 &    32.6 & 59026.1 & 1199 &     806 &  0.18 &    14 &   71.0 &   0.9 &    38.0 &   3.8 &        &        &        &          &        &        \\
 J0045-7319 & 0.9264 &    32.3 & 58624.1 &  514 &         &  0.30 &     4 &  105.4 &       &         &       &        &        &        &          &        &        \\
 J0108-1431 & 0.8076 &    30.8 & 58799.9 &  842 &      33 &  0.24 &   123 &    2.2 &   0.1 &    -0.3 &   2.9 &        &        &        &          &   27.1 &    4.2 \\
 J0111-7131 & 0.6885 &    32.9 & 59026.1 & 1200 &         &  0.19 &     6 &   79.4 &       &     0.0 &  16.4 &        &        &        &          &        &        \\
 J0113-7220 & 0.3259 &    33.7 & 58783.9 &  302 &      81 &  0.37 &    42 &  125.5 &   0.1 &   126.4 &   3.8 &        &        &        &          &  12.66 &   0.70 \\
 J0131-7310 & 0.3481 &    33.2 & 59026.1 & 1200 &     947 &  0.19 &    19 &  205.2 &   0.1 &   -52.3 &   5.6 &        &        &        &          &        &        \\
 J0133-6957 & 0.4635 &    31.7 & 58783.9 &  302 &     141 &  0.39 &    31 &   22.9 &   0.3 &    20.0 &   5.7 &        &        &        &          &        &        \\
 J0134-2937 & 0.1370 &    33.1 & 58774.9 &  152 &       8 &  0.58 &   219 &   21.8 &   0.1 &    13.0 &   2.9 &        &        &        &          &  18.63 &   0.35 \\
 ... \\
    \hline
  \end{tabular}
 \end{table*}

\subsection{The TPA pulsars amongst the known pulsar population}
\label{sample}
\begin{figure}
\includegraphics[width=8.5cm]{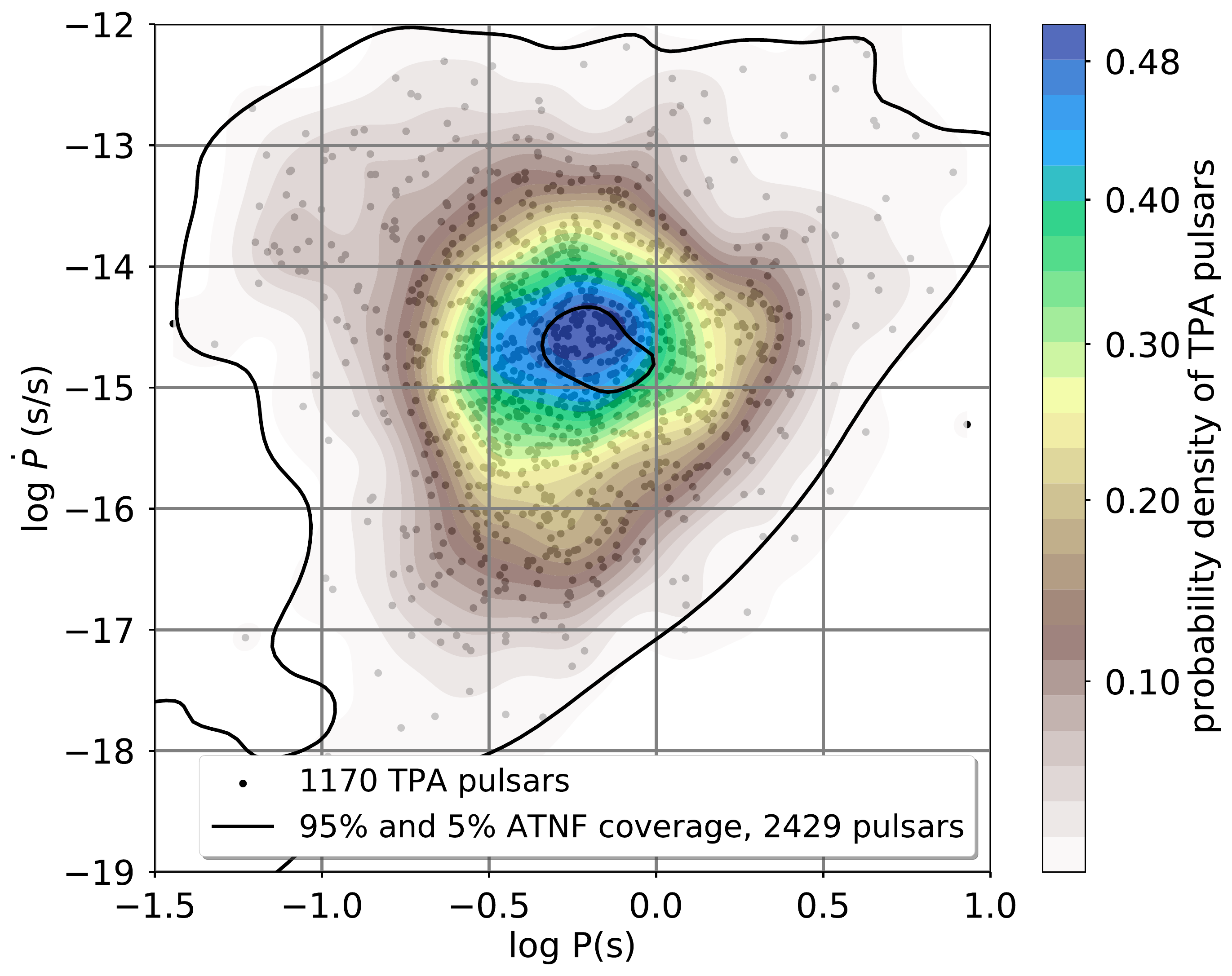}
\caption{The $P-\dot{P}$ coverage of the TPA census sample as a colour probability density plot (smoothed with Gaussian kernel using a smoothing bandwidth according to the reference rule by \citet{Silverman1986}, multiplicatively scaled by a factor 0.8). Individual pulsars are shown as grey points in the background. The contours mark the 95\% and 5\% levels of all pulsars listed in the ATNF (v 1.67) with $P$ and $\dot{P}$ measurements.}
\label{fig:PPdot}
\end{figure}

\begin{figure}
\includegraphics[width=8.6cm]{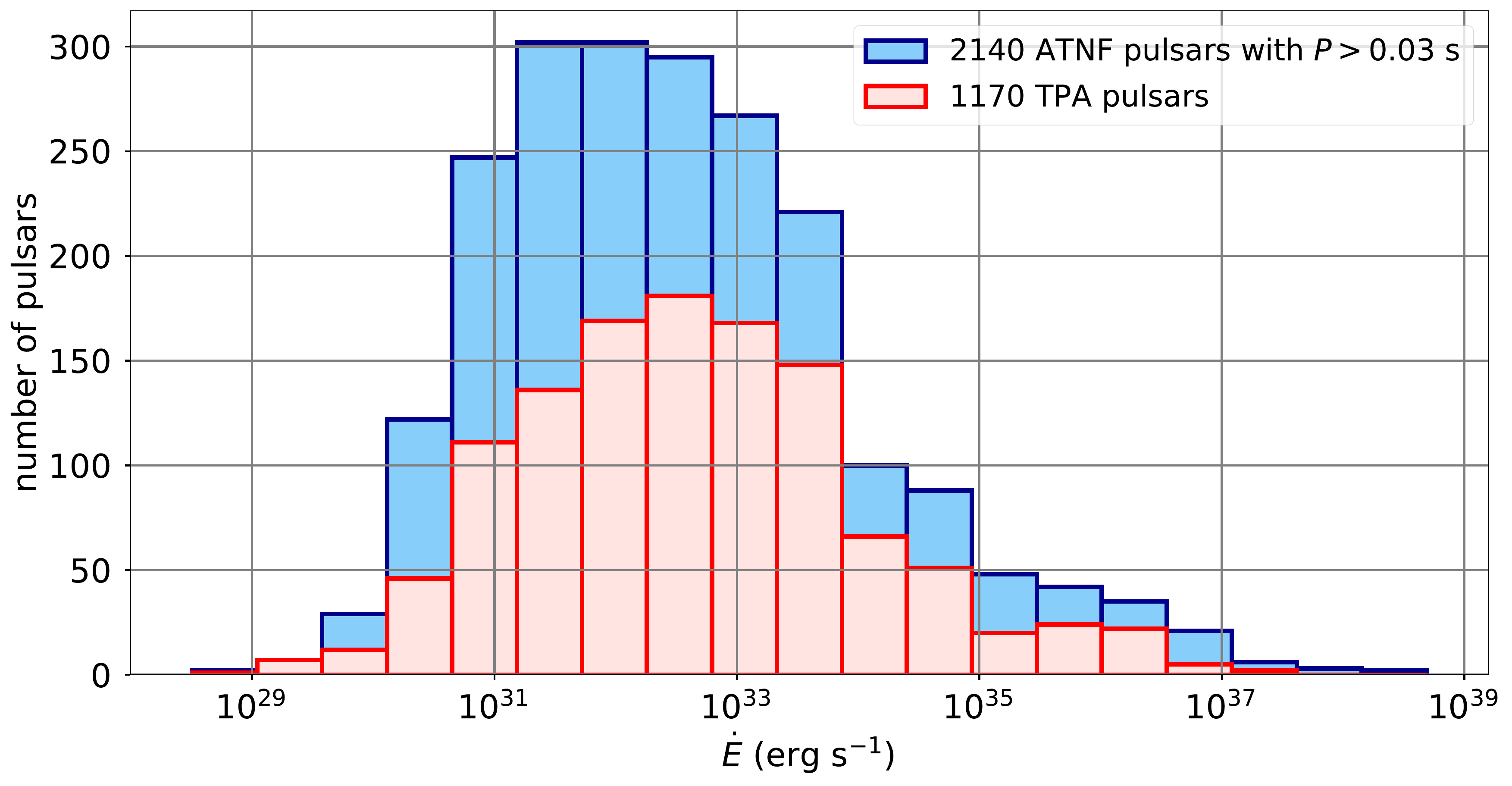}\\
\includegraphics[width=8.6cm]{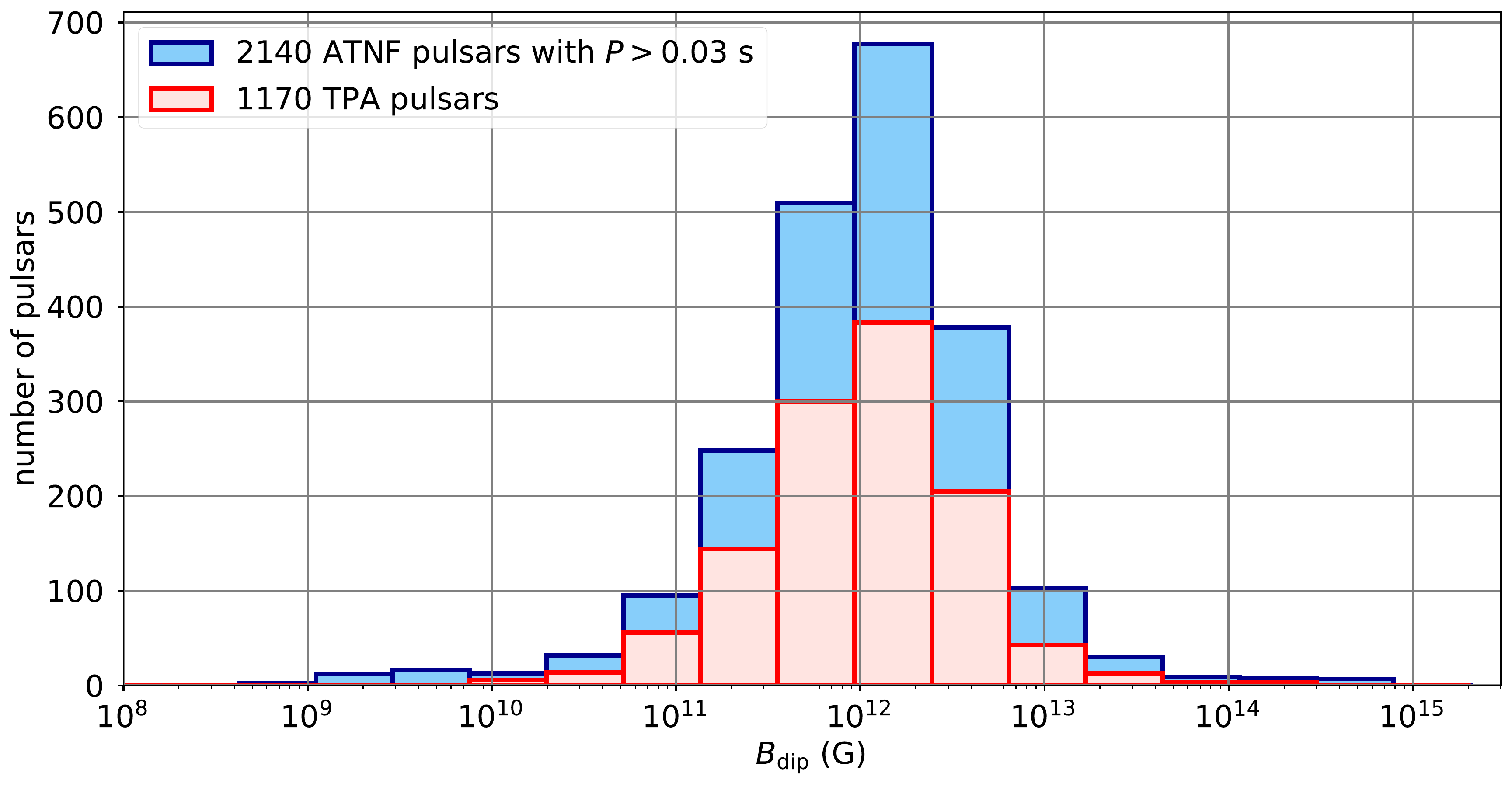}
\caption{The distributions of the TPA Census pulsars (red) with signal-to-noise ratios in the total band $S/N>8$ in comparison to all ATNF-listed pulsars (blue) fulfiling the period cut of the TPA. The upper and lower panels show the 20-bin histograms with respect to the spin-down energy and the inferred magnetic dipole field at the equator. The vertical axis indicates the total numbers in each bin.} 
\label{fig:EdotB}
\end{figure}

\begin{figure}
\hspace{-0.5cm}
\includegraphics[width=9.3cm]{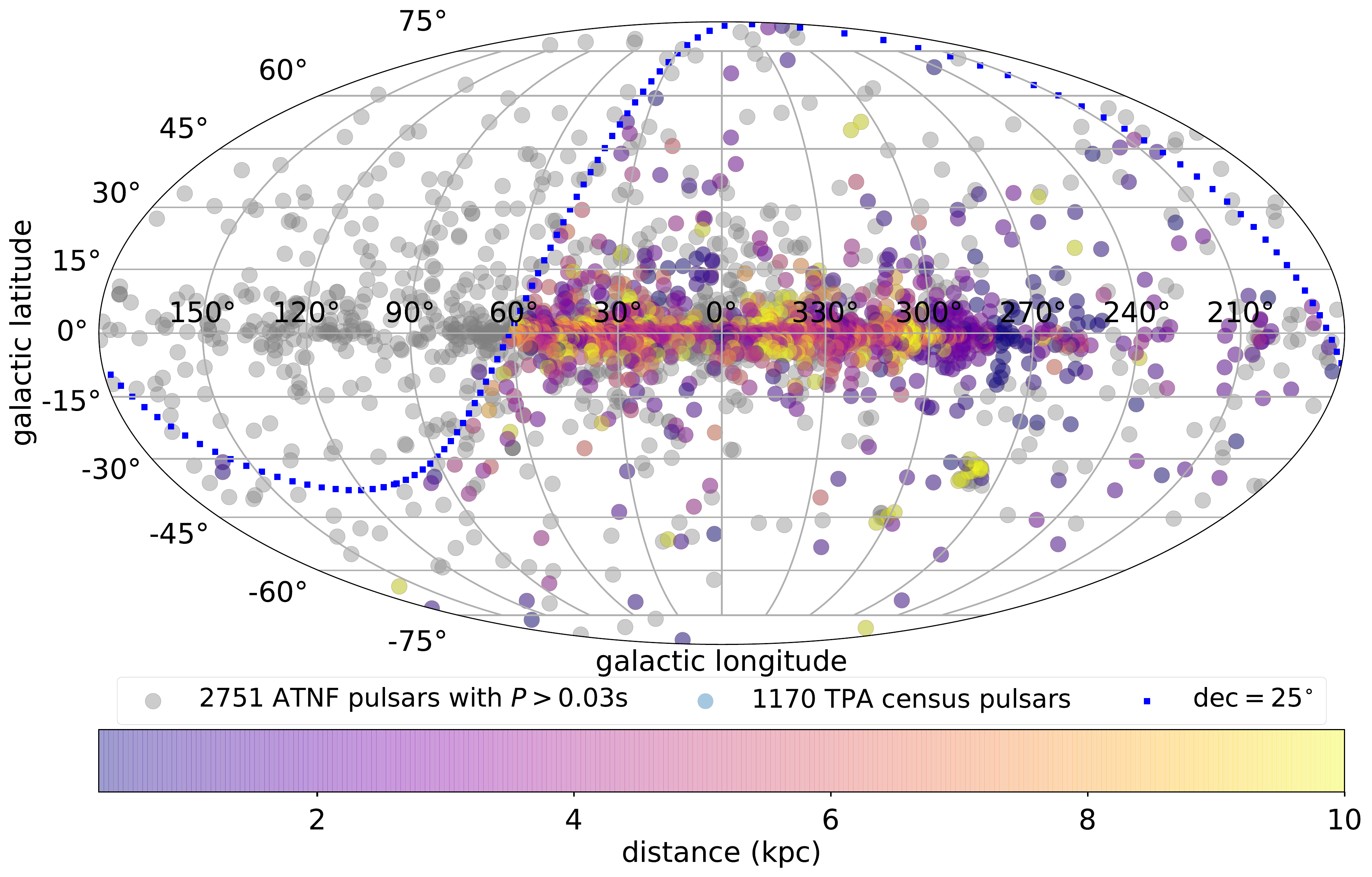}
\caption{The spatial distribution of the TPA census sample. Distances from the ATNF catalog are indicated by colour.}
\label{fig:GlGb}
\end{figure}

Figure~\ref{fig:PPdot} shows the distribution of the TPA census pulsars in the $P-\dot{P}$ diagram. The TPA covers nearly half of the currently known non-recycled pulsar population with measured $P,\dot{P}$. Figure~\ref{fig:PPdot} illustrates that the general $P-\dot{P}$ distribution of the TPA pulsars is similar to the one of all the currently known non-recycled pulsar population. 
The histograms of the spin-down energy $\dot{E}$ (values from $3\times 10^{28}$\,ergs\,s$^{-1}$ to $3\times10^{37}$\,ergs\,s$^{-1}$) and of the inferred magnetic dipole field at the equator $B_{dip}$ (values from $9\times 10^9$\,G to $2\times10^{14}$\,G) show a good sampling of these parameters as well, see Figure~\ref{fig:EdotB}. The TPA lacks most of the magnetars (most of which are not detected at radio wavelengths).
Figure~\ref{fig:GlGb} shows the spatial distribution in Galactic coordinates. Apart from the observabilty cut at declination$<25\deg$, the distribution looks similar to the one of the total pulsar population. Based on Figures~\ref{fig:SNR_P} to \ref{fig:GlGb}, we conclude that the TPA census has no obvious strong biases in  the sampling of the general pulsar properties with respect to the currently known non-recycled pulsar population.\\  

\subsection{Rotation measures and dispersion measures}
\label{sec:dmrm}
\begin{figure}
\includegraphics[width=8.5cm]{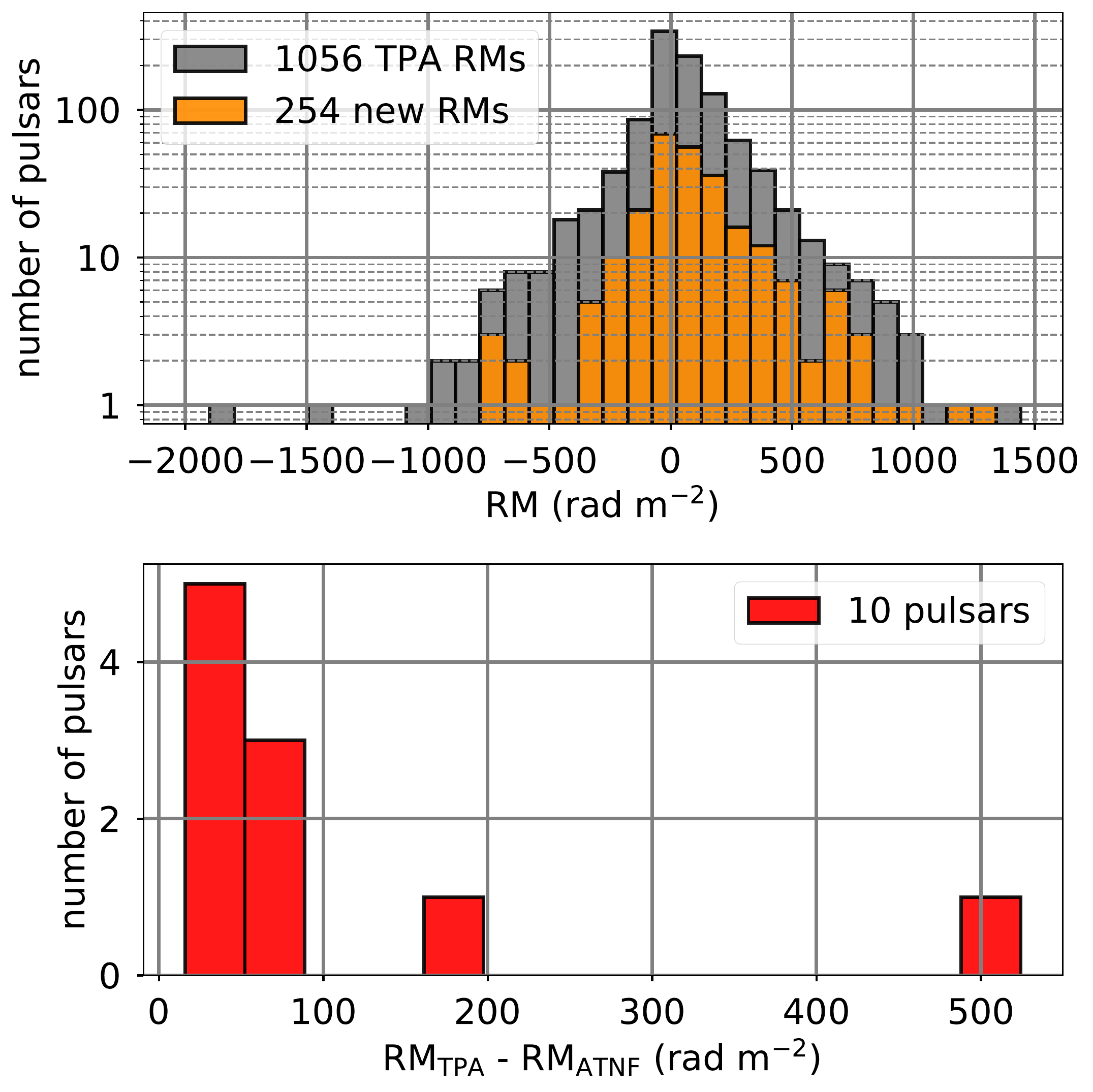}
\caption{\emph{Upper panel:} The distribution of all (grey) accepted RM values for the TPA pulsars that fulfill $S/N>8$ (peak), $S/N>3$ (flux), and $S/N_{LP}>10$. There is one additional pulsar with RM$=-6720$ (not shown). The orange histogram shows the new values (no RM in ATNF catalogue version 1.67).
\emph{Lower panel:} RM difference for the TPA pulsars whose $3\sigma$ RM-uncertainties from TPA and the ATNF do not overlap. \label{fig:RMhistos}}
\end{figure}

When RMs are combined with the respective DMs, the integrated magnetic field strength along the line of sight towards a pulsar can be estimated. 
Since RMs are typically the bottleneck for such estimates, we point out the RMs for more than 1000 TPA pulsars in Table~\ref{tab:PSRs}.  We estimated these RMs following the procedure described in section~\ref{sec:eph}. Visual screening and selection for $S/N>8$ (peak) and $S/N>3$ (flux), and $S/N_{LP}>10$  in the frequency-averaged pulse profile results in 1057  TPA RMs ranging from -6720\,rad\,m$^{-2}$ (for PSR\,J1114$-$6100) to 1442\,rad\,m$^{-2}$ (for PSR\,J1818$-$1607); see Figure~\ref{fig:RMhistos}. 
Among the 1057 RMs are 254 new values, i.e., there were no such measurements reported for these pulsars in the ATNF pulsar catalogue (version 1.67). For easier identification of these new RMs, they are also listed in Table~\ref{tab:onlynewRM}. For 10 pulsars, the $3\sigma$ uncertainties of our RMs and the ATNF do not overlap. The largest difference is found for PSR\,J1849$-$0636, for which we obtain $489 \pm 3$\,rad\,m$^{-2}$, whilst the reference value is $-35 \pm 21$\,rad\,m$^{-2}$ \citep{Han1999}. 
Reasons for the discrepancies can be different measurement methods and used DM values, but also actual RM variations in the cases of small discrepancies.  
The Vela pulsar, for example, is known to show RM (and DM) variations over time \citep{Hamilton1985}.
There were 151 pulsars for which the obtained RM is clearly a bad fit or we could not identify a good RM value at all, for instance if the signal-to-noise (S/N) was very low.
For these pulsars, we do not list RM and we do not include them in the statistical plots of the linear and circular polarization fractions.\\

\begin{table}
\caption{The 254 newly measured RMs and their uncertainties (also included in Table~\ref{tab:PSRs}). Only the first 10 rows are shown,  the full table is available online.}
\label{tab:onlynewRM}
  \begin{tabular}{lrr}
   PSR & RM &  $\sigma_{\rm RM}$ \\
  & rad\,m$^{-2}$ & rad\,m$^{-2}$\\
\hline
 J0038--2501 &         14 &      4 \\
 J0045--7042 &         38 &      4 \\
 J0302+2252 &         $-7$ &      3 \\
 J0418--4154 &         19 &      5 \\
 J0449--7031 &         41 &      4 \\
 J0455--6951 &       $ -24$ &      3 \\
 J0522--6847 &       $-119$ &      5 \\
 J0534--6703 &       $ -35$ &      4 \\
 J0555--7056 &         25 &      4 \\
 J0621+0336 &         46 &      3 \\
 ...\\
\end{tabular}
\end{table}

\begin{figure}
\includegraphics[width=8.5cm]{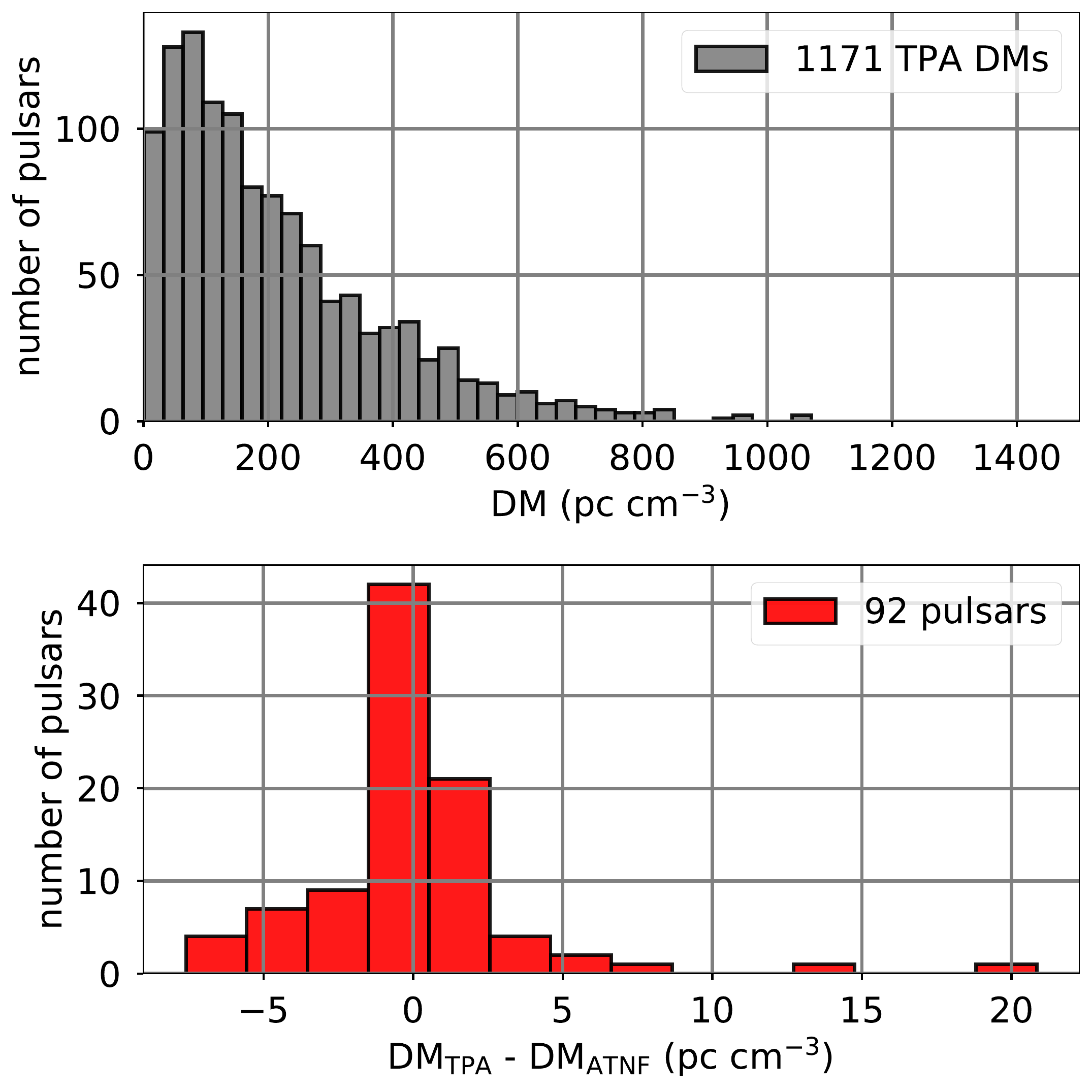}
\caption{\emph{Upper panel:} The distribution of obtained DM values for the TPA pulsars with $S/N>8$ (peak) and $S/N>3$ (flux), including the strongly scattered TPA pulsars investigated by \citet{Oswald2021}. Most DMs agree with values reported in the ATNF pulsar catalogue (version 1.67). \emph{Lower panel:} DM difference for those TPA pulsars whose $3\sigma$ DM-uncertainties from TPA and the ATNF do not overlap (excluding strongly scattered pulsars reported by \citet{Oswald2021}. \label{fig:DMhistos}}
\end{figure}

For completeness, Figure~\ref{fig:DMhistos} shows an overview of the DM-distribution in our sample.
The DM of a pulsar can be used beyond the dedispersion of the radio pulse. 
For example, DM values are also utilized to estimate optical or X-ray extinctions towards pulsars, or to compare such estimates with other extinction measures.
Therefore, we comment briefly on differences of our DM values with respect to the ATNF pulsar catalogue.
For 92 pulsars, our $3\sigma$ DM-uncertainties do not overlap with the value and  $3\sigma$ uncertainties in the ATNF pulsar catalogue (version 1.67). For some, this may simply be due to missing uncertainties in the DM reference values. Our largest difference in terms of $\sigma$, for example, is seen for PSR\,J1704$-$5236 for which we report  $164.0 \pm 0.1$\,pc\,cm$^{-2}$ whilst its ATNF value, 170.0\,pc\,cm$^{-2}$ from \citet{Burgay2019}, does not have an uncertainty.
The largest absolute DM difference is found for PSR\,J1828$+$1359, for which our value is $76.8 \pm 0.3$\,pc\,cm$^{-2}$ in comparison to the ATNF value, $56 \pm 1$\,pc\,cm$^{-2}$ \citep{Navarro2003}.
Whilst some pulsars are known to show actual DM variations (e.g., \citealt{Petroff2013}), we caution again about the dependence of the derived DM on the (here, not considered) frequency-dependent profile evolution and scattering.\\

\subsection{Pulsar spectra}
\label{sec:fluxsi}
\begin{figure}
\includegraphics[width=8.5cm]{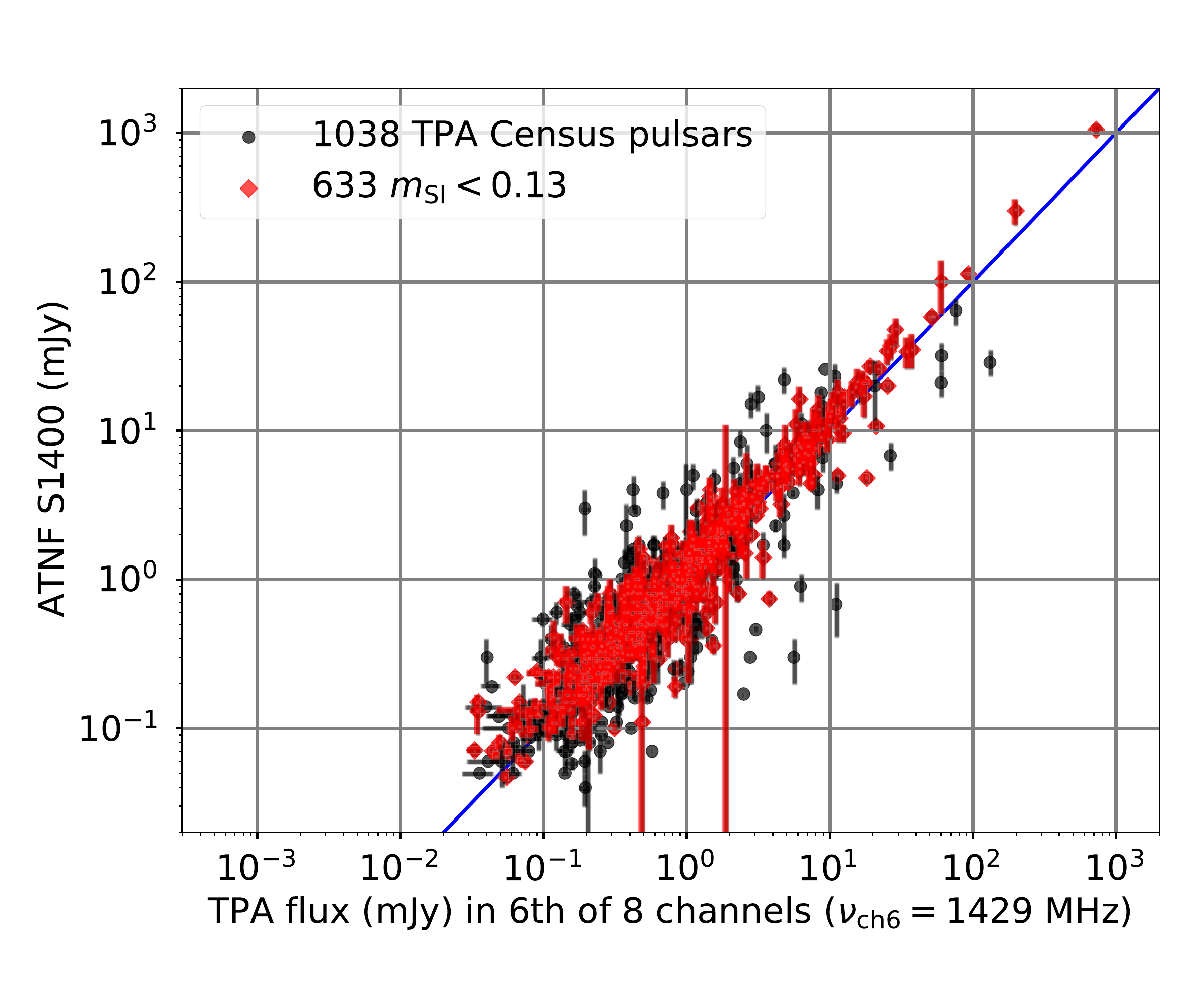}
\caption{The observed TPA fluxes (with $1\sigma$ uncertainties) from one observing epoch in comparison with literature values from the ATNF catalogue at 1.4\,GHz. The plot shows all TPA pulsars with $S/N>8$ (peak) and $S/N>3$ (flux) for which S1400 values are available. The blue line marks equality. The red points illustrate the improvement of the agreement due to the removal of potentially scintillating sources in the TPA census.}
\label{fig:flux1400}
\end{figure}

The 1170 TPA pulsars with $>3\sigma$ significance of the flux measurement span a range from $0.022\pm 0.006$\,mJy (PSR\,J0628+0909) to $979.65 \pm 0.05$\,mJy (PSR\,J0835$-$4510). 1143 TPA pulsars have at least 4 significant flux measurements in the 8 frequency channels. Only these pulsars are used for the weighted PL-fits to derive spectral indices and $m_{SI}$ as described in Section~\ref{profilemeas}. The correponding results are listed in Table~\ref{tab:PSRfluxes}. 
The flux density estimated for a given pulsar at a single epoch, as is the case here, can be affected by diffractive or refractive scintillation. For 1038 pulsars, we can compare our homogenous data set with the $S_{1400}$ values listed in the ATNF catalog, Figure~\ref{fig:flux1400}. The values agree well, in particular if we use $m_{SI}<0.13$ to remove scintillating sources or those with strong deviations from a PL. The TPA uncertainties are typically much smaller than previous literature values. \\

\begin{figure}
\includegraphics[width=8.5cm]{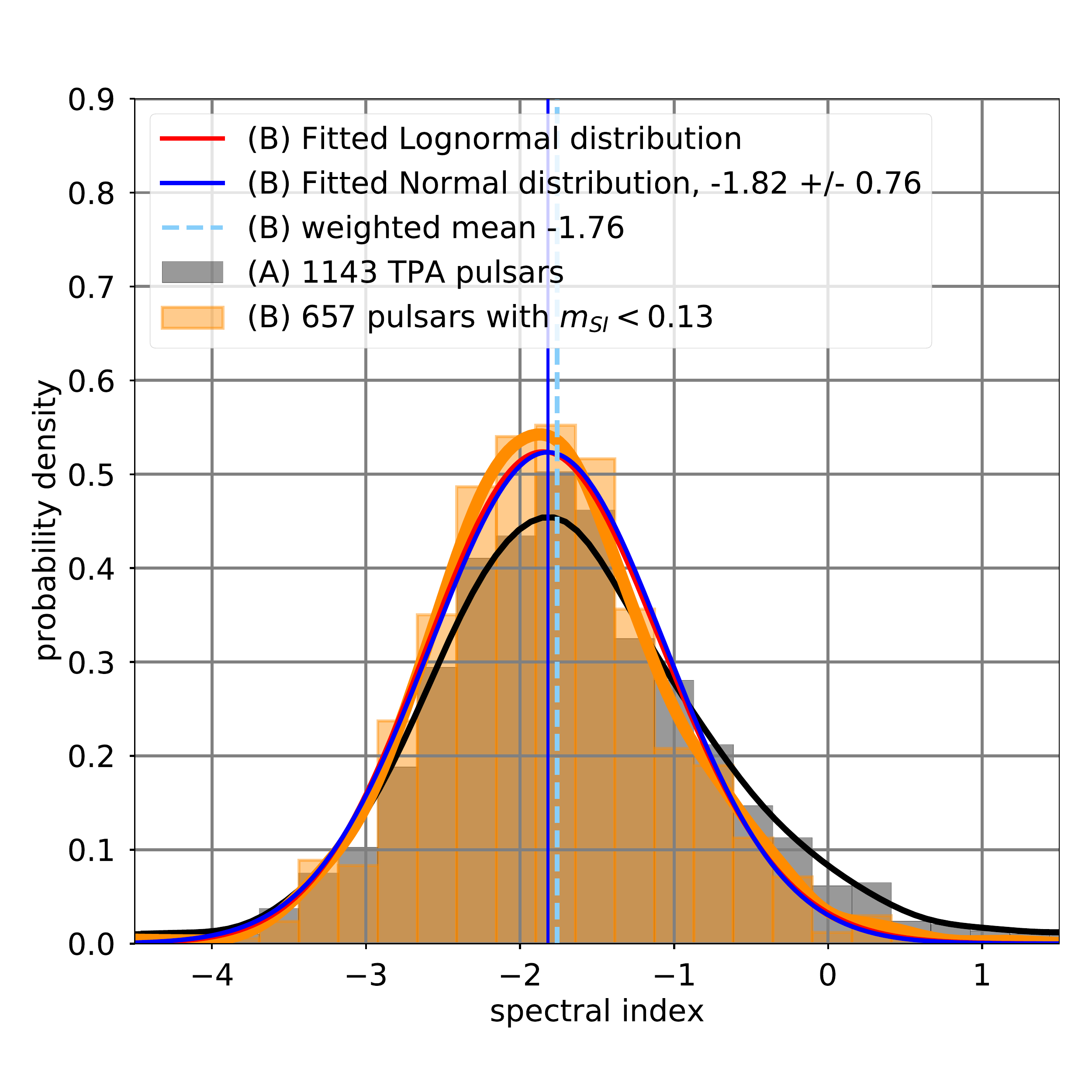}
\caption{The distribution of the spectral index values. The grey histogram and line represent the distribution for all TPA pulsars, (A), for which the spectral index was derived. The sample (B), yellow histogram and line represent the pulsars filtered for the pseudo modulation index. Both histograms employ the same value ranges, 40 bins, and a Gaussian kernel density estimator for the smoothed probability density distribution.}
\label{fig:SIdist}
\end{figure}

Figure~\ref{fig:SIdist} shows the distribution of the derived spectral indices using simple weighted PL-fits. Restricting to the pulsars filtered with $m_{SI}<0.13$ (657 pulsars), we obtain a weighted mean of -1.8. Using a normal distribution, the data can be described with a mean $-1.8$ and $\sigma=0.8$. These results are consistent with previous findings from smaller samples in the literature. \citet{Jankowski2018}, for example, reported a mean and weighted mean of $-1.6$ ($\sigma=0.6$) for 276 pulsars that were found to follow a simple power law in a much larger frequency range (from $\sim 400$\,MHz to $\sim 3$\,GHz).
The normal and the lognormal probability density distributions (blue and red lines in Figure~\ref{fig:SIdist}, respectively) are visually indistinguishable for the shown TPA sample, whilst \citet{Jankowski2018} found a better description with a log-normal fit to their data.
Similarly to their work, we used the Shapiro-Wilk test \citep{Shapiro65} to test for normality of the spectral index distribution and its logarithm (shifted by +5). Whilst for $m_{SI}<0.13$, the $p_{SW}$-value for both is smaller than $0.003$ for both, rejecting a Gaussian distribution, for a stricter filter $m_{SI}<0.1$ (560 pulsars) the test statistics $t_{SW}=0.996$ and $p_{SW}=0.12$ indicate a normal distribution instead of a log-normal distribution ($t_{SW}=0.92$, $p_{SW}=10^{-16}$).

\subsection{Polarization}
\label{sec:polOV}
\begin{figure}
\includegraphics[width=8.5cm]{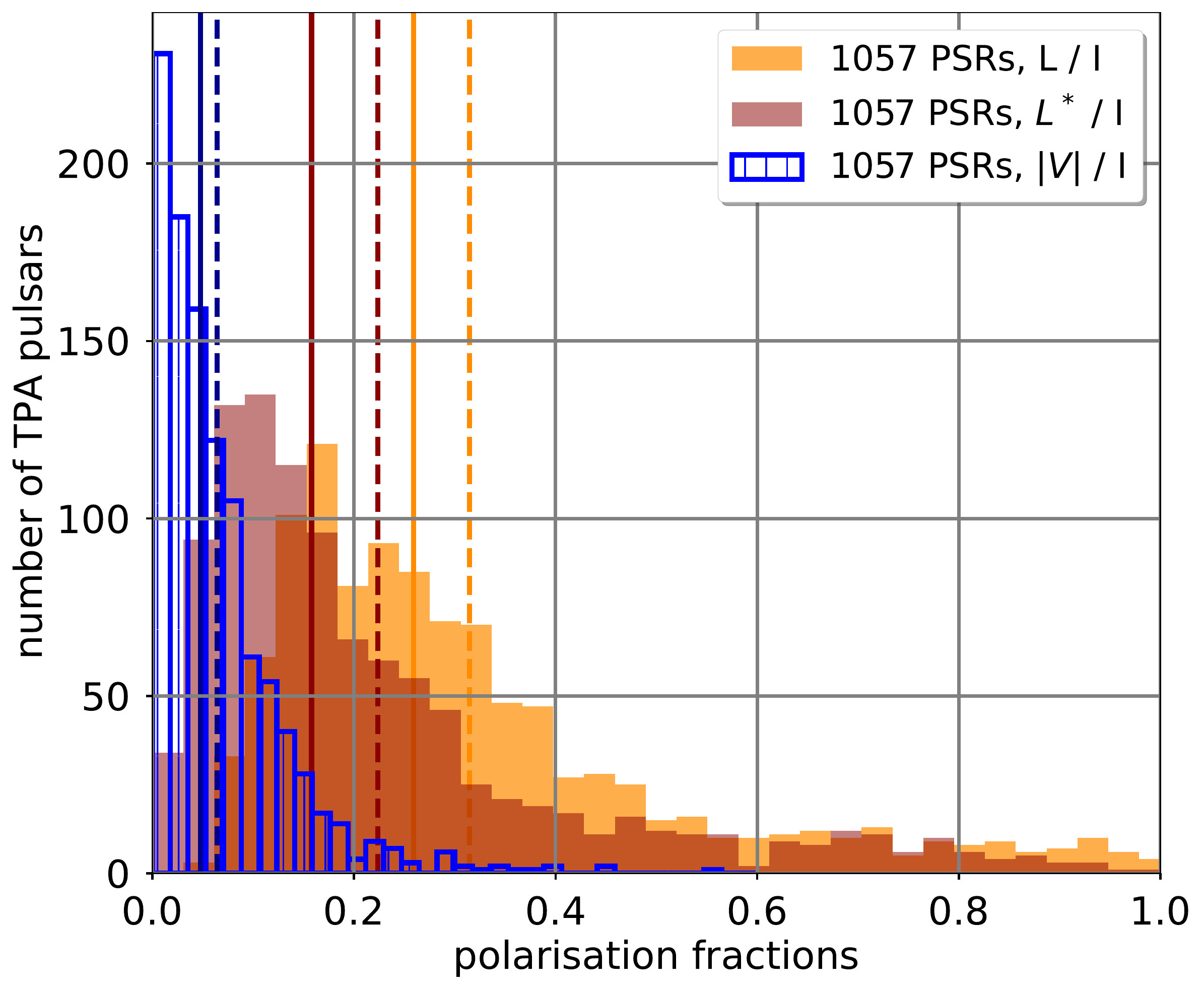}
\caption{The distribution polarization fractions in the TPA data for the frequency-averaged data. All measurements are shown without consideration of their uncertainties or significance. Mean $1\sigma$ uncertainties are 0.032 for LPF and aCPF, and 0.075 for the L$^*$PFs. All histograms have 35\,bins covering the full range, $[0,1]$, for the linear polarization fractions, and $[0,0.6]$ for the absolute circular polarization fraction. The solid and dashed lines indicate the median and mean values in the histogram of the corresponding colour (see also text).}
\label{fig:polfracdist}
\end{figure}

The polarization fraction measurements for the frequency-averaged data and the data set of the eight frequency channels are listed in Table~\ref{tab:PSRpolar}. 
There are 1057 TPA pulsars with computed (debiased) linear polarization fraction (LPF), vector-added linear polarization fractions  (L$^*$PFs), and circular polarization fractions (CPFs, and debiased absolute aCPFs), that fulfil $S/N$(peak)$>8$, $S/N$(flux)$>3$ for the frequency-averaged data, and have valid RM values. The LPFs have a mean value of 0.32 with a 16\%, 50\%, 84\% quantile distribution of 0.15, 0.26, 0.48.
The L$^*$PFs have a mean value of 0.22 with a 16\%, 50\%, 84\% quantile distribution of 0.07, 0.16, 0.38, i.e., this distribution is shifted towards lower values in comparison to that of the LPFs.
The aCPFs have a mean value of 0.06 with a 16\%, 50\%, 84\% quantile distribution of 0.01, 0.05, 0.11.
All three polarization fraction distributions for the frequency-averaged data are shown in Figure~\ref{fig:polfracdist}.  
\begin{figure}
\includegraphics[width=8.5cm]{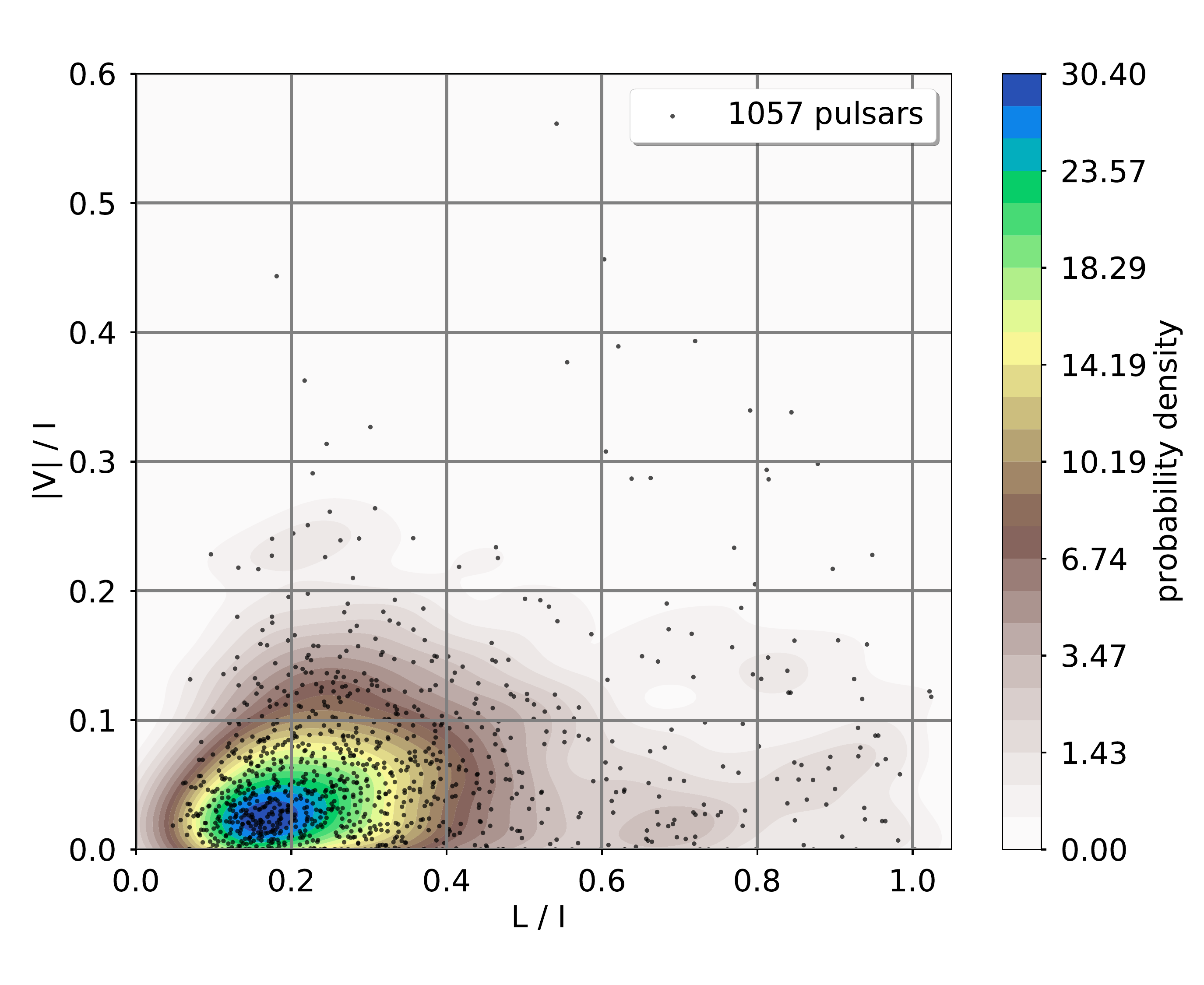}
\caption{The distribution of the absolute circular polarization fractions versus the linear polarization fractions. The probability density is indicated by the color bar (smoothing parameters are chosen similarly as in Figure~\ref{fig:PPdot}). Individual measurements are shown with black dots. }
\label{fig:LPaCP}
\end{figure}
As already known from previous studies, the LPFs are typically much higher than the aCPFs -- 50 TPA pulsars have LPFs above 80\%, but only 3 pulsars (PSRs\,J1222$-$5738, J1410$-$7404, J1907+0918) have aCPFs above 40\%.
Figure~\ref{fig:LPaCP} shows the (debiased) aCPFs over LPFs for the 998 pulsars that have values $>0$ for both. 
\begin{figure}
\includegraphics[width=8.5cm]{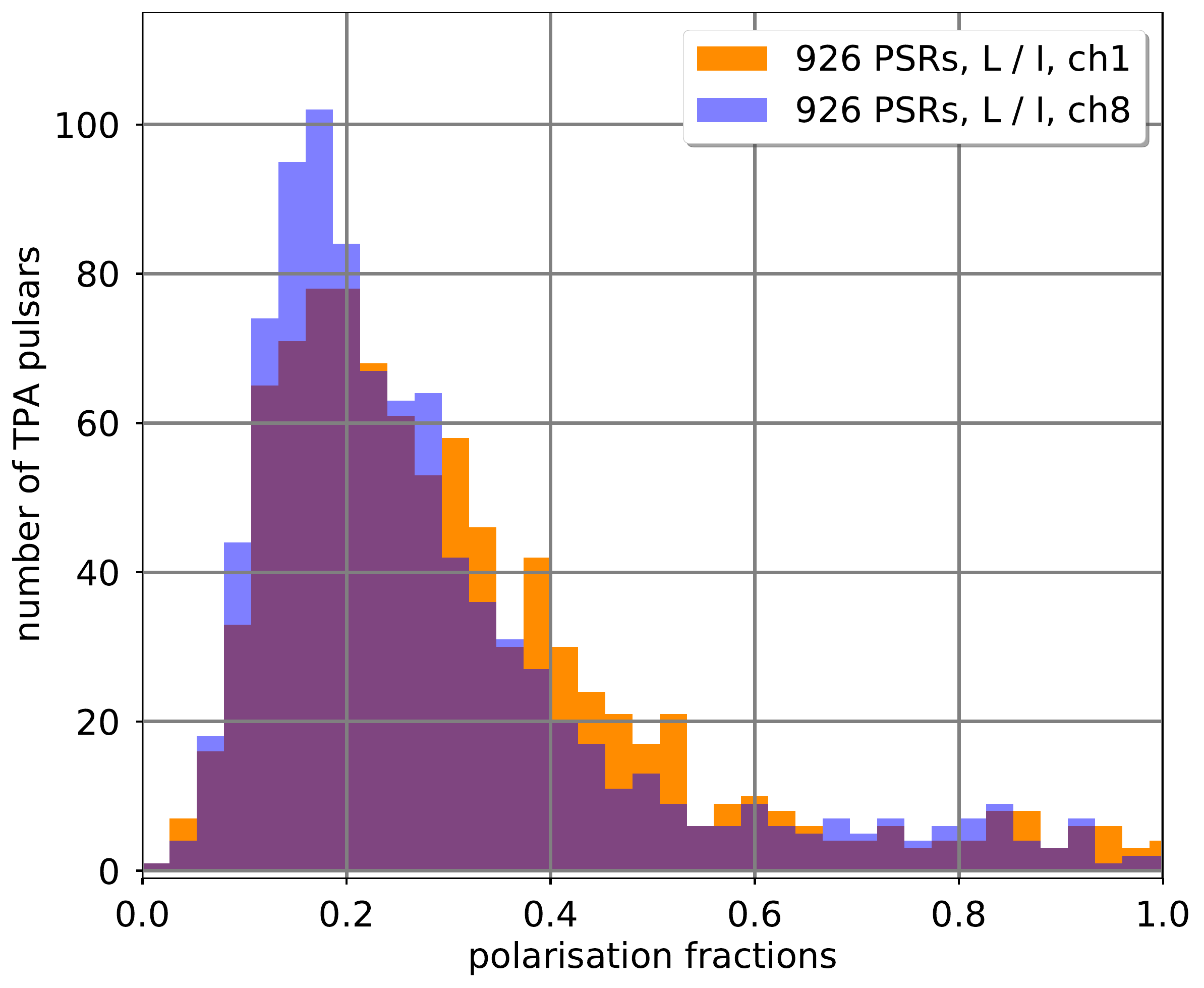}
\caption{The distribution of all the linear polarization fraction for the \emph{same} pulsars in the first frequency channel (944\,MHz) and the eight frequency channel (1626\,MHz). The histograms have 40 bins. Mean uncertainties for the LPF in the first and eight channel are 0.034 and 0.039. }
\label{fig:LPch1ch8}
\end{figure}
\begin{figure}
\includegraphics[width=8.5cm]{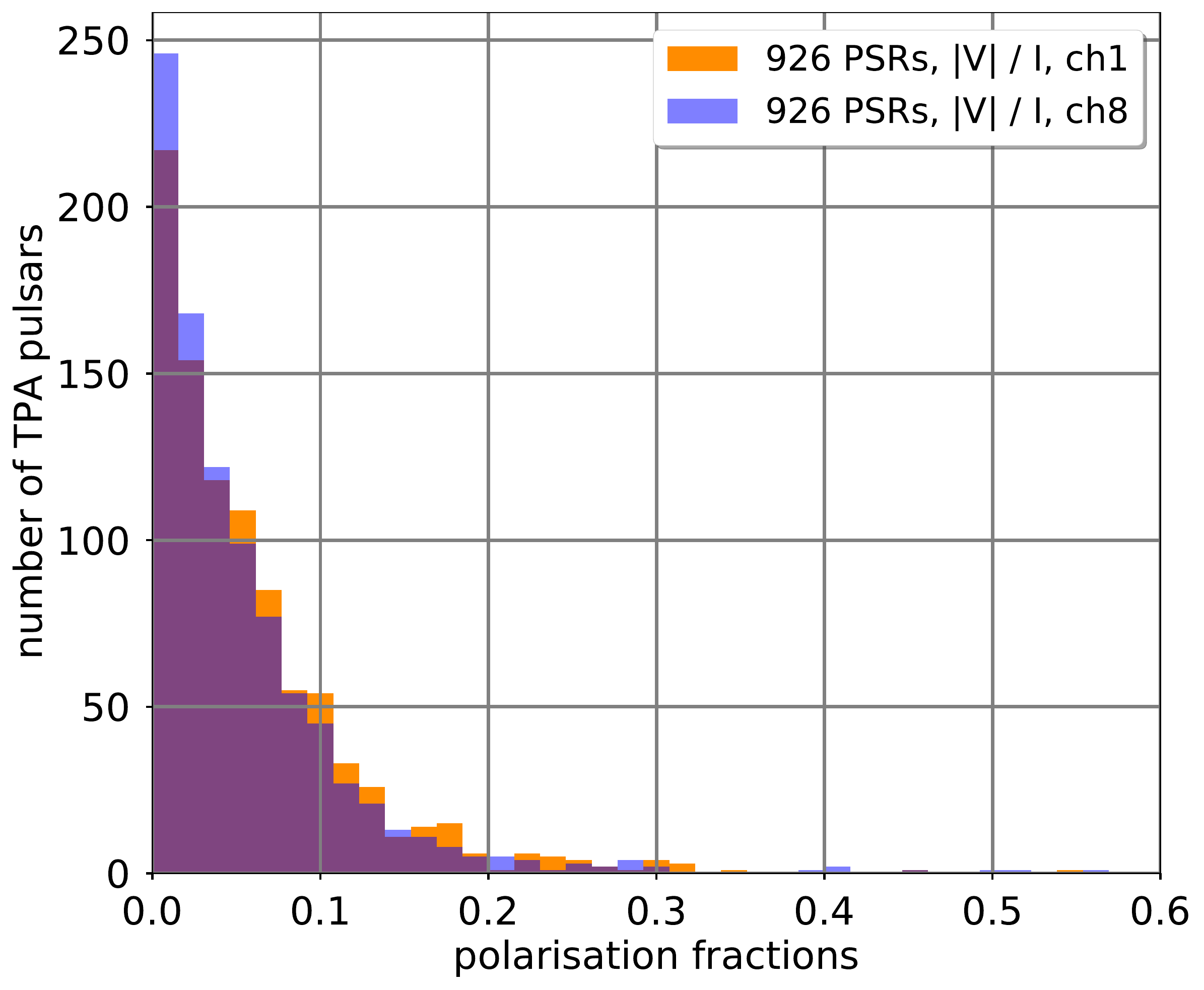}
\caption{The distribution of the absolute circular polarization fraction for the \emph{same} pulsars in the first frequency channel (944\,MHz) and the eight frequency channel (1626\,MHz). The histograms have 40 bins. Mean uncertainties for the aCPF in the first and eight channel are 0.034 and 0.038.}
\label{fig:aCPch1ch8}
\end{figure}
The wide bandwidth of the TPA data is sufficient to illustrate the known systematic (e.g., \citealt{Hoensbroech1998b} and more recent simulations by \citealt{Wang2015}) tendency for decreasing LPFs with increasing frequency, see Figure~\ref{fig:LPch1ch8} where we compare LPFs for the lowest (944\,MHz) and highest (1.626\,GHz) frequency channels. 
For the aCPF (Figure~\ref{fig:aCPch1ch8}), no obvious trend is seen.

\begin{figure}
\includegraphics[width=8.5cm]{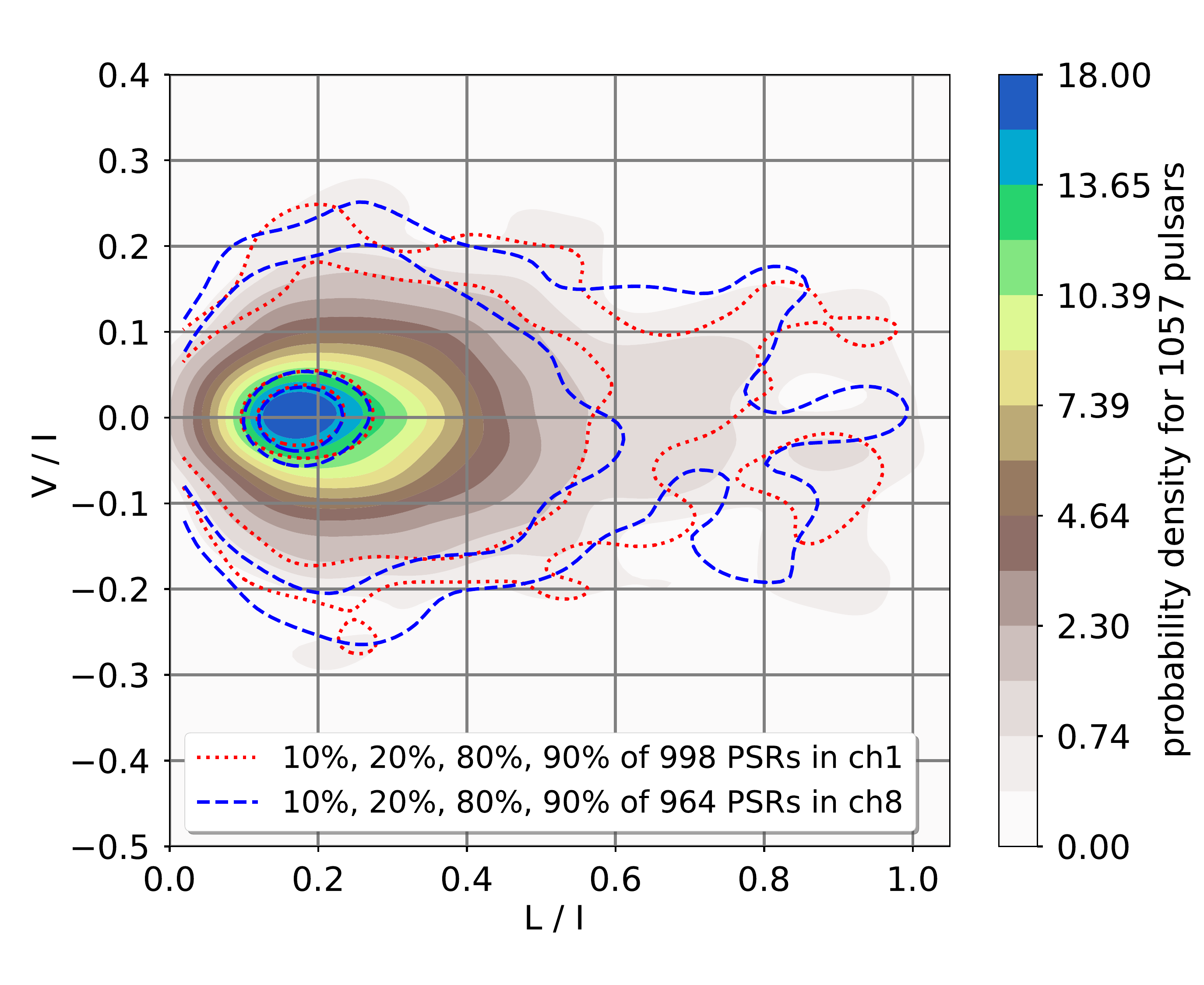}
\caption{The distribution of the circular polarization fractions versus the linear polarization fractions. The shaded area shows the probability density of the frequency-averaged data (smoothing parameters are chosen similarly as in Figure~\ref{fig:PPdot}). The red dotted and the blue dashed lines indicate the 10\%, 20\%, 80\%, 90\% levels of all the measurements in the first frequency channel (944\,MHz) and all the eight frequency channel (1626\,MHz). }
\label{fig:LPCP}
\end{figure}

Considering all circular polarizations, see Figure~\ref{fig:LPCP}, the LPF - circular polarization fraction distribution becomes completely dominated by its peak at low values. It is also rather symmetric for positive and negative circular polarization fractions.
 In Figure~\ref{fig:LPCP}, contour levels of the respective distributions measured at the lowest (944\,MHz) and highest (1.626\,GHz) frequency channels do not show any obvious population-wide frequency dependency for the correlation between these two polarization fractions.   

\section{Discussion} 
\label{sec:discussion}
Above, we presented TPA measurements for pulsar fluxes, spectra, and polarizations, with $\sim 1000$ TPA pulsars having \emph{all} these quantities measured, many of them for the first time. Our tables and the associated pulse profiles are very useful to identify, for example, extreme sources for individual, detailed studies of key aspects of pulsar emission.
Most importantly, however, our homogeneous, big data set also enables a fresh and unbiased look onto the known and suspected dependencies of a large variety of pulsar parameters. 
Here, we present such an evaluation of correlations for three examples of TPA pulse profile measurements -- the spectral index, the (pseudo) radio luminosity, and the linear polarization fraction.
These three pulse profile properties have all been discussed in the literature to depend on age or on its evolutionary proxy, the spin-down power. 
Instead of solely restricting to the spin-down power, or its equivalent $P^{-3} \dot{P}$ proportionality, we aim for broader correlation tests with our large new data set. We use a common procedure to explore all three pulse profile measures. For possible dependencies on $P$, $\dot{P}$, $\dot{E}$, $\tau$, $B$, we use the $P^a \dot{P}^{b}$ approach as explained in Section~\ref{sec:cor}. 
The $a-b$ space is well sampled with a Spearman rank correlation analysis, whilst individual OLS fits directly assess the constraints in linear or logarithm space. We show the results for all these spin-related parameters together in one overview plot (via the inferred $a,b$ values, e.g., Figure~\ref{fig:SIfits}) for each of the three test quantities. This allows for an easy overview of all fit constraints within the broad (and symmetric) trends from the Spearman rank analysis, highlighting what is the dominant dependency amongst $P$, $\dot{P}$, $\dot{E}$, $\tau$, and $B$.  
We also explored other pulsar parameters such as the pulse widths, DMs, or locations, with a combination of Spearman rank analysis, and, if strong indications of correlations are found, OLS fits.
Table~\ref{tab:polarSpear} provides a summary of some notable results from the former, whilst  we discuss individually constrained dependencies in the following. 

\begin{table*}
  \caption{The Spearman rank correlation coefficients and $p$-values (separated by ``$|$'')  are listed for the analysis of selected measured TPA quantities (first column) with other pulsar parameters. The sixth column lists the number of pulsars $N_{\rm PSR}$ that were considered for the columns two to five (period, period derivative, spin-down energy, and charactersistic age). The $N_{\rm PSR}$ considered for the pulse widths, $W_{10}$, and spectral indices, SI, are listed in brackets in the respective columns.}
  \label{tab:polarSpear}
  \begin{tabular}{lccccccc}
    \hline
       & $P$ & $\dot{P}$ & $\dot{E}$ & $\tau$ & $N_{\rm PSR}$ & $W_{10}$ & SI \\
    \hline
spectral index &  $-0.24|1 \times 10^{-9}$ & $0.19|2 \times 10^{-6}$ &  $0.29|2\times 10^{-14}$ &  $-0.25|1\times 10^{-10}$ & 657 & $0.20|8\times 10^{-6}$\,(499) & $\dots$ \\
luminosity &  $-0.15|3 \times 10^{-7}$ &   $0.14|1 \times 10^{-6}$ &  $0.24|2\times 10^{-16}$ &  $-0.20|1\times 10^{-11}$ & 1170 & $0.28|6\times 10^{-17}$\,(874) & $-0.04|0.3$\,(657) \\
LPFs &  $-0.25|4\times 10^{-16}$ &   $0.21|1\times 10^{-11}$ &  $0.29|7\times 10^{-22}$ &  $-0.27|2\times 10^{-18}$ & 1054 & $0.28|1\times 10^{-16}$\,(839) & $0.34|2\times 10^{-17}$\,(589) \\
LPFs ($\dot{E}>10^{32}$\,erg\,s$^{-1}$) &  $-0.35|2\times 10^{-21}$ &   $0.28|2\times 10^{-13}$ &  $0.49|7\times 10^{-42}$ &  $-0.40|5\times 10^{-27}$ & 684 & $0.21|6\times 10^{-7}$\,(534) & $0.37|2\times 10^{-16}$\,(464) \\
LPFs ($\dot{E}\le 10^{32}$\,erg\,s$^{-1}$) &  $0.07|0.2$ &   $-0.07|0.2$ &  $-0.20|0.0002$ &  $0.12|0.02$ & 370 & $0.39|2\times 10^{-12}$\,(305) & $0.17|0.06$\,(125) \\
L$^*$PFs &  $-0.28|2\times 10^{-21}$ &   $0.25|6\times 10^{-16}$ &  $0.35|2\times 10^{-31}$ &  $-0.31|2\times 10^{-25}$ & 1054 & $0.23|1\times 10^{-11}$\,(839) & $0.31|2\times 10^{-14}$\,(589) \\
    \hline
  \end{tabular}
 \end{table*}

\subsection{Correlations of the spectral index}
\label{sec:SI}
\begin{figure}
\includegraphics[width=8.5cm]{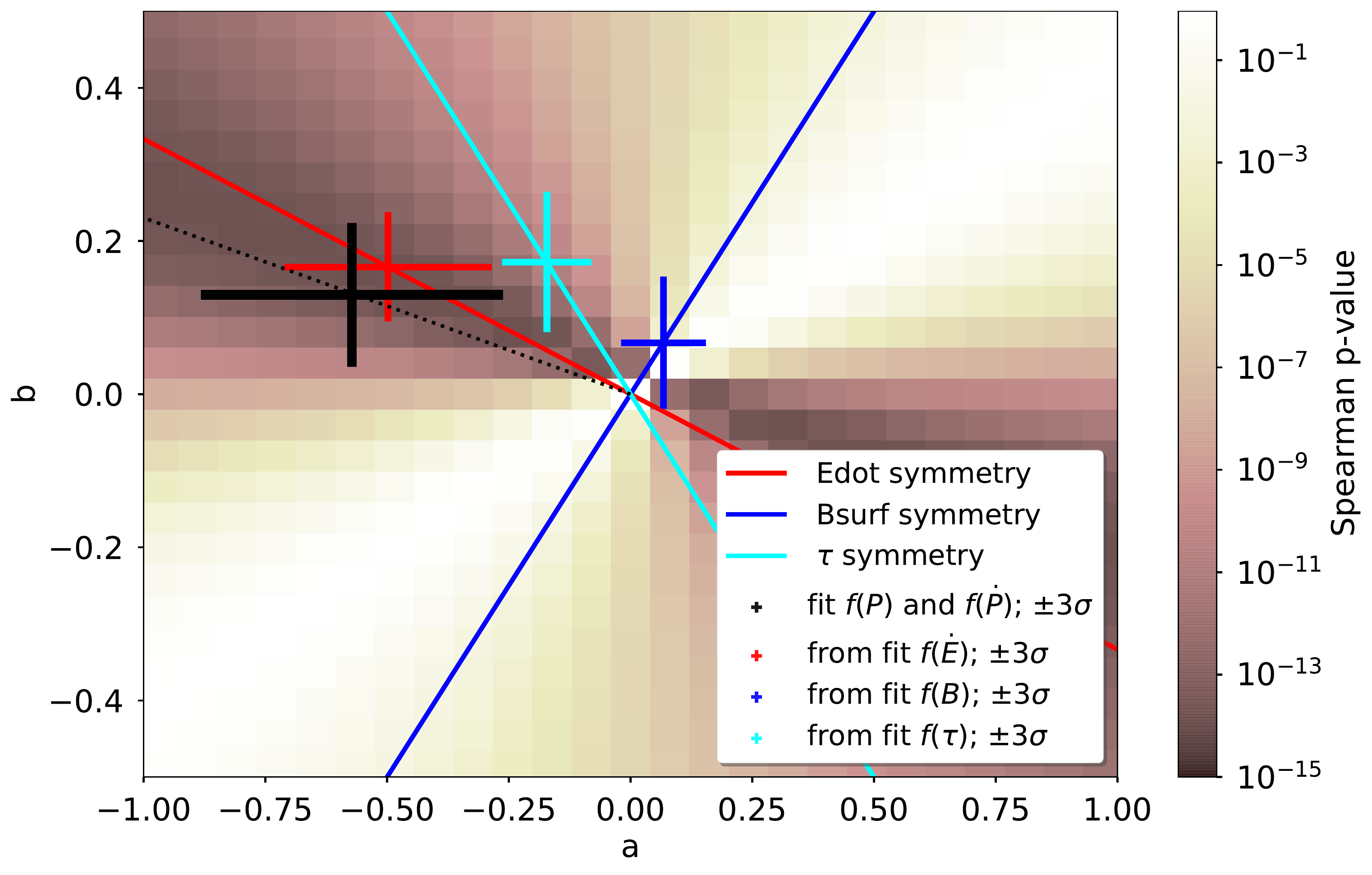}
\caption{The dependency of the spectral index of the TPA pulsars from the (logarithm of the) rotational parameters. The investigated sample only includes pulsars with $m_{SI}<0.13$. The $x$-axis shows a range of powers $a$ in the rotation period $P$, $b$ indicates the power in the period derivative $\dot{P}$. The colour background image shows the Spearman $p$-value for each $a$-$b$ combination with respect to the spectral index. The red, cyan, and blue lines indicate lines dependencies for the spin-down power, characteristic age and magnetic dipole field. Crosses with the same colours correspond to the respective fits to the spectral index, translated into the $a$-$b$ space. The black cross was obtained from separate fits of the spectral index to ${P}$ and $\dot{P}$. All uncertainties in this plot are $3\sigma$ uncertainties.}
\label{fig:SIfits}
\end{figure}

Figure~\ref{fig:SIfits} shows the $a$ and $b$ plane with results from the non-weighted linear OLS fits and the Spearman rank $p$-values for the correlation of the spectral index with $P^a \dot{P}^{b}$. It is evident that the independent fits in $\log P$ and $\log \dot{P}$ give $a$ and $b$ values that are close to the ones implied by the  $\log \dot{E}$-fit. This illustrates that the dependency on $\dot{E}$ is the dominating dependency, in line with the Spearman rank correlation analysis (Table~\ref{tab:polarSpear}) which has the highest rank /lowest $p$-value for $\dot{E}$. The general trend is an increasing value of the spectral index  with an increasing spin-down power.\\

For this analysis, we only use pulsars filtered with $m_{SI}<0.13$.
The derived values from the OLS fits of the spectral index with dependencies $a \log P$, $b \log \dot{P}$, ${c_{\dot{E}}} \log \dot{E}$, ${c_{\tau}} \log \tau$, and ${c_B} \log B$ are 
$a=-0.6 \pm 0.1$, $b= 0.13 \pm 0.03$, 
$c_{\dot{E}}=0.17 \pm 0.02$\footnote{As, an example, $c_{\dot{E}}=0.17 \pm 0.02$ translates into $a=-0.51 \pm 0.06$, $b= 0.17 \pm 0.02$ in Figure~\ref{fig:SIfits}}, 
$c_{\tau}=-0.17 \pm 0.03$, and 
${c_B}= 0.13 \pm 0.06$, respectively (all with $1\sigma$ uncertainties.
The slope for the magnetic field is insignificant which agrees with a low Spearman rank (0.1).\\

\begin{figure}
\includegraphics[width=8.5cm]{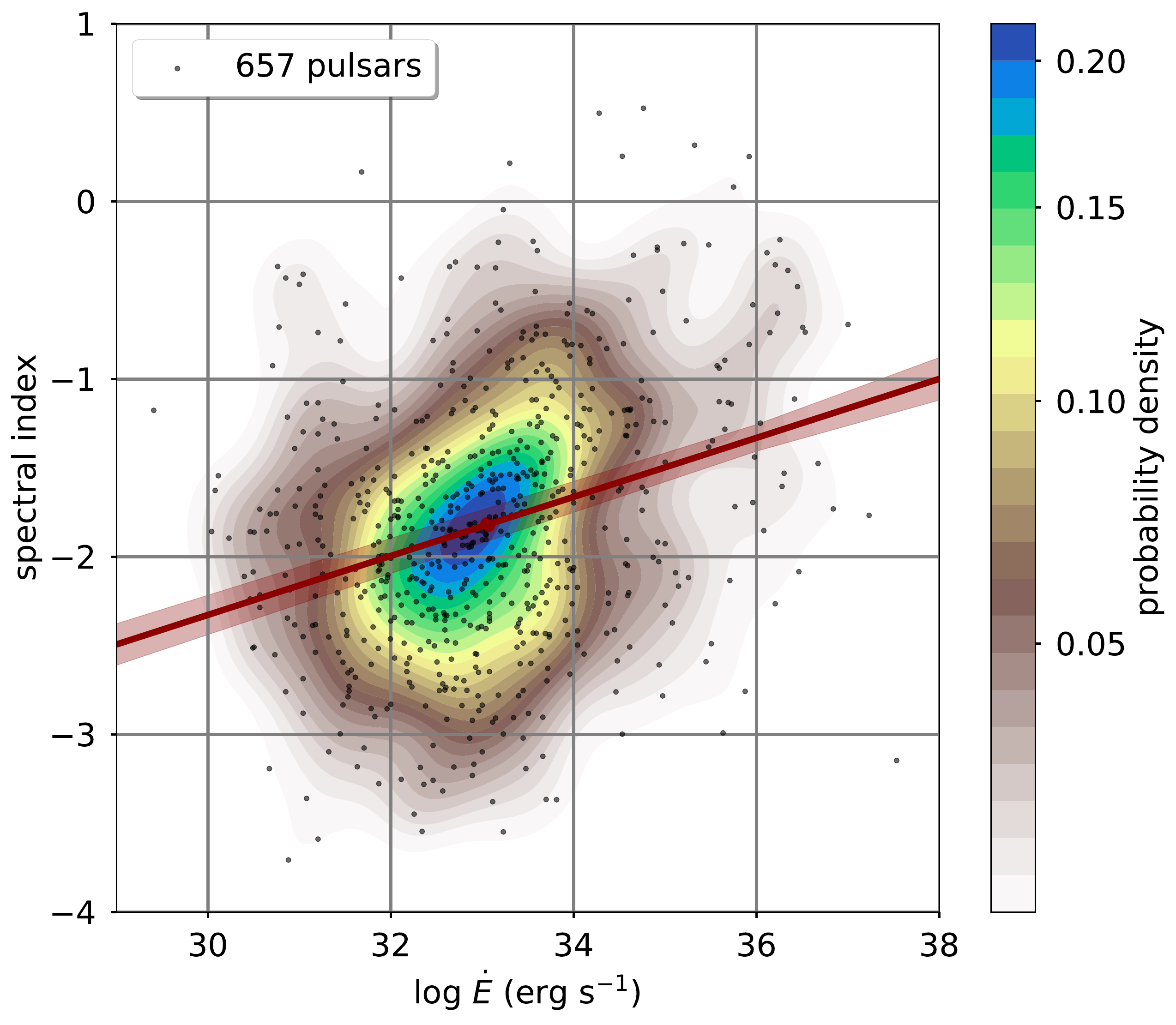}
\caption{The distribution of the spectral index and the logarithm of the spin-down power of the TPA pulsars that fulfil $m_{SI}<0.13$. The probability density is indicated by the color bar (smoothing parameters are chosen similarly as in Figure~\ref{fig:PPdot}). The OLS-fit and its $1\sigma$ uncertainty region are shown with lines and shaded area in darkred.}
\label{fig:SIEdotdist}
\end{figure}

The OLS fit in $\log \dot{E}$ does not follow the region of highest densities of the spectral index distribution (see Figure~\ref{fig:SIEdotdist}) due to a wide and rather non-Gaussian distribution of values along the $y$-axis for individual $\log \dot{E}$-slices (i.e., pear-shape of the overall density distribution). 
Nevertheless, a significant $\log \dot{E}$ dependency is obvious.
The spectral index is getting flatter (increasing its value) with increasing spin-down power. 
This has already been previously noticed in the literature for several pulsars. 
\citet{Johnston2006} concluded that the pulsar beams of high $\dot{E}$ pulsars  consist of a hollow cone, and they pointed out that the conal beam components have in general flatter spectral indices than the core components (as reported by, e.g., \citealt{Lyne1988}), i.e, in line with the hollow cone interpretation.
\citet{Kramer1994} showed that the spectral index tends to be flatter for \emph{any} outer components.
More recently, \citet{Jankowski2018} also reported flatter spectral indices for individual  pulsars that are on average young or energetic which is consistent with our results. Two selection effects were discussed as possible causes of such findings in the past. Firstly, preferably young pulsars constituted the past target samples. Young pulsars tend to be closer to the Galactic plane, where pulsars with
flat spectra are easier to see than ones with steep spectra.
Secondly, low-frequency flux densities of pulsars in the Galactic plane might be affected by hotspots of the radio emission in the Galactic plane. 
Neither did \citet{Jankowski2018} nor did we find any correlation of the spectral index with Galactic latitude ($Gb$),  or longitude.
Despite the known strong correlation between $\dot{E}$ and $|Gb|$, the correlations of the spectral index with  $|Gb|$,  $DM \,\cos|Gb|$ are too weak 
to explain the much stronger correlation of the spectral index with $\dot{E}$. 
Figures~\ref{fig:EdotB} and \ref{fig:SIEdotdist} further demonstrate the wide distribution in $\dot{E}$ and a significant fraction of older pulsars that are considered in our analysis.   
Overall, the dependency of the spectral index on  $\dot{E}$ is a robust result.\\

\begin{figure}
\includegraphics[width=8.5cm]{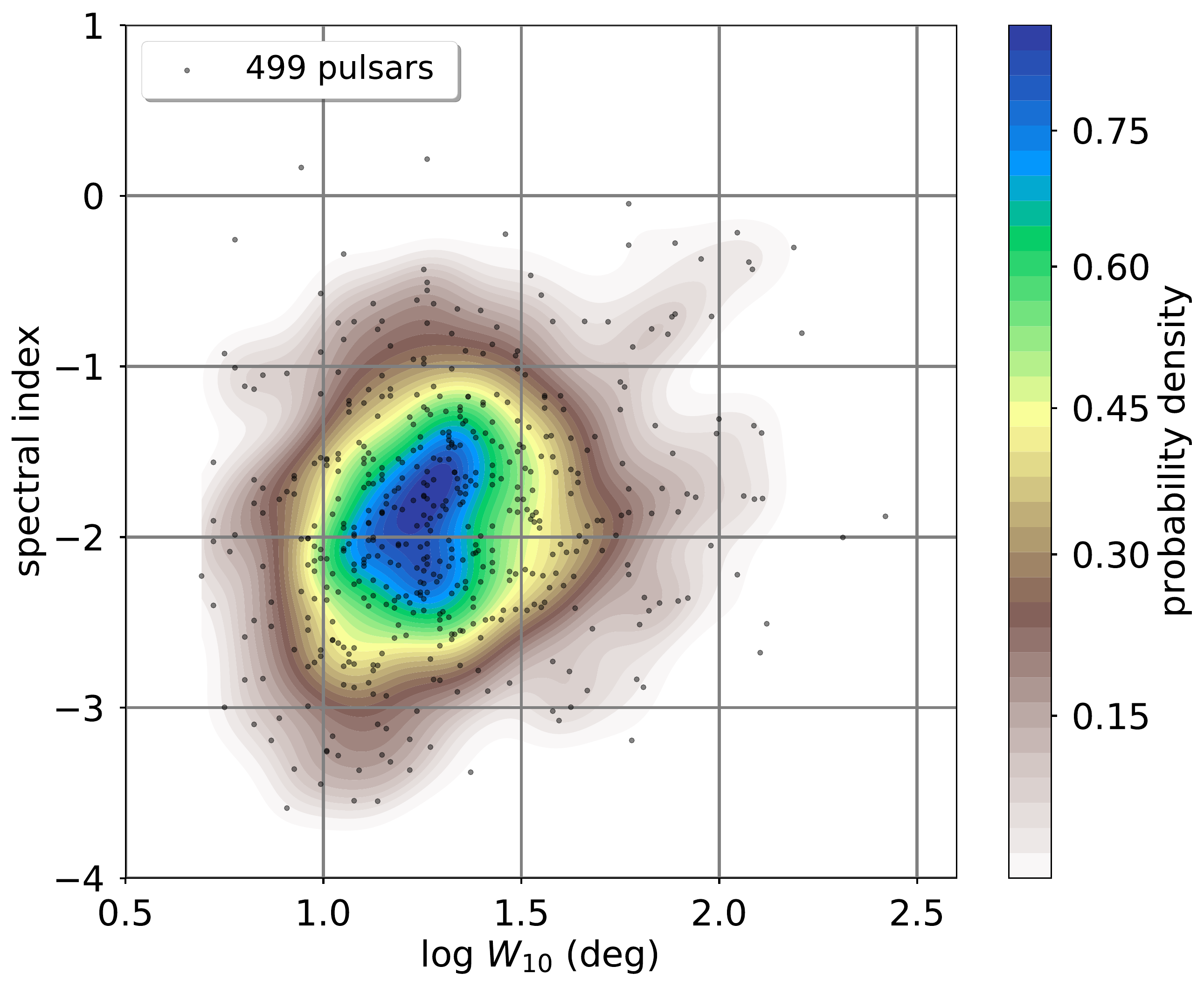}
\caption{The distribution of the spectral index and the logarithm of the pulse width $W_{10}$. The considered TPA pulsars that fulfil $m_{SI}<0.13$ and have a $W_{10}$ significance larger than three. The probability density is indicated by the color bar (smoothing parameters are chosen similarly as in Figure~\ref{fig:PPdot}).}
\label{fig:SIw10dist}
\end{figure}

\citet{Lorimer1995} reported an indication for an anti-correlation between the spectral index and the characteristic age. The latter's anti-correlation with $\dot{E}$ in turn suggested a correlation of the spectral index with $\dot{E}$ according to \citet{Hoensbroech1998a}.
The TPA data now confirms this with the constraining statistics of a large sample. 
Within the Goldreich-Julian model of the pulsar magnetosphere \citep{Goldreich1969}, the potential drop between the magnetic pole and last open field line (the polar cap) is responsible for accelerating the charges creating the radio emission. 
A higher $\dot{E}$ can accelerate more particles to higher energies, flattening their energy distribution  and thus the spectral index.
Whilst our ansatz of $\propto \log \dot{E}$ captures some of the underlying dependencies, 
Figure~\ref{fig:SIEdotdist} illustrates that further work is needed to better describe the spectral index distribution, likely requiring additional parameters. 
One such parameter could be, for instance,  a preferred misalignment angle between rotation and magnetic axes.
Based on the TPA data, another parameter could be 
the pulse width $W_{10}$ whose distribution is shown in Figure~\ref{fig:SIw10dist} and for which the Spearman rank analysis (Table~\ref{tab:polarSpear}) hints at a possible correlation. We leave a more thorough exploration of this potential dependency to a future work.
  
\subsection{Luminosity correlations}
\label{sec:lumos}
\begin{figure}
\includegraphics[width=8.5cm]{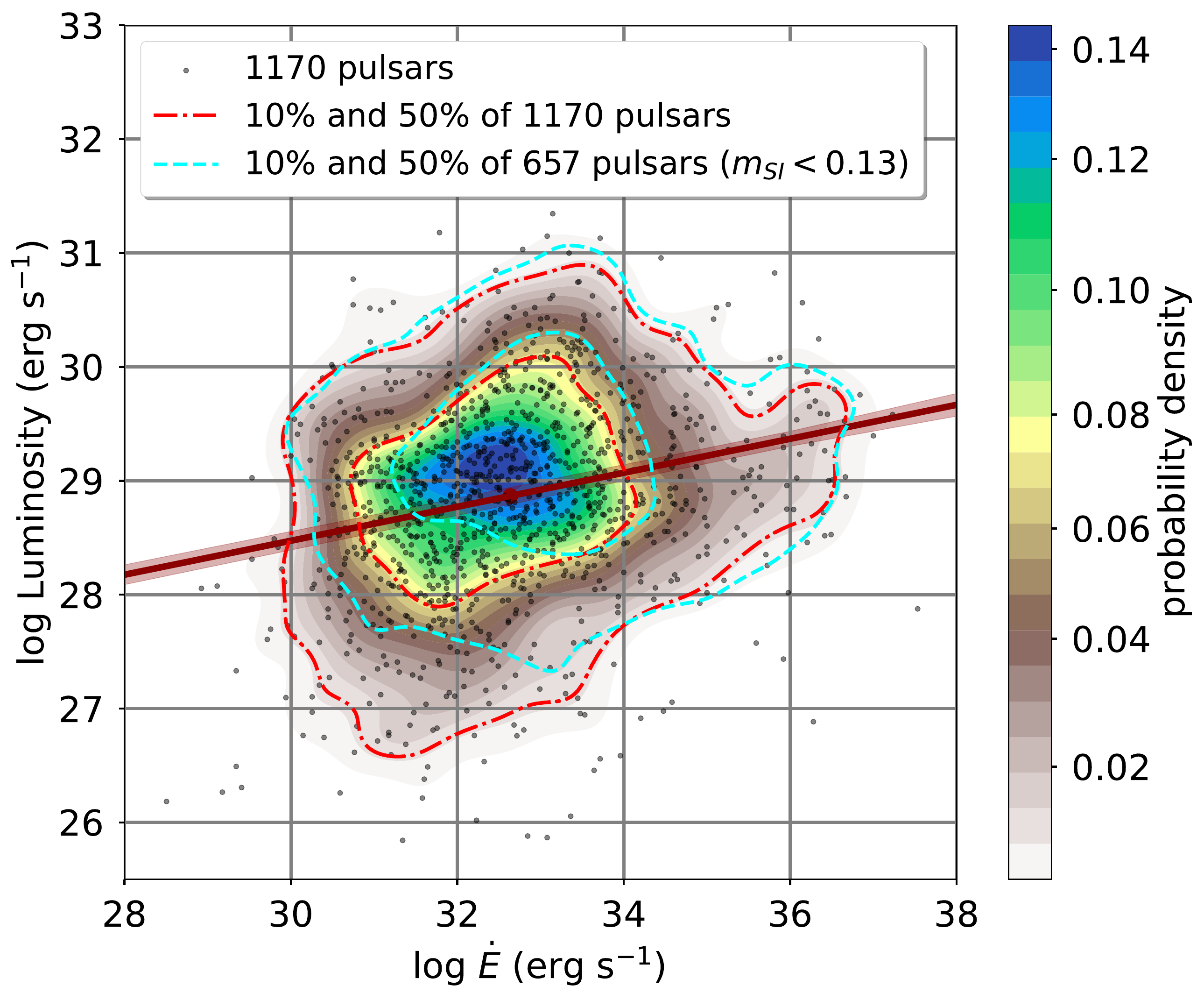}
\caption{The distribution of the luminosities and spin-down power of 1170 TPA pulsars. The  probability density is indicated by the color bar (smoothing parameters are chosen similarly as in Figure~\ref{fig:PPdot}), the 10\% and 50\% probability levels are shown with the dashed-dotted line. The OLS-fit and its $1\sigma$ uncertainty are shown with lines and shaded area in darkred. The cyan dashed line indicates the 10\% and 50\% probability levels of the distribution if the TPA pulsars are filtered in addition for the pseudo modulation index $m_{SI}$.}
\label{fig:LumEdotdist}
\end{figure}
\begin{figure}
\includegraphics[width=8.5cm]{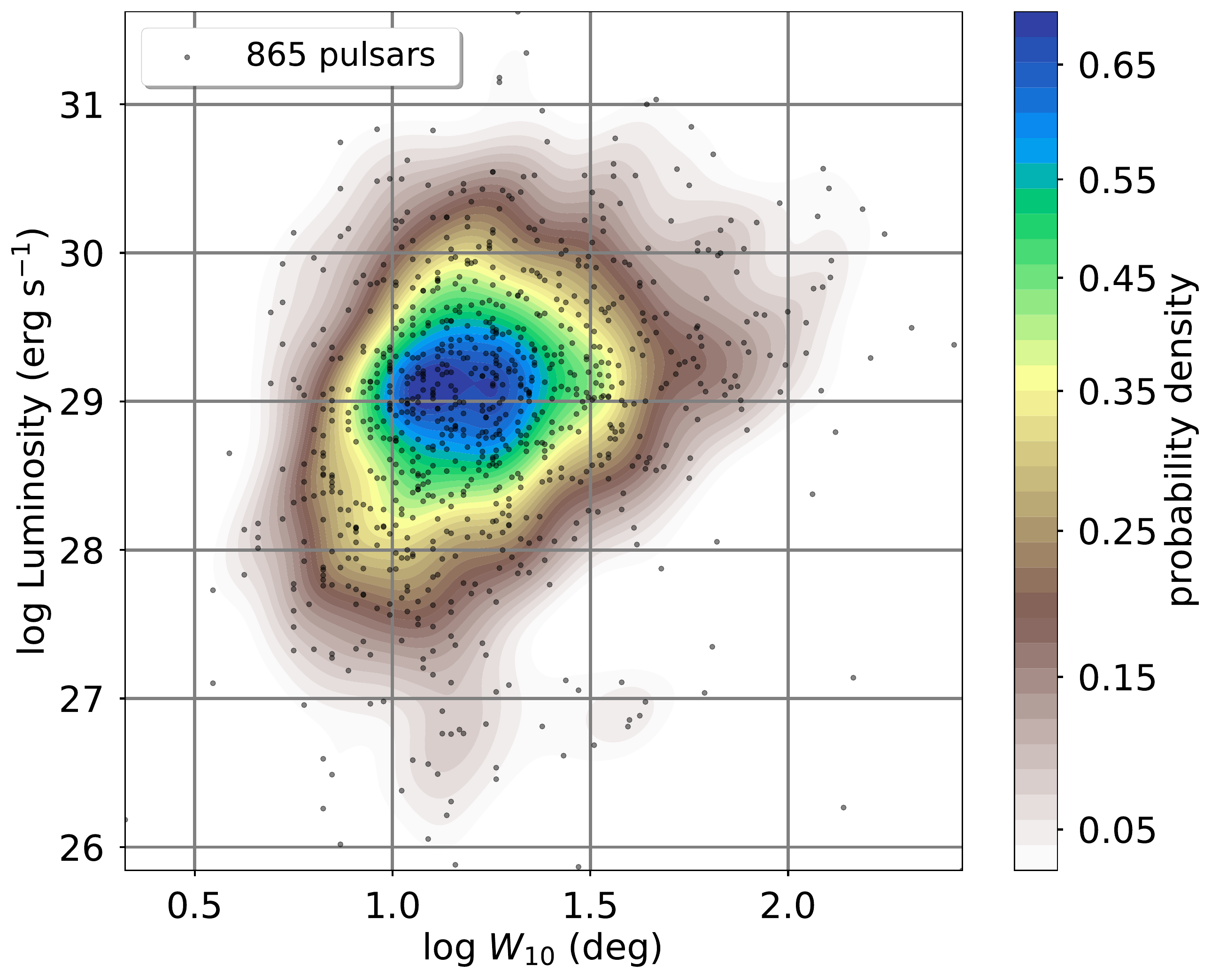}
\caption{The distribution of the luminosities and pulse width $W_{10}$ (with significance $>3$). The  probability density is indicated by the color bar (smoothing parameters are chosen similarly as in Figure~\ref{fig:PPdot}).}
\label{fig:LumW10dist}
\end{figure}
\begin{figure}
\includegraphics[width=8.5cm]{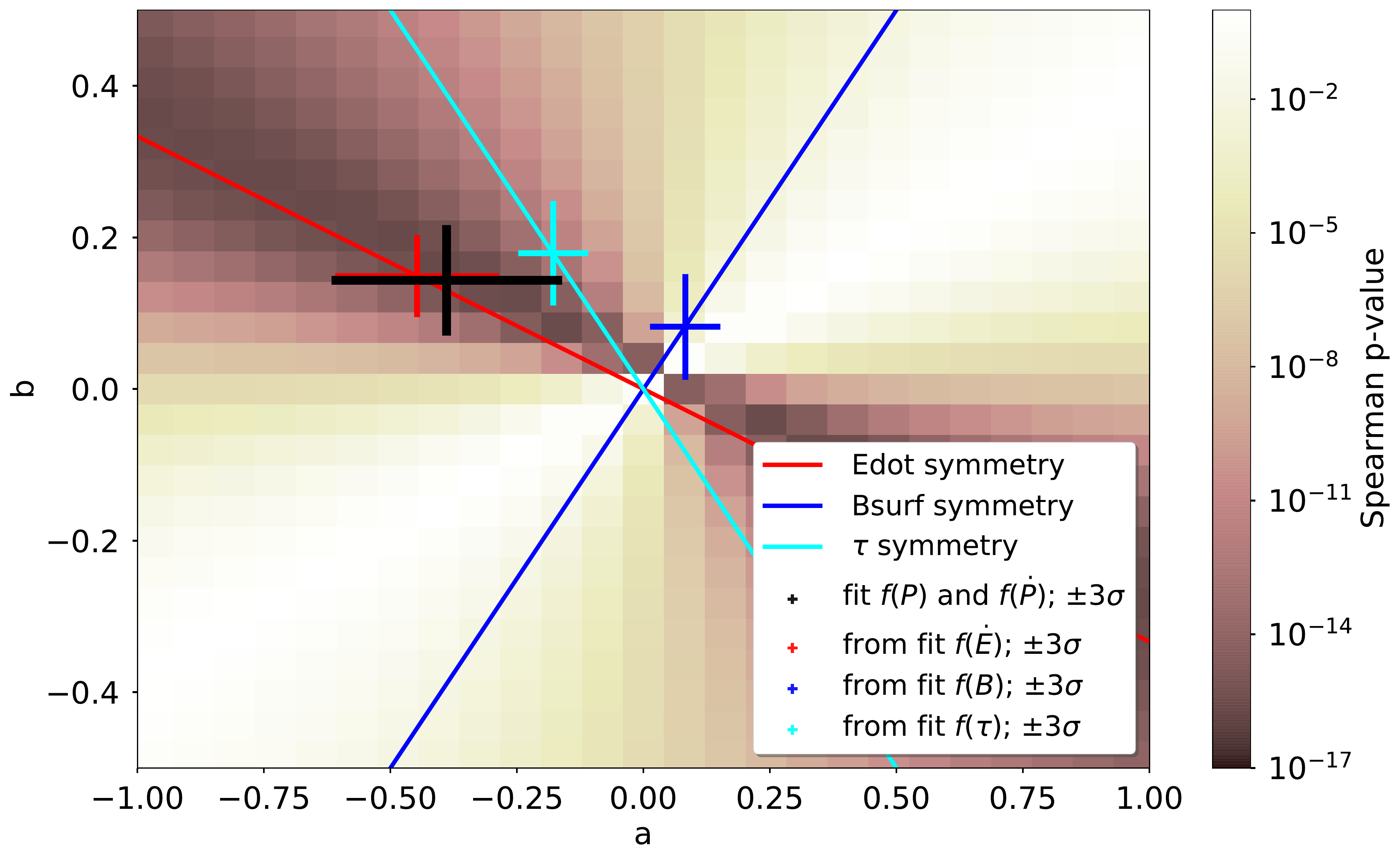}
\caption{The power law dependency of the TPA luminosity from the rotational parameters, see the similar Figure~\ref{fig:SIfits} for detailed description of what is shown.}
\label{fig:LumEdotfit}
\end{figure}

We find the strongest indications for correlations of the luminosity with $\dot{E}$ (pulsar distribution shown in Figure~\ref{fig:LumEdotdist}) and with $W_{10}$ (pulsar distribution shown in Figure~\ref{fig:LumW10dist}) based on the Spearman rank correlation analysis (Table~\ref{tab:polarSpear}).
A correlation with $W_{10}$ is not suprising since more of the radio beam is seen for larger $W_{10}$. 
Note that we use mean fluxes for our luminosity estimates, i.e., the duty cycle is already taken into account. 
The luminosity is found to increase with increasing spin-down energy, as illustrated by the derived PL from the $\dot{E}$ fit in Figure~\ref{fig:LumEdotdist}. 
Figure~\ref{fig:LumEdotfit} demonstrates the dominating dependency of the luminosity on the spin-down power with a plot of the OLS fit results in the $a-b$ plane (i.e., $\propto P^a \dot{P}^b$), representing the PL-fits for $P$, $\dot{P}$, $\dot{E}$, $\tau$, $B$. Spearman $p$-values for the $a-b$ plane are shown in the background.
However, Figure~\ref{fig:LumEdotdist} also shows that for similar $\dot{E}$ values, there is a large spread in luminosities, covering up to four orders of magnitude, indicating that additional pulsar or viewing parameters influence the luminosity distribution.\\

As summarized by \citet{Bagchi2013}, for the rotation period and its derivative, a PL-fit approach such as ours ($\propto P^a \dot{P}^b$) is common practice in order to evaluate the radio luminosity distribution. In Table~\ref{tab:lumosRelations}, we list respective recent references. 
The exponent derived for the magnetic field is nearly insignificant $c_B=0.16 \pm 0.05$, whilst for the other rotational parameters the PL exponent has a significance of 5 or larger. Figure~\ref{fig:LumEdotfit} illustrates that the $a=-0.39 \pm 0.08$ and $b=0.14 \pm 0.02$ from the independent $P$ and $\dot{P}$ fits 
agree well with the separate $\dot{E}$ fit ($c_{\dot{E}}=0.15\pm 0.02$). Thus, in terms of the considered rotational parameters, the radio luminosity distribution can be expressed as a function of $\dot{E}$ only.\\
 
\begin{table*}
  \caption{A selection of recent luminosity relations assuming the  PL-dependencies with $P$, $\dot{P}$, $\dot{E}$, $\tau$. $\#$ PSRs indicates the number of pulsars used (if known).} 
  \label{tab:lumosRelations}
  \begin{tabular}{lcc|cc| cc}
   Reference & $a$ (for $P$) & $b$ (for $\dot{P}$) & $c_{\dot{E}}$  & $c_{\tau}$ & $\#$ PSRs & frequency\\
    \hline
    this work (YMW16-based distances) & $-0.39 \pm 0.08$  & $0.14 \pm 0.02$ & $0.15\pm 0.02$ & $-0.18\pm 0.02$ & 1170 & 1.3\,GHz\\
    this work (NE2001-based distances) & $-0.43 \pm 0.07$  & $0.21 \pm 0.02$ & $0.19\pm 0.02$ & $-0.25\pm 0.02$ & 1170 & 1.3\,GHz\\
    \hline
\citet{Wu2020}  & & & $0.06 \pm 0.01$&  &  & 1.4\,GHz\\
\citet{Johnston2017} & $-0.75$ & 0.25 & & &  & 1.4\,GHz\\
\citet{Bates2014} & $-1.39 \pm 0.09$ & $0.48 \pm 0.04$ & & & & $0.4-6.6$\,GHz\\
\citet{Gullon2014} & $[-1.5,-1.2]$ & $[0.4,0.5]$ & $[0.42,0.46]$ & & & 1.4 GHz\\
\citet{Szary2014} & & & 0.10 &  & 1436 & 1.4\,GHz\\
\citet{Ridley2010} & $-1.0$ & 0.5 & & & & 1.4 GHz\\
\citet{Faucher2006} & $-1.5$ & 0.5 & & & & 1.4 GHz\\
\citet{Arzoumanian2002} (for $n=3$)& $-1.3$ & 0.4 & & & & true $L$ \\
  \end{tabular}
 \end{table*}

\begin{figure*}
\includegraphics[width=8.5cm]{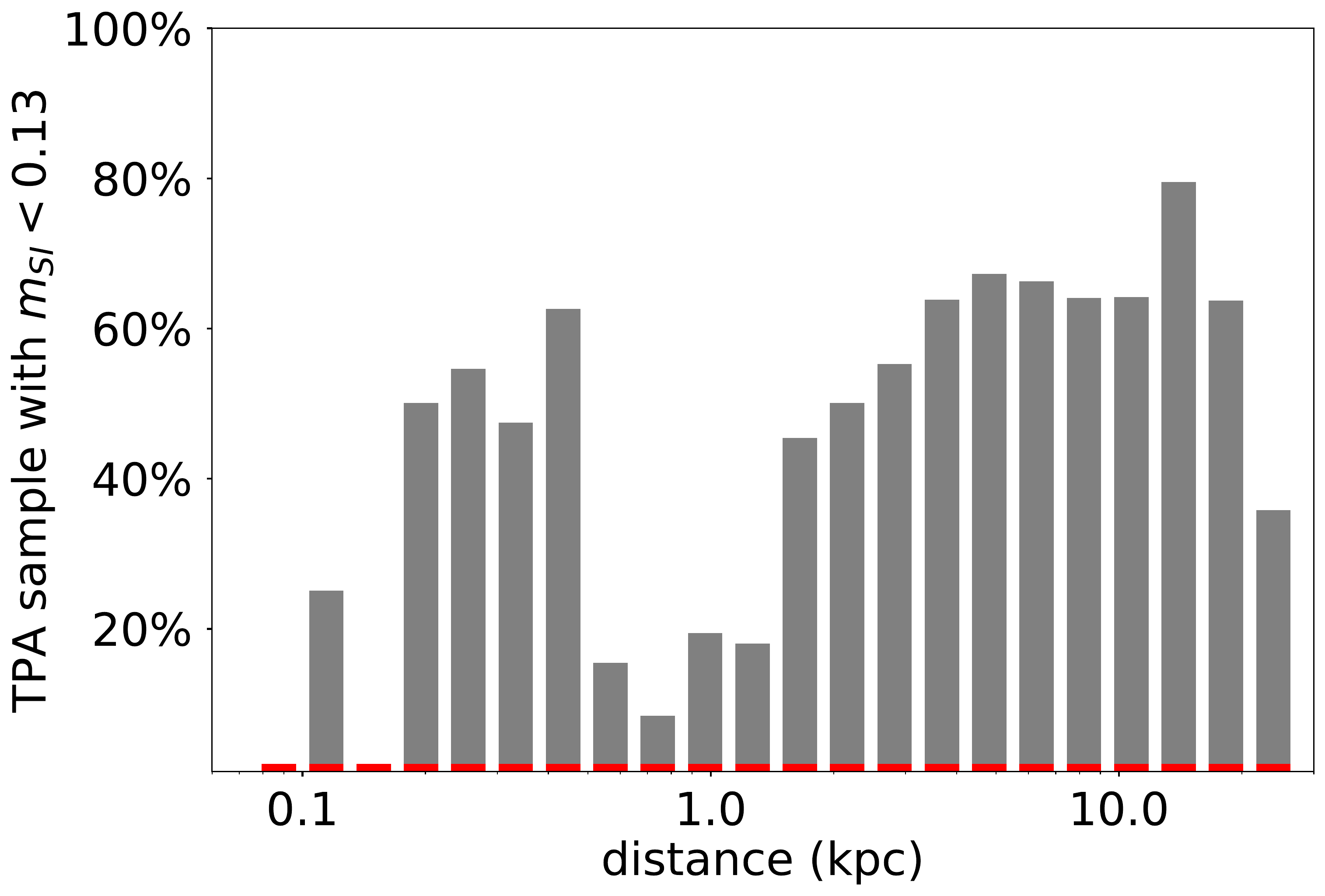}
\includegraphics[width=8.5cm]{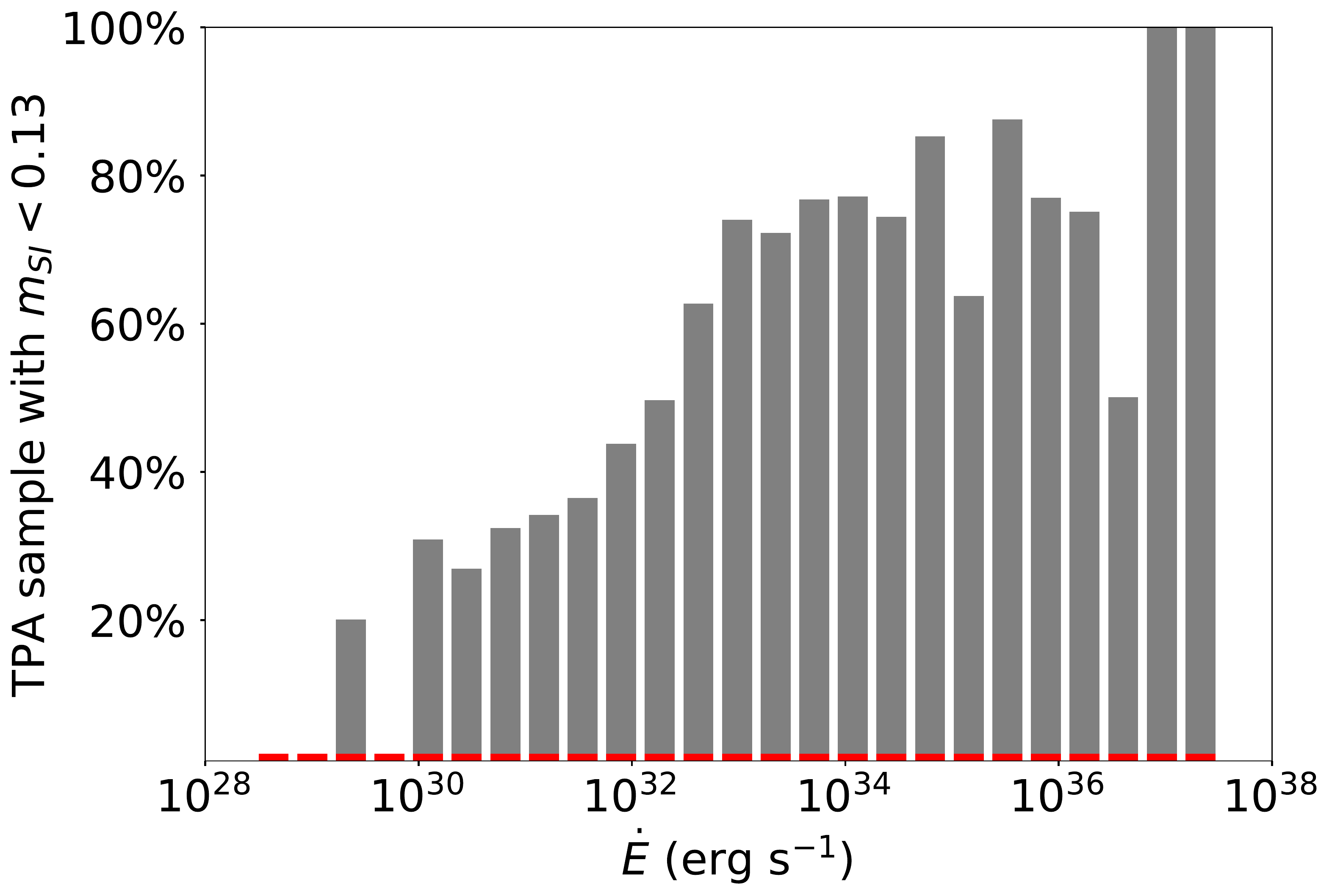}
\caption{The selection effects of a pseudo modulation index $m_{SI}$ restriction which was used to remove potentially scintillating pulsars in the TPA sample. The grey histograms (left panel: distance histogram, right panel: histogram of the spin-down power) show the percentages of the total sample \emph{per bin}, once $m_{SI}<0.13$ has been employed. The red lines indicate the bins with TPA pulsars before filtering.}
\label{fig:DistEdotSI}
\end{figure*}

Since the luminosity depends on the distance, and many distances are derived from the DMs, thus are dependent on the applied electron density model (YMW16, \citealt{Yao2017}, the standard choice in the ATNF pulsar catalog), we checked whether our results change if a different electron density model (NE2001, \citealt{Cordes2002}) is used, see Table~\ref{tab:lumosRelations}.  
For the $\dot{E}$ fit, $c_{\dot{E}}$ slightly increases from $0.15\pm 0.02$ (YMW16) to $0.19\pm 0.02$ (NE2001). Although these exponents (as well as those the other rotational parameters) are consistent within  $3\sigma$ uncertaintites, they also indicate some systematic uncertainty (on the order of 0.02) due to choice of the electron density model\footnote{Interestingly, the exponent for $B$ changed from $c_B=0.16 \pm 0.05$ to $0.27\pm 0.04$.}. 
Excluding all DM-based distances, and restricting to the 73 pulsars that either have a parallax or other alternative listed distances lead to a consistent, but statistically insignificant result ($c_{\dot{E}}=0.18\pm 0.09$). Overall, we find $c_{\dot{E}}=0.15\pm 0.04$ (statistical and systematic $1\sigma$ uncertainty) to be a robust result for the $\dot{E}^{c_{\dot{E}}}-$luminosity dependency.\\

Table~\ref{tab:lumosRelations} shows a large spread of the literature values of PL-dependencies of the radio luminosity for $P$, $\dot{P}$, $\dot{E}$. Our obtained best-fit parameters do not agree with the earlier estimates, for example they are inconsistent with the luminosity law assumed by \citet{Faucher2006}. 
Our results agree better with the more recent publications, although the consistency of the values is often difficult to assess due to the lack of reported error estimates.\\ 

Finding a relation of the radio luminosity with the spin-down power is not surprising, based on observational as well as theoretical arguments. 
 If the luminosity were independent of  $\dot{E}$, the observed population of pulsars in the $P-\dot{P}$ diagram would show a ``pileup'' in  the number of observed objects at low $\dot{E}$ values, yet the observed radio pulsar population is clearly peaked in the ``centre'' of the population (Figure~\ref{fig:PPdot}). 
Since the potential drop over the polar cap scales with $\dot{E}^{0.5}$ (e.g., \citealt{Lorimer2012}), an $\dot{E}$-dependency of the radio luminosity can be expected. According to our results from the consistent TPA data, the \emph{population-wide} radio luminosity distribution scales with a shallower $\dot{E}$-exponent than 0.5.
The shallower luminosity distribution implies that not only high-$\dot{E}$ pulsars can be seen at larger distances. Rather, the  likelihood for finding new pulsars over a larger $\dot{E}$ range should be increased for estimates of expectations for future surveys, e.g. with the SKA.\\ 

We note, that if we filter the TPA pulsars for those with good PL-fits using $m_{SI}<0.13$ (removing, for example, scintillating sources), we find that the $\dot{E}$-dependency of the radio luminosity disappears ($c_{\dot{E}}=0.04\pm 0.02$, 657 pulsars). A clue to understanding this finding comes from Figure~\ref{fig:LumEdotdist}. It shows the 10\% and 50\% contours (in red) of the $\dot{E}-$luminosity distribution of the 1170 pulsars used for our fits, and it also shows  the cyan 10\% and 50\%  contours of the 657 TPA pulsars that fulfil $m_{SI}<0.13$. The comparison of the 10\% level contours (i.e., 90\% of the pulsars are within this contour) illustrate that a large chunk of low-$\dot{E}$, low-luminosity pulsars are removed. 
Figure~\ref{fig:DistEdotSI} shows the fraction of pulsars that remain after  $m_{SI}<0.13$ is applied, both with respect to the distance as well as to the $\dot{E}$ distribution. From Figure~\ref{fig:DistEdotSI}, it is clear that the filter removes preferably low-distance pulsars, sometimes up to 100\% in a distance bin. Since scintillation is more noticable for low-distance pulsars (the scintillation bandwidth decreases with $d^{-2.2}$ \citealt{Lorimer2012}), this behaviour of our pseudo modulation index filter is expected. However, because of an observational selection bias in the known pulsar population, there are many low-$\dot{E}$ pulsars at relatively close distances and an increasing fraction of brighter high-$\dot{E}$ pulsars at larger distances. The net result is a substantial removal of low-$\dot{E}$ pulsars (Figure~\ref{fig:DistEdotSI}) and low-luminosity pulsars. The inclusion of the low-$\dot{E}$ pulsars is crucial to allow us to constrain the relatively weak $\dot{E}-L$ dependency. The seen effect of the modulation index filter calls for caution with respect to other selection biases.\\

\begin{figure}
\includegraphics[width=8.5cm]{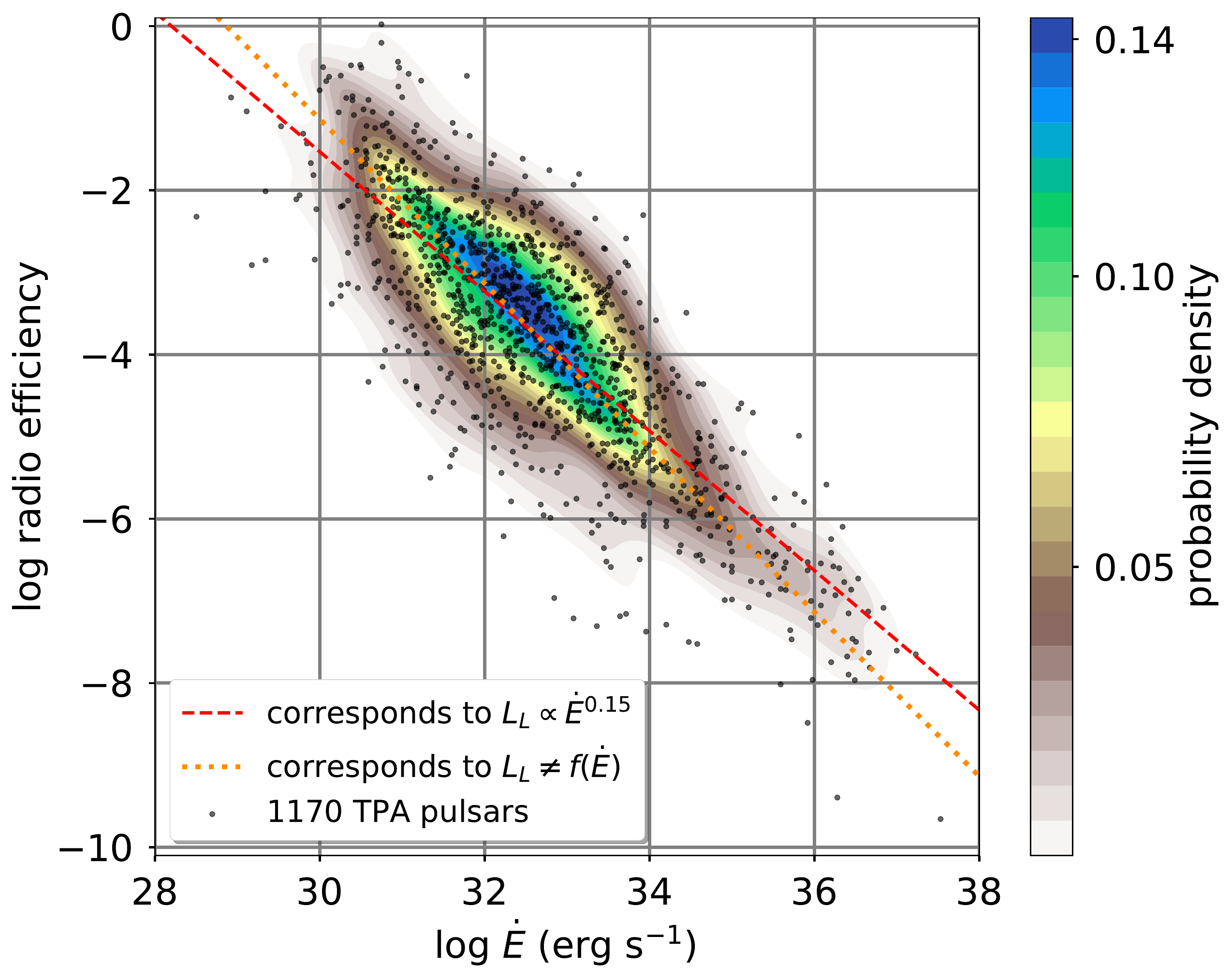}
\caption{The radio efficiency distribution of the TPA pulsars. The probability density is indicated by the color bar (smoothing parameters are chosen similarly as in Figure~\ref{fig:PPdot}). Since our luminosities are pseudo luminosities, the actual radio efficiency for a specific pulsar can be different. For illustration, the slope of the orange line represents a luminosity independent of $\dot{E}$, and the slope of the red line represents the $L_L \propto \dot{E}^{0.15}$ dependency derived in this work (Table~\ref{tab:lumosRelations}). The two lines are centered on the average logarithm values of $\dot{E}$ and $L_L$ of the TPA sample.}
\label{fig:effidist}
\end{figure}

For completeness, we use our (pseudo) luminosities to calculate the radio efficiencies, $L_{\rm L} / \dot{E}$. The distribution of these radio efficiencies is shown for 1170 pulsars in Figure~\ref{fig:effidist}. It is very similar to the distribution of normal pulsars that was presented and discussed in detail by \citet{Szary2014}. 
The radio efficiency is $\propto {\dot{E}}^{-0.85}$, a result of the very shallow luminosity dependence on $\dot{E}$.
This is in contrast to the GeV $\gamma$-ray efficiency that has a proportionality of $\propto {\dot{E}}^{-0.54}$ (\citealt{Szary2014}; for $\approx 50$ normal pulsars) due to the steeper dependency of the  $\gamma$-ray luminosity on $\dot{E}$ as measured with \emph{Fermi} (e.g., \citealt{Abdo2013}).
While the bulk of the TPA radio pulsars have spin-down powers between $10^{31}$\,erg\,s$^{-1}$ and $10^{34}$\,erg\,s$^{-1}$, most $\gamma$-ray detected normal pulsars have $\dot{E}>10^{34}$\,erg\,s$^{-1}$. The scatter range (for similar $\dot{E}$) is typically four orders of magnitude for the radio pulsars and three orders of magnitude for the $\gamma$-ray pulsars.   
Most X-ray detections are also reported for $\dot{E}>10^{34}$\,erg\,s$^{-1}$. However, the (less filled) coverage extends down to $10^{30}$\,erg\,s$^{-1}$ thanks to a few detected old ($>1$\,My) pulsars. The non-thermal X-ray efficiency at $1-10$\,keV (dominated by emission from the magnetosphere) does not appear to follow a simple power law, but shows some indications of higher efficiencies at both ends of the $\dot{E}$ distribution. Although number statistics is still low, this trend becomes apparent for $\dot{E}<10^{34}$\,erg\,s$^{-1}$  where efficiencies of $\sim 0.01$ are reached in comparison to $\sim 10^{-4}$ for the bulk of X-ray detected pulsars (e.g., \citealt{Posselt2012} and \citealt{Vahdat2022}; for $\approx 100$ pulsars). The scatter range covers about two orders of magnitude.  
At optical-UV wavelengths, pulsars are faint and if they are detected, it can be very difficult to differentiate between thermal and magnetospheric emission components. 
Figure~\,4 of \citet{Shibanov2016} may indicate for the $V$-band a similar trend as in X-rays. However, with just 13 objects in this plot, the number statistics are very poor. A recent measurement\footnote{The value of $\dot{E}$ is corrected for the Shklovskii effect.} for the old, $\dot{E}=5 \times 10^{30}$\,erg\,s$^{-1}$, pulsar J0108$-$1431 found a record high optical-UV efficiency of $1-6 \times 10^{-4}$ \citep{Abramkin2021} in comparison to the typical optical efficiency range of $10^{-7} -10^{-5}$ for normal pulsars, supporting the hint of increasing efficiencies at lower $\dot{E}$.
The emission efficiency from radio to $\gamma$-ray is complex and connections across the wavelengths are non-trivial and clearly require more parameters than $\dot{E}$.  Different emission sites can complicate the interpretation, and different pulsar parameters such as the angle between magnetic and rotation axes can influence the luminosities \citep{Philippov2015}.
Different emission mechanisms could explain the different efficiency dependencies, as \citet{Szary2014} speculated in the case of the radio and the high-energy emission. \citet{Kisaka2017}, considering synchrotron emission for the optical$-\gamma$-ray range, discussed non-dipolar field structures to explain the trend of higher efficiencies in X-rays at low $\dot{E}$ values.    

\subsection{Correlations of the polarization fractions}
\label{sec:pol}
\begin{figure}
\includegraphics[width=8.5cm]{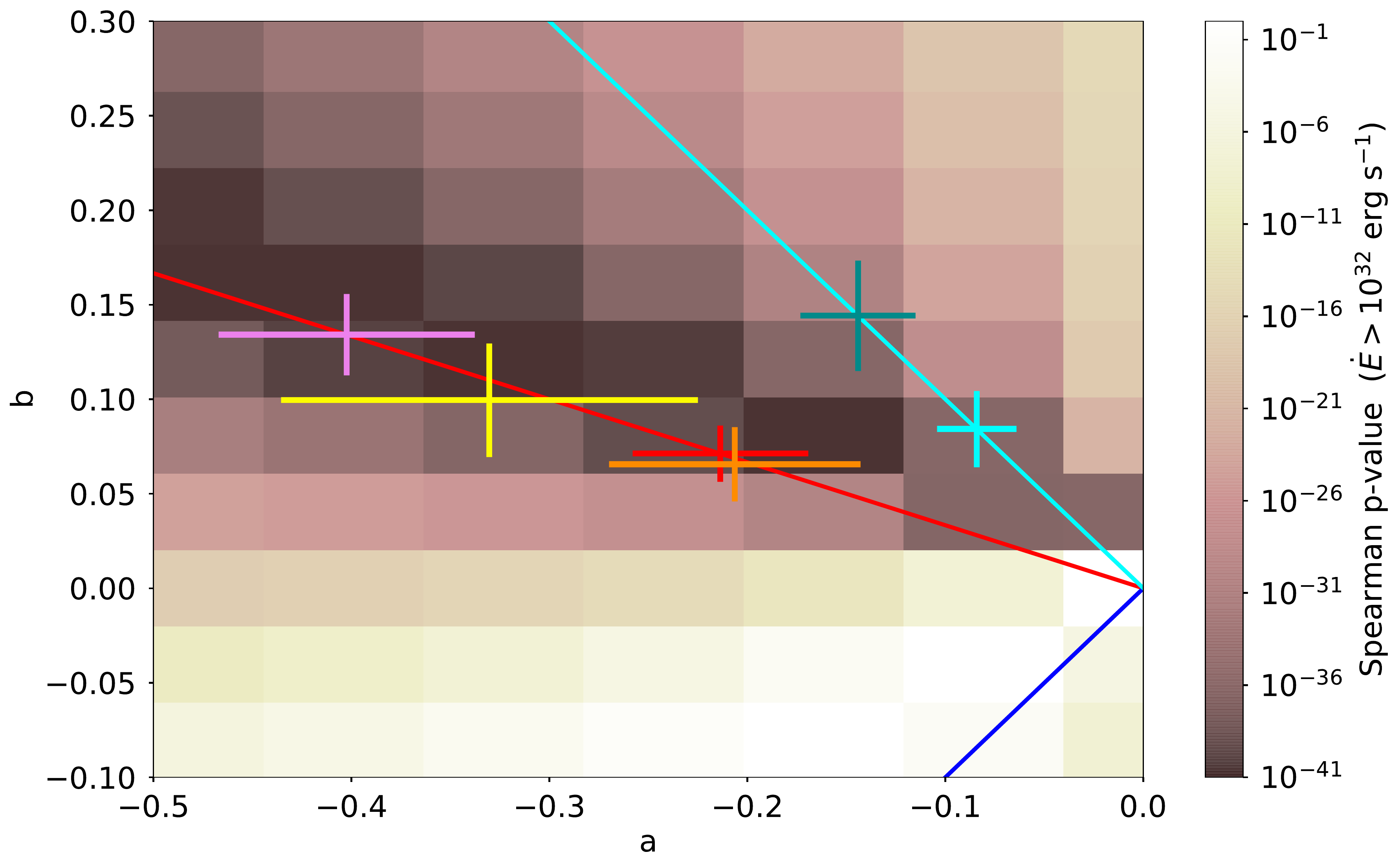}
\caption{The dependency of the LPF from the rotational parameters. The $x$ and $y$-axes show a range of powers: $a$ for $P$, and $b$ for $\dot{P}$. The colour background image shows the Spearman p-value for each $a$-$b$ combination with respect to the LPFs for the 684 TPA pulsars with $\dot{E}>10^{32}$\,erg\,s$^{-1}$. The red, cyan, and blue lines indicate lines of PL-dependencies for the spin-down power, characteristic age and magnetic dipole field.
Crosses with the same colours correspond to the respective PL fits to the LPFs, translated into the $a$-$b$ space, but for \emph{all 1054} TPA pulsars with valid LPF measurements and all rotational parameters. 
The violet and darkcyan crosses correspond to fits using 684 TPA pulsars with $\dot{E}>10^{32}$\,erg\,s$^{-1}$
The orange (1054 pulsars) and yellow (684 pulsars) crosses were obtained from separate fits of the LPFs to ${P}$ and $\dot{P}$. 
This plot shows $3\sigma$ uncertainties.}
\label{fig:LPEdotfit}
\end{figure}

Previous works reported high linear polarization for high $\dot{E}$ pulsars (e.g., \citealt{Weltevrede2008b, Johnston2006}), and this correlation is even stronger at higher frequencies (few GHz, \citealt{Hoensbroech1998b}). 
Indeed, we also find a strong positive correlation of the LPFs with $\dot{E}$ in the TPA data. 
The non-weighted linear fit results of the $\log P$ and $\log \dot{P}$ (orange cross in Figure~\ref{fig:LPEdotfit}) agree with the $\log \dot{E}$ fit result (red cross in Figure~\ref{fig:LPEdotfit}), demonstrating the 
 $\dot{E}$ dependency is the actual underlying and dominating dependency amongst the rotational parameters.
From the Spearman rank correlation analysis for the LPF, we derive high ranks and low $p$ values for $P$, $\dot{P}$, $\dot{E}$, and  $\tau$ as expected, but also for $W_{10}$, and the spectral index (Table~\ref{tab:polarSpear}).\\

The increase of LPFs with increasing $\dot{E}$ was predicted for natural wave
modes in the cold plasma approximation \citep{Hoensbroech1998b}. In particular, the two plasma modes (X and O-modes) are expected to have an increasing difference of their refractive indices with increasing $\dot{E}$. This leads to an easier angular separation, preventing the modes from mixing, and therefore preventing depolarization. For high $\dot{E}$ one thus may see only one or the other polarization mode. 
In addition, \citet{Johnston2006} noted that emission heights (where the radio emission leaves the pulsar magnetosphere) are larger for high-$\dot{E}$ pulsars. 
Larger emission heights mean stronger rotation effects aiding a separation of the two polarization modes \citep{Wang2015}.\\

\begin{figure*}
\includegraphics[width=5.8cm]{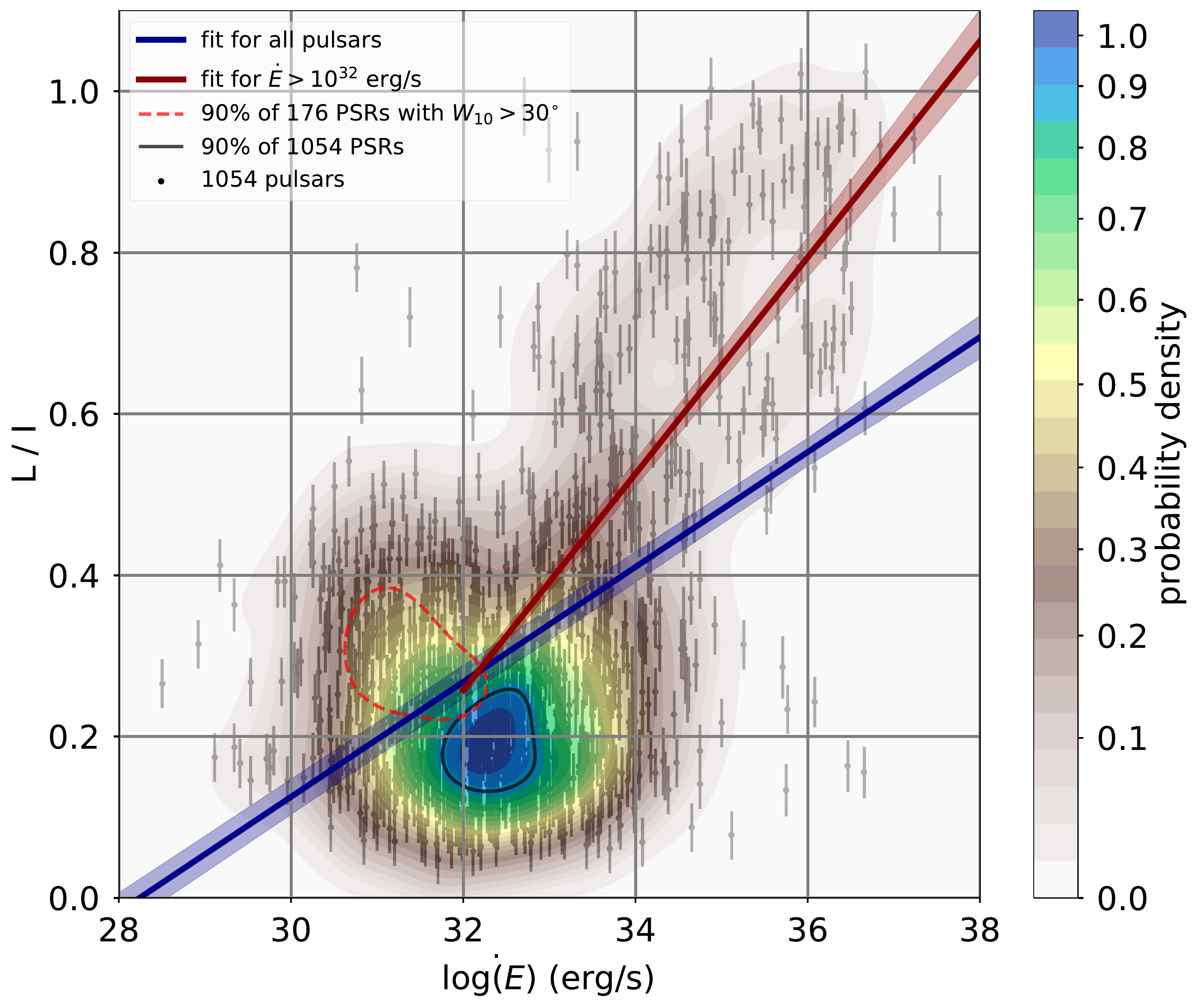}
\includegraphics[width=5.8cm]{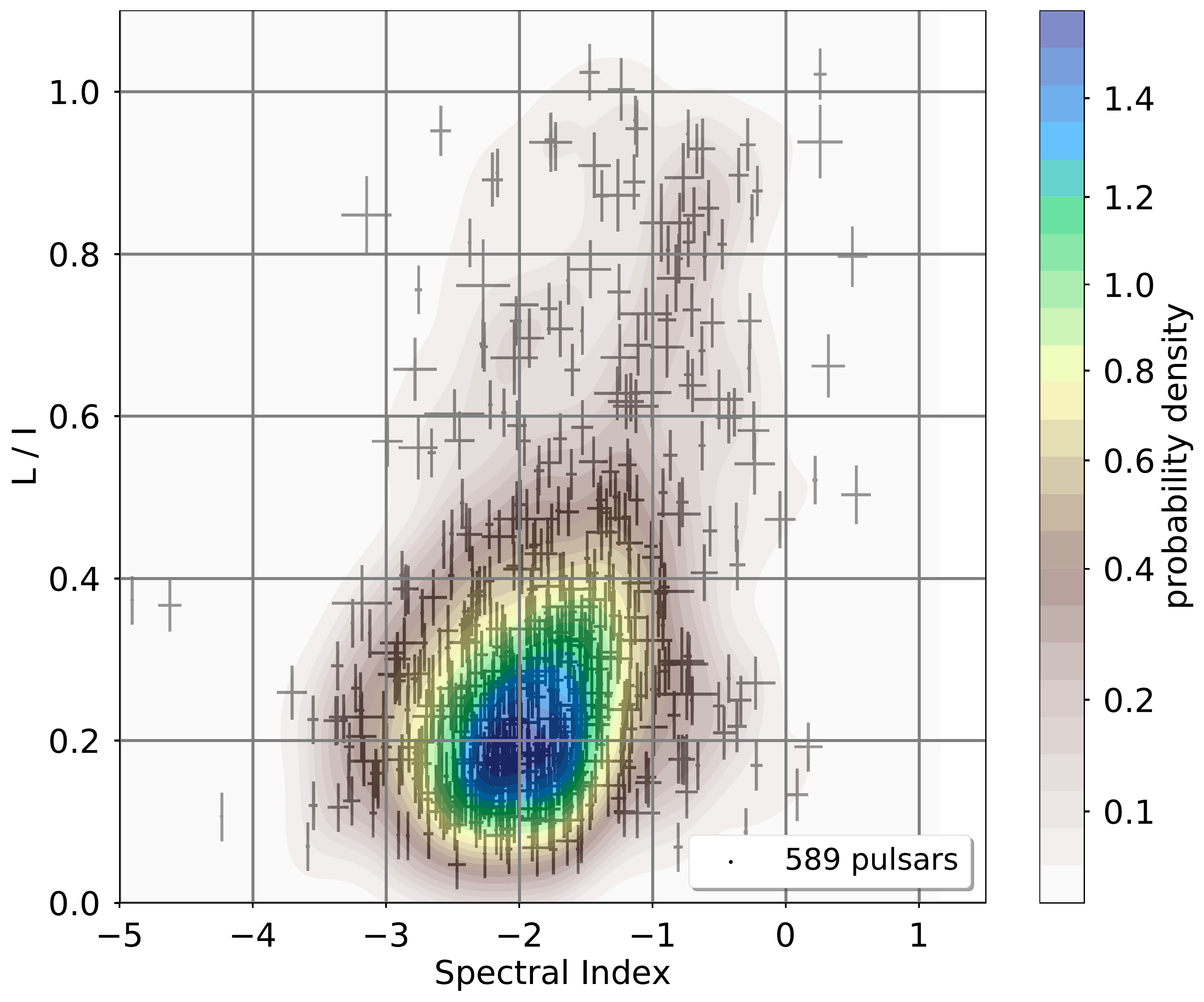}
\includegraphics[width=5.8cm]{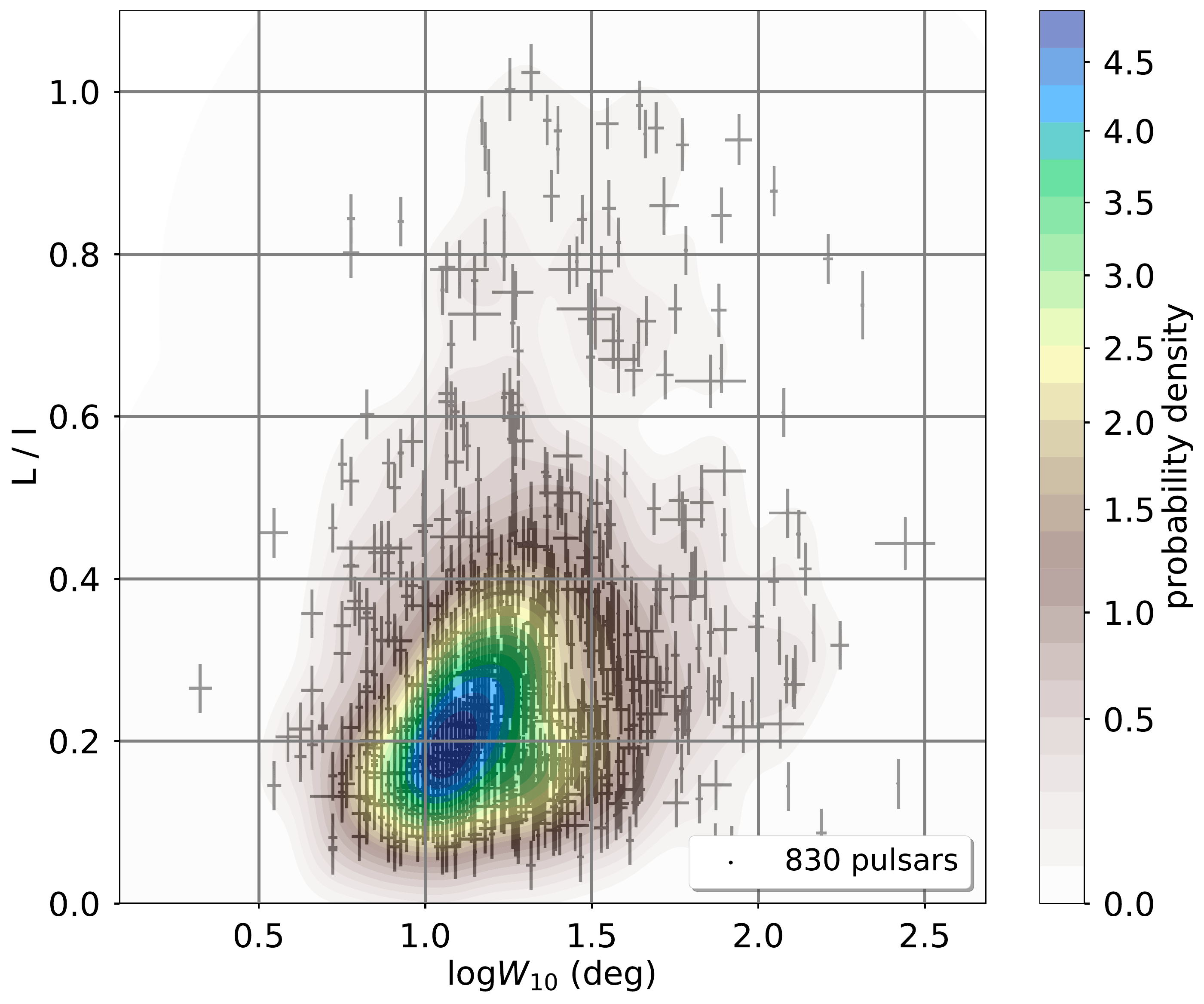}
\caption{The distribution of the LPFs and their correlations with spin-down energy (left panel), spectral index (middle panel), and pulse width $W_{10}$ (right panel). 
The probability densities are indicated by the color bars (smoothing parameters are chosen similarly as in Figure~\ref{fig:PPdot}).
For the spectral index, only pulsars with $m_{SI}<0.13$, for $W_{10}$ only widths with signifiance above three are considered. 
The black solid and red dashed contour in the $\dot{E}-L/I$ plot mark the 90\% level of the population for all pulsars and those that fulfill $W_{10} >30$\,$\deg$, respectively. The OLS-fit of the LPF with $\propto c_{\dot{E}} \log \dot{E}$ and its $1\sigma$ uncertainty are shown with lines and shaded area in darkblue (all pulsars) and darkred (only pulsars with $\dot{E} >10^{32}$\,erg\,s$^{-1}$), see text. }
\label{fig:LPedotW10SI}
\end{figure*}

\begin{figure}
\includegraphics[width=8.5cm]{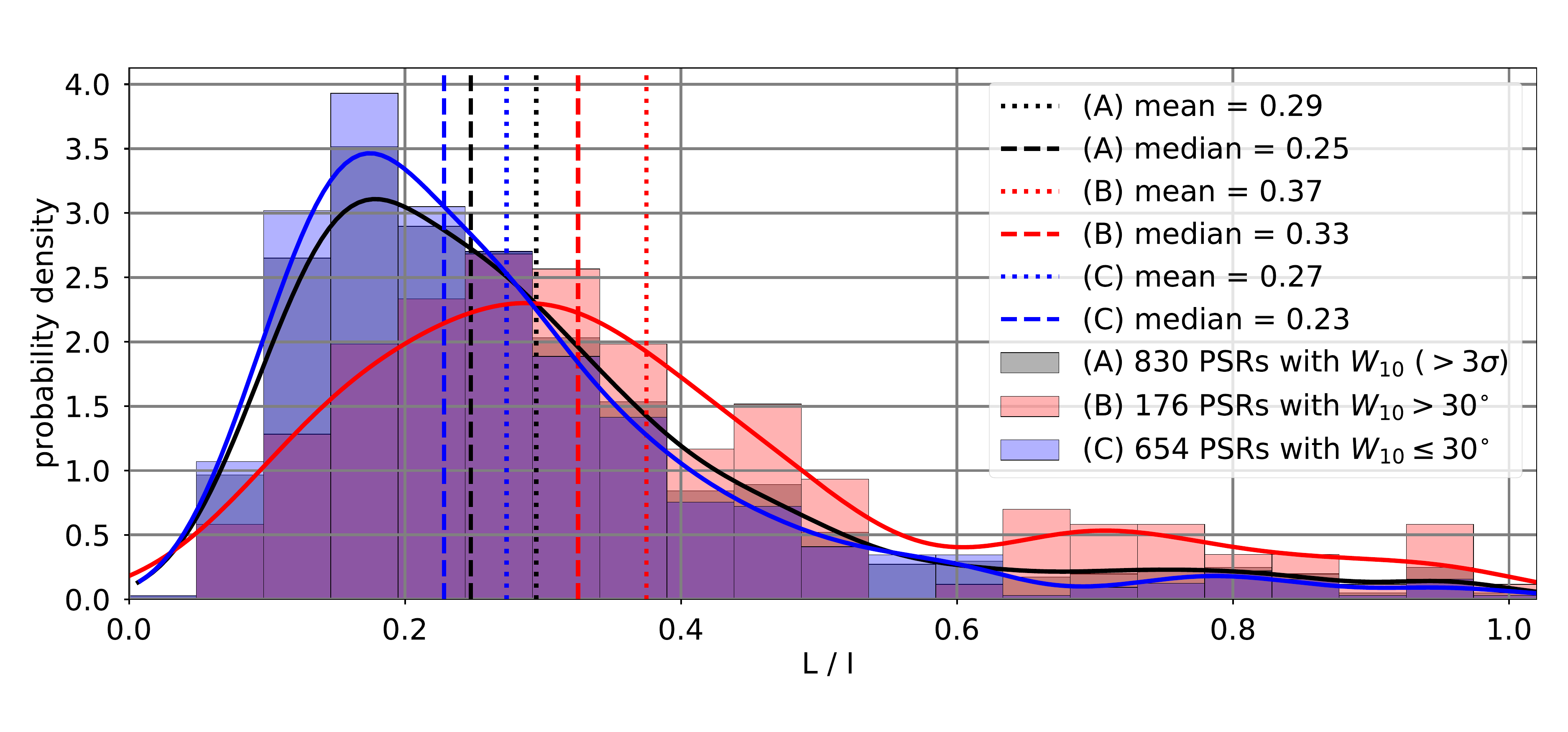}
\caption{The LPF distribution for pulsars filtered by their $W_{10}$ pulse widths according to the figure legend. All histograms are shown with 22 bins for easier comparison. Smoothed curves are produced with a gaussian kernel (bandwidth according to \citet{Silverman1986}, scale factor 0.8). Dashed and dotted lines show the median and mean values of each distribution. }
\label{fig:LPw10}
\end{figure}
\citet{Weltevrede2008b} found a transition from low to high LPFs which happens around $\dot{E}\sim 10^{34} - 10^{35}$\,erg\,s$^{-1}$, and our data allow us to test whether the transition is smooth or abrupt.
We varied our sample with changing $\dot{E}$ selection with a sampling of 0.5 dex. Our Spearman rank correlation of the LPF with $\dot{E}$ is maximized  for $\dot{E} >10^{32}$\,erg\,s$^{-1}$. Table~\ref{tab:polarSpear} lists rank and $p$-values for such selection. 
This $\dot{E}$ selection appears to be in the middle of two ``ears'' (with a much longer  right ``ear'') of the $\dot{E}-$LPF distribution shown in the left panel of Figure~\ref{fig:LPedotW10SI}. The existence of the left ``ear'' (which is due to the width dependency as we discuss below) is the likely reason for the obtained rank maximum for pulsars with $\dot{E} >10^{32}$\,erg\,s$^{-1}$.  
Using only pulsars fulfilling that criterion, we repeated the non-weighted linear fits in log space, resulting in the violet cross ($\dot{E}$ fit) and yellow cross ($\log P$ and $\log \dot{P}$ fits) in Figure~\ref{fig:LPEdotfit}.
The dependency $LPF \propto c_{\dot{E}} \log \dot{E}$ changes from $c_{\dot{E}} = 0.071 \pm 0.005 (1\sigma) $ (1054 pulsars) to $c_{\dot{E}} = 0.134 \pm 0.007 (1\sigma) $ (684 pulsars with $\dot{E} >10^{32}$\,erg\,s$^{-1}$).
Whilst the high LPFs with high $\dot{E}$ appeared to be happen relatively sudden in the data from \citet{Weltevrede2008b}, the distribution in our more comprehensive data set looks relatively smooth if the left ``ear'' of the distribution is ignored (Figure~\ref{fig:LPedotW10SI}).
Interestingly, the correlation with $\dot{E}$ appears to be even stronger for the vector-added L$^*$ (see Table~\ref{tab:polarSpear}).\\

There is also a correlation of the LPF with the spectral index (middle panel in Figure~\ref{fig:LPedotW10SI} and  Table~\ref{tab:polarSpear}). 
The ``tilt'' in that distribution, in particular of its densest part, (Figure~\ref{fig:LPedotW10SI}) indicates that this is genuine additional dependency and not due to the common dependency of the LPF and the spectral index on $\log \dot{E}$. However, unfortunate sample selection effects could, in principle, also mimic such a dependency. As a visual check, we simulate 100 expected pulsar distributions assuming the two parameters (LPF and spectral index) are each only dependent on the respective observed $\log \dot{E}$ probability distributions (see Appendix~\ref{app:lpfsi}). No ``tilt'' is found in these TPA-sample based simulations, strengthening the argument for genuine relation where a flatter spectral index is correlated with a higher LPF. 
A high LPF can be interpreted as the visibility of only one of the two polarization modes, which can be caused by little mixing of the OPMs, for example, because of quickly diverging propagation vectors due to different refractive indices, as noted above. Another explanation for one-mode prevalence could be intrinsically discrepant intensities of the two modes in addition to them having slightly different spectral indices \citep{Karastergiou2005}. Such a combination could explain a high LPF as well as a flatter spectral index in Stokes I (due to the additive  contribution of both modes accross the frequency band).
The investigation of expected polarization fraction changes over frequency are the subject of a detailed study by Oswald et al. (in preparation).\\

In addition to the correlations between the LPF and $\dot{E}$ and the spectral index, the  Spearman rank correlation coefficient and $p$-value indicate another correlation with the pulse width, $W_{10}$ (right panel in Figure~\ref{fig:LPedotW10SI} and  Table~\ref{tab:polarSpear}).
If one selects $W_{10} >30$\,$\deg$ (176 pulsars), the peak density in the $\dot{E}$-LPF distribution in Figure~\ref{fig:LPedotW10SI} shifts from ($\sim 2\times 10^{32}$\,erg\,s$^{-1}$, $\sim 0.18$) to ($\sim 10^{31}$\,erg\,s$^{-1}$, $\sim 0.3$), illustrated by the respective black and red dashed 90\% contour levels.
Figure~\ref{fig:LPw10} shows the shifted probability density for $W_{10} >30$\,$\deg$ in form of a histogram, where the median and mean LPF values are larger by 10\% in comparison to the sample with $W_{10} \leq 30$\,$\deg$. 
The $W_{10}$-LPF correlation gets weaker if $\dot{E} >10^{32}$\,erg\,s$^{-1}$, it gets stronger for $\dot{E} <10^{32}$\,erg\,s$^{-1}$ (Table~\ref{tab:polarSpear}). 
Since pulsars have decreasing pulse widths with increasing periods (e.g., \citealt{Posselt2021}), i.e. with (on average) decreasing $\dot{E}$, this indicates that wider pulse profiles tend to have larger linear polarization \emph{regardless of $\dot{E}$}. 
This could point towards the influence of refraction or scattering in these pulsars' magnetospheres. Refraction would be expected to (i) widen pulsar beams, and to (ii) influence only the ordinary plasma mode of the radio emission coming from an emission height which is different for different field lines (e.g., \citealt{Weltevrede2008b}). Depending on where the two emission modes leave the influence of the magnetosphere, this could lead to the prevalence of one polarization mode (the O-mode), i.e., a larger LPF, even for low $\dot{E}$ pulsars.\\

\section{Summary}
\label{sec:sum}
In this work, we presented comprehensive and homogeneous pulse profile measurements for over 1200 non-recyled pulsars based on the time-averaged TPA data obtained with MeerKAT. 
This unprecedented large data set of radio pulsar parameters includes 
(i) 254 new rotation measures,   
(ii) flux measurements at eight frequency bands as well as in the total 775\,MHz band, centered at 1.3\,GHz,
(iii) spectral indices and pseudo-modulation indices to filter for scintillating sources, 
(iv) luminosities, 
(v) polarization fractions (in the total frequency band and in the eight frequency bands), in particular continuum-added and vector-added linear polarization, circular and absolute circular polarization. We also present  estimates of what fraction of the on-pulse region has polarization above specific levels. 
We show that the TPA pulsars constitute a representative sample of the currently known non-recycled rotation powered radio pulsar population.
Our catalogue of measurement values accompanies the public data release of the TPA census.
The public data encompass the time-averaged data of the TPA census observations with eight frequency channels and full polarization information.\\

The outstanding potential of the TPA data for pulsar population studies is demonstrated by our correlation study of three example properties.
We show that the spectral index, the (pseudo) luminosity, and LPF are all clearly dependent on the spin-down power. We describe these dependencies in form $\propto c_{\dot{E}} \times \log{\dot{E}}$, where $c_{\dot{E}}=0.17 \pm 0.02$,  $0.15 \pm 0.02$, $0.134 \pm 0.07$ for the spectral index, the natural logarithm of the luminosity, and the LPF, respectively (all $1\sigma$ statistical uncertainties, the LPF value is for pulsars with $\dot{E} > 10^{32}$\,erg\,s$^{-1}$).
The found general behaviours are consistent with the understanding of the canonical pulsar for which one expects a flatter energy distribution of the emitting particles and higher emission heights with increasing $\dot{E}$. 
However, the found $\dot{E}$-dependency of the luminosity, for example, is much shallower that what one might  expect from conventional formulas describing the $\dot{E}$-dependency of the potential drop over the polar cap region. Our value for the luminosity is  also different from what is typically assumed in population synthesis models.
The broad distributions of pulsars in the considered parameter spaces indicate the influences of other pulsar properties. For the LPF, we show that there are also correlations with the spectral index and with the pulse width. The former can be explained with slightly different spectral indices and intensities of the two orthogonal emission modes. The latter can be summarised as the presence of large LPFs for large widths.
It can be explained by the influence of refractive processes, preventing depolarization. Refraction affects the O-mode more strongly, widening its pulse profile, preventing mixing, and resulting in the observed pulse being dominated by the O-mode.
Although the TPA frequency band of 750\,MHz is wide, we emphasize that the frequency evolution of pulsars can change the observed relationships depending on the chosen observing frequencies.  
On-going TPA observations use the MeerKAT Ultra High Frequency band ($0.58-1.02$\,GHz), expanding the frequency range for future studies of such effects, and allowing additional ISM-related studies.\\

As we illustrated by means of the filtering for the pseudo-modulation index for the luminosity, selection effects can alter the interpretation of the TPA measurements and need to be carefully accounted for. 
Care is also necessary in the interpretation of the flux densities given they are measured at one epoch per pulsar.
Our example correlations just scratched the surface of the possibilities the TPA data provides for the study of pulsar radio emission on a population-level, e.g., with population synthesis models.
In addition, the TPA data enable the identification and potential detailed investigation of individual pulsars with ``exceptional'' properties such as, for instance, the very high absolute circular polarization fraction of 0.56 for the low-$\dot{E}$ pulsar\,J1222-5738.\\     

\section{Data Availability}
\label{sec:thedata}
The online tables~\ref{tab:PSRs}, \ref{tab:onlynewRM},  \ref{tab:PSRfluxes}, and \ref{tab:PSRpolar}, are available as supplementary material in csv format with names Census$\_$Table1.csv, Census$\_$Table2.csv, Census$\_$Table5.csv, and Census$\_$Table6.csv, respectively.\\

We provide the folded pulse profiles of the 1271 pulsars with a peak S/N in the total power $ > 3$ in the frequency-averaged data. The files are in PSRFITS format and contain full polarization, 8 frequency channels, and 1024 bins, together with the ephemerides we used here\footnote{For the few pulsars listed in Section~2.3 where the true period was found to be harmonically related to the ephemeris value, the original ephemerides are not corrected}. 
The complete dataset is available under the DOI\,10.5281/zenodo.7272361.
The same data set also includes the observational properties, as well as the measurement results of this paper, namely Table~\ref{tab:PSRs}, \ref{tab:PSRfluxes}, and \ref{tab:PSRpolar} in csv format.

\section*{Acknowledgements}
The MeerKAT telescope is operated by SARAO, which is a facility of the National Research Foundation, an agency of the Department of Science and Innovation.
MeerTime data are housed and processed on the OzSTAR supercomputer at Swinburne University of Technology, funded
by Swinburne University of Technology and the National Collaborative Research Infrastructure Strategy (NCRIS). 
LO acknowledges the support of Magdalen College, Oxford. AK and BP acknowledge
funding from the STFC consolidated grant to Oxford Astrophysics,
code ST/000488.\\
We acknowledge the use of the following Python packages: 
\texttt{ASTROPY}  \citep{astropy}, 
\texttt{MATPLOTLIB} \citep{matplotlib}, 
\texttt{PANDAS} \citep{pandas,reback2020pandas},
\texttt{SCIPY} \citep{2020SciPy-NMeth}, and 
\texttt{SEABORN} \citep{seaborn}.

\begin{landscape}
\begin{table}
  \caption{The properties of the TPA census pulsars (I) -- fluxes, logarithm of the luminosities, and spectral indices as measured in the data at the observing epoch listed in Table~\ref{tab:PSRs}. Here, $\nu_t$ and $\nu_{\rm chX}$ indicate the frequency of the frequency-averaged and the frequency channel X where the continuum-equivalent fluxes $F_t$ and $F_{\rm chX}$ are obtained, respectively.  The spectral index, SI, is derived from our weighted PL fits, $m_{\rm SI}$ is our pseudo modulation index defined in equation~\ref{eq:modSI1} and indicating the scatter around the PL-fit. For clarity, only values for the first and last frequency channels are shown. $\log L_L$ is the calculated (pseudo) luminosity in the the L-band. The full table is available online.}
\label{tab:PSRfluxes}
  \begin{tabular}{lcccc|ccc|c|ccc|ccc}
    \hline
    PSRJ & $\nu_t$ & $F_t$ & $\sigma_{F_t}$ &
    $\log L_L$ &  
    $\nu_{\rm ch1}$ & $F_{\rm ch1}$ & $\sigma_{F_{\rm ch1}}$ &
    ch2 to ch7 &
    $\nu_{\rm ch8}$ & $F_{\rm ch8}$ & $\sigma_{F_{\rm ch8}}$ & 
    SI & $\sigma_{\rm SI}$ & $m_{\rm SI}$ \\
         & MHz & mJy & mJy
        & erg\,s$^{-1}$ & 
MHz &  mJy & mJy & & MHz & mJy & mJy & 
& &\\
    \hline
J0034-0721 & 1276 & 5.3101 &  0.0060 &   28.6 &     944 &  3.8804 &     0.0193 &  ... &    1634 &  3.3447 &     0.0178 &   -0.14 &       0.01 &      0.356 \\
J0038-2501 & 1299 & 0.3444 &  0.0074 &   27.0 &     944 &  0.9859 &     0.0241 &  ... &   1625 &  0.2003 &     0.0212 &   -3.46 &       0.11 &      0.324 \\
J0045-7042 & 1281 & 0.1050 &  0.0034 &   30.4 &     943 &  0.0928 &     0.0113 &  ... &   1627 &  0.0786 &     0.0090 &   -0.57 &       0.19 &      0.331 \\
J0045-7319 & 1285 & 0.0680 &  0.0056 &   30.3 &     945 &  0.0749 &     0.0160 &  ... &        &         &            &    1.03 &       0.34 &      0.222 \\
J0108-1431 & 1294 & 1.2641 &  0.0074 &   26.6 &     945 &  1.7012 &     0.0215 &  ... &   1627 &  0.6290 &     0.0210 &   -2.11 &       0.04 &      0.268 \\
J0111-7131 & 1281 & 0.0232 &  0.0014 &   29.8 &     943 &  0.0231 &     0.0038 &  ... &   1627 &  0.0274 &     0.0038 &   -0.06 &       0.30 &      0.136 \\
J0113-7220 & 1276 & 0.2527 &  0.0082 &   30.8 &     944 &  0.4112 &     0.0272 &   ... &  1625 &  0.1540 &     0.0231 &   -2.13 &       0.19 &      0.100 \\
J0131-7310 & 1285 & 0.0492 &  0.0013 &   30.1 &     943 &  0.0902 &     0.0042 &  ... &   1625 &  0.0350 &     0.0035 &   -2.12 &       0.14 &      0.049 \\
J0133-6957 & 1276 & 0.1152 &  0.0113 &   27.7 &     944 &  0.7650 &     0.0351 &  ... &        &         &            &         &            &            \\
J0134-2937 & 1292 & 3.0392 &  0.0115 &   31.1 &     944 &  4.5078 &     0.0405 &  ... &   1625 &  1.8754 &     0.0282 &   -1.28 &       0.02 &      0.099 \\
...\\
    \hline
  \end{tabular}
 \end{table}

\begin{table}
  \caption{The properties of the TPA census pulsars (II) - polarizations  measurements (see also Section~\ref{sec:thedata}). The (debiased) linear, vectorial linear, circular, and absolute circular polarization fractions and their uncertainties for the frequency-averaged (``$_t$'') data are listed in columns $2-8$. Note that all fraction uncertainties include a 3\% systematic error and that the polarization fractions are \emph{not} filtered for significance. The following 6 columns list the percentages of the pulse that have linear, LP, (or absolute circular, aCP) polarization fractions above 30\%, 60\%, 90\% indicated by PFLP30, PFLP60, PFLP90, respectively (or 10\%, 25\%, 40\% indicated by PFaCP10, PFaCP25, PFaCP40). Their uncertainties are obtained from considering the $1\sigma$ \emph{statistical} uncertainties of the polarisation fractions. Since the latter can be very small, often the same PF$-$value is measured and this is indicated for completeness by a formal ``0'' uncertainty. For each of the eight frequency channels, the full online table also includes the respective measurements corresponding to columns 2-8 if these measurements were successful.}
  \label{tab:PSRpolar}
  \begin{tabular}{lccccccc|ccc|ccc c}
    \hline
    PSRJ & ${L/I}_t$ & $\sigma_{{L/I}_t}$ & ${L^*/I}_t$ & $\sigma_{{L^*/I}_t}$ & ${V/I}_t$ & $\sigma_{{V/I}_t}$ & ${|V|/I}_t$ &   PFLP30$_t$ & PFLP60$_t$ & PFLP90$_t$ & PFaCP10$_t$ & PFaCP25$_t$ & PFaCP40$_t$ & ch1 to ch8 \\
    \hline
     &  &  &  &  &  &  &   & \% & \% & \% &  \% & \% & \%
\\
J0034-0721 &0.156 &0.030 &0.068 &0.031 &0.044 &0.030 &0.037  & 30$_{-0}^{+0}$ &  \\
J0038-2501 &0.174 &0.031 &0.078 &0.119 &0.031 &0.031 &0.014  &  \\
J0045-7042 &0.402 &0.036 &0.200 &0.132 &0.021 &0.036 &0.011  & 100$_{-0}^{+0}$ & \\
J0108-1431 &0.781 &0.030 &0.767 &0.046 &0.114 &0.030 &0.097  & 100$_{-0}^{+0}$ &81$_{-3}^{+4}$ & \\
J0113-7220 &0.171 &0.031 &0.073 &0.061 &-0.043 &0.031 &0.021 &  \\
J0131-7310 &0.178 &0.036 &0.159 &0.096 &-0.027 &0.036 &0.032 &  \\
J0133-6957 &0.136 &0.034 &0.072 &0.107 &-0.093 &0.034 &0.048 &  \\
J0134-2937 &0.501 &0.030 &0.444 &0.034 &-0.206 &0.030 &0.194 & 86$_{-6}^{+0}$ &34$_{-5}^{+22}$ & &74$_{-0}^{+2}$ &62$_{-3}^{+0}$ &\\
J0151-0635 &0.367 &0.030 &0.132 &0.034 &0.002 &0.030 &0.059  &71$_{-2}^{+0}$ & \\
J0152-1637 &0.150 &0.030 &0.129 &0.031 &-0.016 &0.030 &0.060 & & & &14$_{-3}^{+3}$ & \\ 
...\\
\hline
\end{tabular}
 \end{table}
\end{landscape}




\bibliographystyle{mnras}
\bibliography{TPApops}



\appendix

\section{The LPF and the spectral index}
\label{app:lpfsi}
\begin{figure*}
\includegraphics[width=15cm]{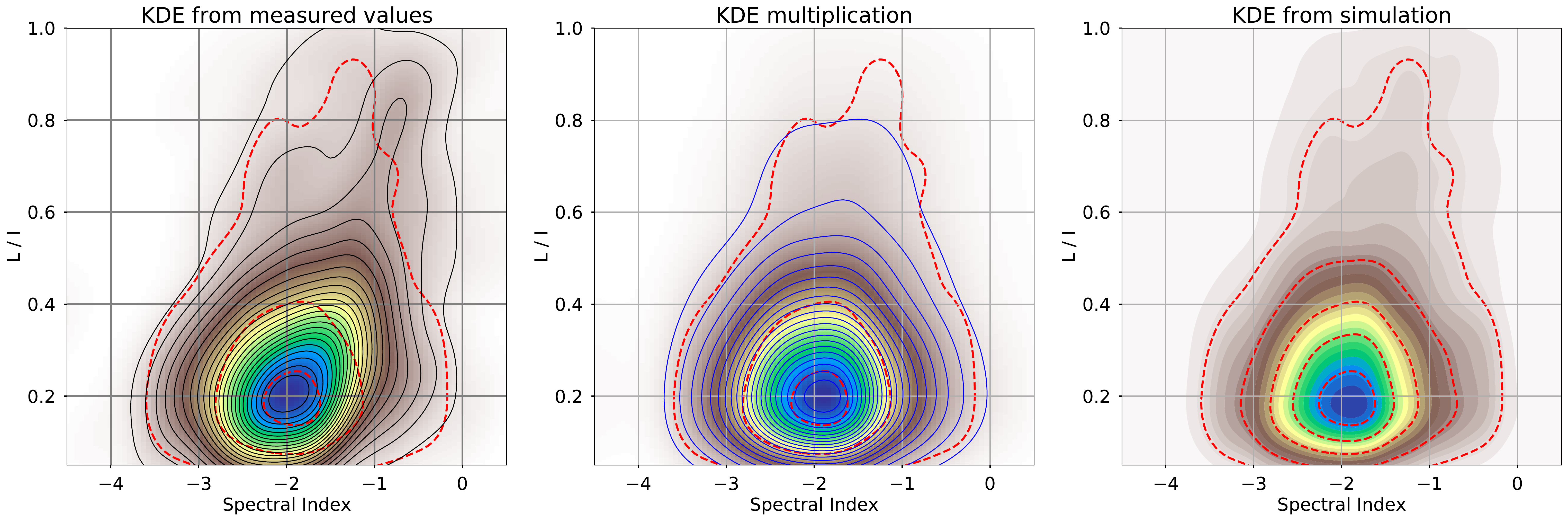}
\caption{The correlation of the LPF--spectral index distribution. 
The left panel is the probability density plot of the measured values from Figure~\ref{fig:LPedotW10SI} with 25 of its levels marked in black (all smoothed with Gaussian kernel, smoothing bandwidth according to \citet{Silverman1986}, scale factor 0.8). 
The middle panel is the product of the by-90\,degrees-rotated (smoothed)  $\log \dot{E}-$ spectral index distribution from Figure~\ref{fig:SIEdotdist} and the (smoothed) $\log \dot{E}-LPF$ distribution from  Figure~\ref{fig:LPedotW10SI} with 15 of its levels marked in blue.
The right panel is the smoothed (Gaussian kernel, smoothing bandwidth according to \citet{Silverman1986}, scale factor  1.6) result of using $\sim 100,000$ independent $\log \dot{E}-$ spectral index and  $\log \dot{E}-LPF$ samples drawn from the probability distributions of the measured values (see text) together with the 10\%, 30\%, 50\%, 70\% and 90\% contour levels. The 10\%, 50\% and 90\% contour levels of this simulated pulsar sample are also overplotted in the left and middle panels with the red dashed contours.}
\label{fig:LPSI}
\end{figure*}

The LPF is found to be correlated with $\log \dot{E}$ as well as with the spectral index (Section~\ref{sec:pol} and Figure~\ref{fig:LPedotW10SI}). At the same time, the spectral index is also correlated with $\log \dot{E}$ (Section~\ref{sec:SI} and Figure~\ref{fig:SIEdotdist}).
As both LPF and spectral index depend on $\log \dot{E}$, one could suspect that the spectral index-LPF correlation is merely an unlucky coincidence due to a selection effect and the common dependency. 
Assuming this is the case means implicitly assuming that the two distributions, $\log \dot{E}$--LPF, $\log \dot{E}$ -- spectral index, are independent. We tested this visually in two different ways. Firstly, we multiplied the two (resampled) probability density distributions (one rotated), making sure that the same sampling in $\log \dot{E}$ (250 resample points) is used. The result is shown in the middle panel of Figure~\ref{fig:LPSI}. Although a ``tail'' towards high LPF is visible, it is much more averaged over the spectral index, and the whole distribution is not ``tilted'', i.e., there is no obvious correlation.
Secondly, we used the observed distributions to draw independent samples. Dividing the $\log \dot{E}$ range into 9 bins (with interval borders of 27.6, 30.6, 31.6, 32.6,33.1, 33.6,34.1,34.6, 35.6, 37.6), we produced \emph{empirical} probability density distributions for the LPFs and for the spectral indices enabling independent samples of these two parameters. 
We shifted the $\log \dot{E}$ interval borders after each TPA-sized sample by tiny amounts to avoid aliasing effects. The final sample has 
105,288 independent LPF--spectral index pairs that are smoothed to produce the right panel in Figure~\ref{fig:LPSI}. The respective contours indicate a similar shape as the product of the two smoothed probability density distributions, whilst the actual, measured and smoothed LPF--spectral index distribution appears tilted in comparison. 

\bsp	
\label{lastpage}
\end{document}